\pdfoutput=1
\documentclass[fleqn,10pt]{wlscirep}

\usepackage{multicol}
\usepackage[version=3]{mhchem}
\usepackage{multicol,lipsum,graphicx,float}
\usepackage{xcolor, colortbl}
\newcounter{firstbib}
\usepackage{footnote}
\usepackage{lineno}
\makesavenoteenv{tabular}
\makesavenoteenv{table}
\definecolor{LightCyan}{rgb}{0.88,1,1}
\newcolumntype{a}{>{\columncolor{LightCyan}}c}

\makeatletter
\newcommand\footnoteref[1]{\protected@xdef\@thefnmark{\ref{#1}}\@footnotemark}
\makeatother
\title{Detection of a Westward Hotspot Offset in the Atmosphere of a Hot Gas Giant CoRoT-2b}

\author[1,2,3,*]{Lisa Dang}
\author[1,2,3,4]{Nicolas B. Cowan}
\author[1,2,3,4]{Joel C. Schwartz}
\author[5]{Emily Rauscher}
\author[6]{Michael Zhang}
\author[7]{Heather Knutson}
\author[8]{Michael Line}
\author[9]{Ian Dobbs-Dixon}
\author[10]{Drake Deming}
\author[11]{Sudarsan Sundararajan}
\author[12]{Jonathan J. Fortney}
\author[13]{Ming Zhao}

\affil[1]{Department of Physics, McGill University, 3600 University St, Montr\'eal, QC H3A 2T8, Canada}
\affil[2]{McGill Space Institute (MSI), McGill University, 3550 rue University, Montr\'eal, QC H3A 2A7, Canada}
\affil[3]{Institut de Recherche sur les Exoplan\`etes (iREx), Universit\'e de Montr\'eal, C.P. 6128 Succ. Centre-ville, Montr\'eal, QC H3C 3J7, Canada}
\affil[4]{Department of Earth and Planetary Sciences, McGill University, 3450 rue University, Montr\'eal, QC H3A 2A7, Canada}
\affil[5]{Department of Astronomy, University of Michigan, 311 West Hall, 1085 South University, Ann Arbor, MI 48109, USA}
\affil[6]{Division of Physics, Mathematics \& Astronomy, California Institute of Technology, 1200 E California Blvd MC 249-17, Pasadena, CA 91125 USA}
\affil[7]{Division of Geological \& Planetary Sciences, California Institute of Technology, 1200 E California Blvd MC 150-21, Pasadena, CA 91125 USA}
\affil[8]{School of Earth and Space Exploration, Arizona State University, 781 South Terrace Road, Tempe, AZ 85281, USA}
\affil[9]{New York University Abu Dhabi, PO Box 129188, Abu Dhabi, UAE}
\affil[10]{Department of Astronomy, University of Maryland, College Park, MD 20742-2421 USA}
\affil[11]{Department of Computer Science and Engineering, Amrita University, India}
\affil[12]{Department of Astronomy and Astrophysics, University of California, Santa Cruz, CA 95064, USA}
\affil[13]{Astronomy \& Astrophysics, Pennsylvania State University, University Park, PA 16802, USA}

\affil[*]{Corresponding author:  \href{mailto:lisa.dang@physics.mcgill.ca}{lisa.dang@physics.mcgill.ca}}


\begin{abstract}

Short-period planets exhibit day--night temperature contrasts of hundreds to thousands of degrees K. They also exhibit eastward hotspot offsets whereby the hottest region on the planet is East of the substellar point \cite{2007Natur.447..183K}; this has been widely interpreted as advection of heat due to eastward winds \cite{2002A&A...385..166S}. We present thermal phase observations of the hot Jupiter CoRoT-2b obtained with the IRAC instrument on the Spitzer Space Telescope. These measurements show the most robust detection to date of a \emph{westward} hotspot offset of $23 \pm 4$ degrees, in contrast with the nine other planets with equivalent measurements \cite{2012ApJ...747...82C, 2012ApJ...754...22K, 2013MNRAS.428.2645M, 2014ApJ...790...53Z, 2015ApJ...811..122W, 2016ApJ...823..122W, 2016MNRAS.455.2018D, 2017AJ....153...68S}. 
The peculiar infrared flux map of CoRoT-2b may result from westward winds due to non-synchronous rotation \cite{2014ApJ...790...79R} or magnetic effects \cite{2014ApJ...794..132R, 2017NatAs...1E.131R}, or partial cloud coverage, that obscures the emergent flux from the planet's eastern hemisphere \cite{2013ApJ...776L..25D, 2016ApJ...828...22P, 2016A&A...594A..48L, 2017arXiv170907459R}. Non-synchronous rotation and magnetic effects may also explain the planet's anomalously large radius \cite{2011Guillot,2014ApJ...794..132R}. On the other hand, partial cloud coverage could explain the featureless dayside emission spectrum of the planet \cite{2013ApJ...763...25M, 2014ApJ...783..113W}. If CoRoT-2b is not tidally locked, then it means that our understanding of star--planet tidal interaction is incomplete. If the westward offset is due to magnetic effects, our result represents an opportunity to study an exoplanet's magnetic field. If it has Eastern clouds, then it means that our understanding of large-scale circulation on tidally locked planets is incomplete.
\end{abstract}

\begin{document}

\flushbottom
\maketitle

\thispagestyle{empty}
\begin{multicols}{2}

\section*{Main Text}

Amongst the plethora of known hot Jupiters, the CoRoT-2 system stands out from the rest for three reasons: its remarkably active host star, its unusual inflated radius, and its puzzling emission spectrum. In addition to these anomalous features, previous observations of the CoRoT-2 system show a gravitationally bound stellar companion candidate, 2MASS J19270636+0122577.

CoRoT-2b's optical phase curve obtained by the \textit{CoRoT} mission has previously been studied \cite{2009Alonso, 2010A&A...513A..76S} and yielded an upper limit on the planet's geometric albedo of 0.12. Later near-infrared (NIR) and mid-infrared (mid-IR) observations, acquired with ground-based \cite{2010Alonso} and space-based \cite{2010Gillon, 2011Deming, 2014ApJ...783..113W} instruments, have shown that the planet's emission spectrum could not be explained by conventional solar composition spectra or by a blackbody. Several scenarios were invoked to interpret the perplexing spectrum including the presence of silicate clouds affecting the mid-IR emission of the planet \cite{2013ApJ...763...25M} and optically thick dayside clouds or a vertically isothermal atmosphere to explain the lack of features in the data acquired by the Wide Field Camera 3 (WFC3) on board of the Hubble Space Telescope (\textit{HST})\cite{2014ApJ...783..113W}.

\begin{figure*}[!htpb]
	\includegraphics[width=0.95\textwidth]{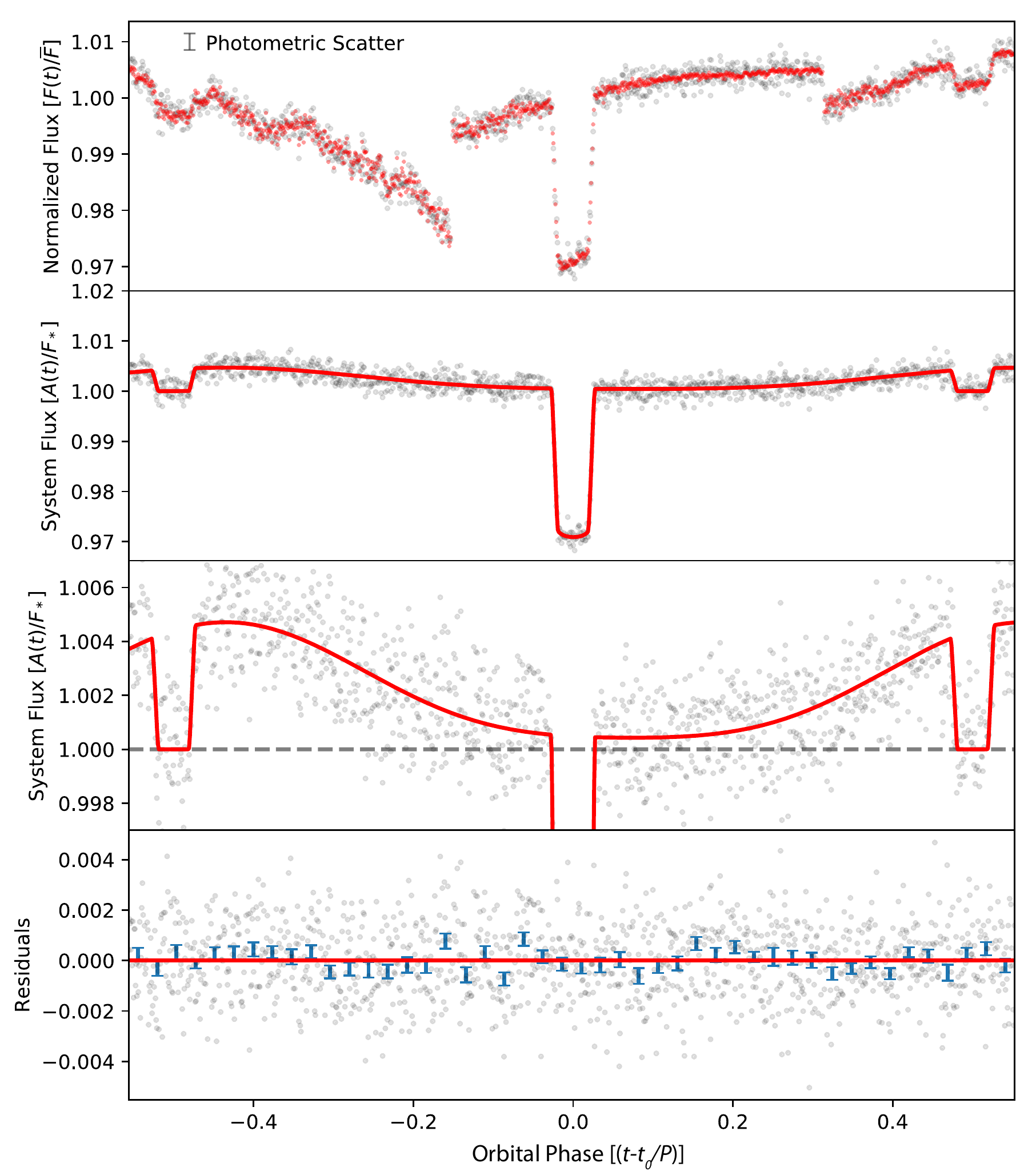}
	\caption{\textbf{Fit model to Spitzer phase observation of CoRoT-2b.} The top panel shows the normalized raw photometry obtained from \textit{Spitzer} observations of the CoRoT-2 system (gray dots) and the fit with greatest Bayesian Evidence, instrumental systematics modeled as a 2$^{nd}$ order polynomial and with no stellar variability (red dots). The error on the photometry measurements presented (top-left) is the photometric scatter, $\sigma _F$, which is estimated using a Markov Chain Monte Carlo (MCMC). The second panel shows the photometry corrected for detector systematics (gray dots) and the most probable astrophysical signal (red line). The third panel is a zoomed-in version of the second panel to better show the phase variation--we can see the peak of the phase variation occurring after the secondary eclipse. This corresponds to a \textit{westward} offset of the brightest longitude on CoRoT-2b. The bottom panel shows the residuals obtained from subtracting the most probable astrophysical model from the corrected photometry (gray dots) and the binned residuals with a bin size of $\sim$ 1 hour (blue points) and the errorbars are the error on the mean of each bin. }
    \label{Fig. 1 Phase Curve}
\end{figure*}

We present new phase observations of the CoRoT-2 system (PID 11073; PI Cowan) acquired with the Infrared Array Camera (IRAC) on the Spitzer Space Telescope with the 4.5 $\mu m$ channel on January 3--5, 2016. To minimize the impact of the visual companion in our analysis, we subtract it from our images. We combine the data into bins of 64 frames and detrend the lightcurve for detector systematics using various detrending strategies explained in more details in Methods.

We experiment with various decorrelation methods and fit for both the astrophysical models and the time-correlated systematics simultaneously. Most importantly, we find that the phase curve exhibits a westward hotspot offset. The offset is detected regardless of the planetary phase variation model, with and without imposing priors on the phase variation coefficients, and using the trimmed and untrimmed data. Additionally, we find the westward offset to be robust to the different photometry extraction schemes.

Our analysis shows a phase curve peak occurring $2.7 \pm 0.4$ hours after the time of secondary eclipse and a phase variation amplitude, from peak to trough, of $(4.3 \pm 0.2) \times 10^{-3}$. Using our observations, including two secondary eclipses and one transit, we measure a secondary eclipse depth, and transit depth of $(4.3 \pm 0.2) \times 10^{-3}$ and $(2.87 \pm 0.03) \times 10^{-2}$, respectively (see Supplemetary Tables for the complete list of parameter values). We find a smaller eclipse depth than previously reported using channel 2 \textit{Spitzer} IRAC data \cite{2010Gillon,2011Deming}. Our new measurement decreases the abnormally deep 4.5 $\mu m$ planet-star contrast previously reported. Fitting the emission spectrum of the planet, we infer an optical geometric albedo of $0.08 \pm 0.04$, which is consistent with the published upper limit using \textit{CoRoT} data \cite{2009Alonso, 2010A&A...513A..76S}.

\begin{figure}[H]
	\centering
	\includegraphics[width=\linewidth]{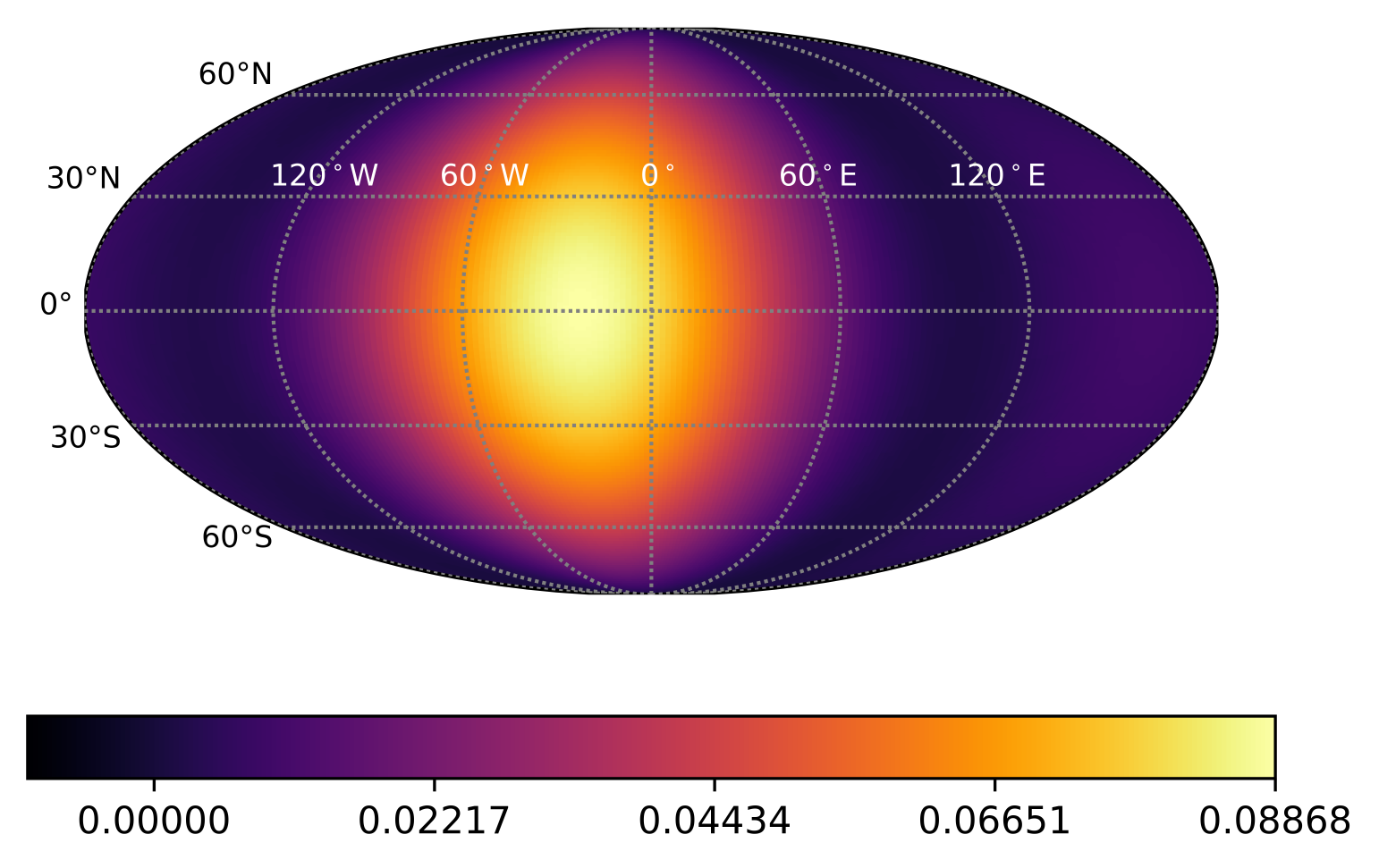}
	\caption{\textbf{Surface brightness map of CoRoT-2b.} This is the 1D longitudinal brightness obtained from the phase variations converted into a surface brightness map of CoRoT-2b. The surface brightness is scaled in units of stellar flux. The peak of the phase variation after the secondary eclipse shown in Fig. \ref{Fig. 1 Phase Curve} corresponds to the westward offset of the brightest longitude on the planet.}
    \label{Fig2 Surface Brightness}
\end{figure}

Full-orbit phase curves at 4.5~$\mu$m have so far been published for nine exoplanets on circular orbits ---all of them exhibit phase offsets consistent with an eastward hotspot offset or no offset: WASP-12b \cite{2012ApJ...747...82C}, HD 189733b \cite{2012ApJ...754...22K}, WASP-18b \cite{2013MNRAS.428.2645M}, HD 209458b \cite{2014ApJ...790...53Z}, WASP-14b \cite{2015ApJ...811..122W}, WASP-19b \cite{2016ApJ...823..122W}, HAT-P-7b \cite{2016ApJ...823..122W}, 55~Cancri~e \cite{2016MNRAS.455.2018D}, and WASP-43b \cite{2017AJ....153...68S}. The westward hotspot offset of $23 \pm 4$ degrees we measure for CoRoT-2b in the mid-IR therefore makes it unique. We note that another westward offset has previously been observed for Kepler-7b in optical \textit{Kepler} data, attributed to reflected light from inhomogeneous clouds \cite{2013ApJ...776L..25D}. We derive the longitudinal 4.5 $\mu$m brightness map of CoRoT-2b shown in Figure \ref{Fig2 Surface Brightness}. 

\begin{figure}[H]
	\centering
	\includegraphics[width=\linewidth]{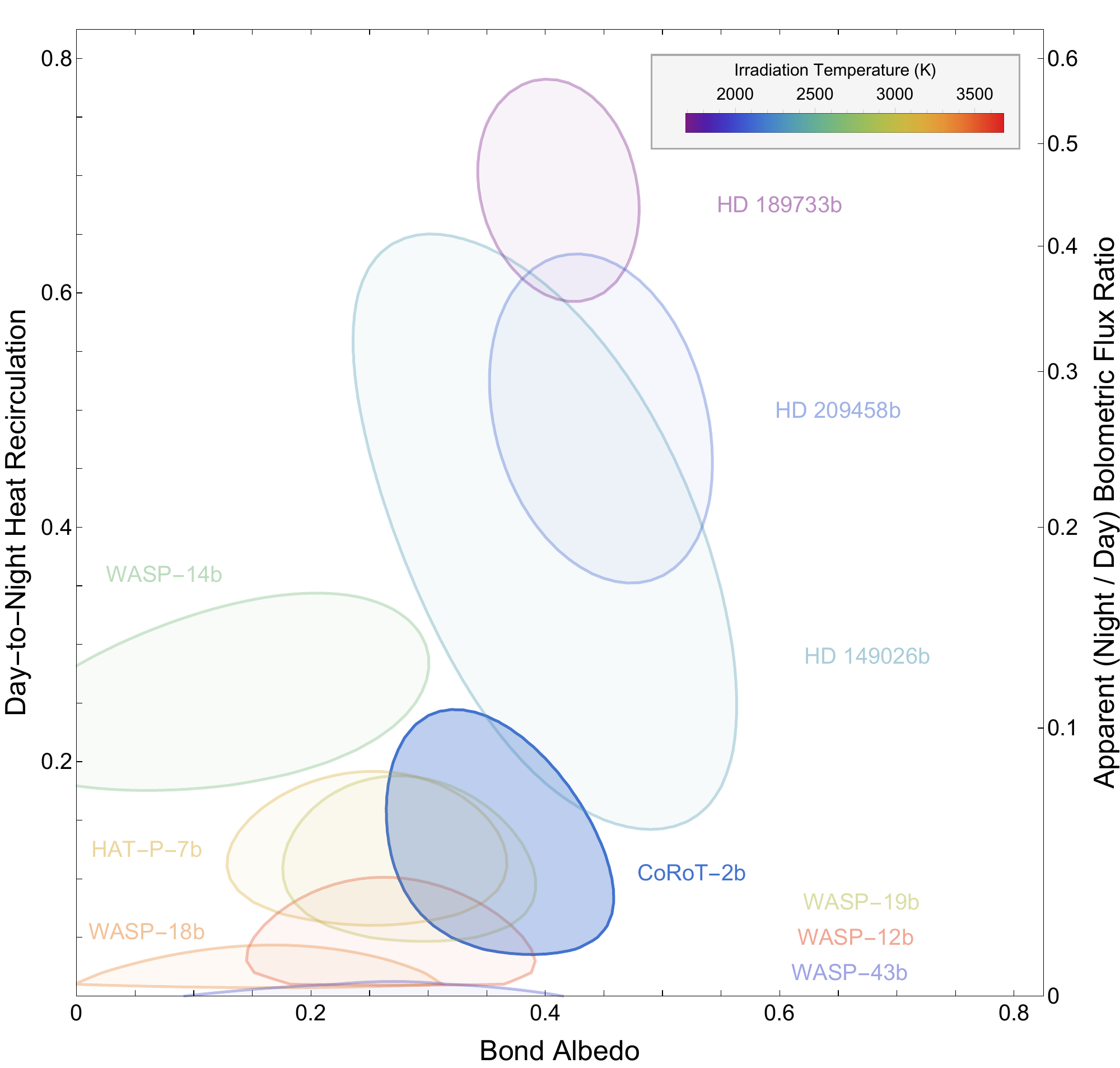}
	\caption{\textbf{Energy budget of CoRoT-2b and other hot Jupiters.} The 1$\sigma$ confidence region for Bond albedo and day-to-night heat recirculation efficiency of CoRoT-2b and other hot Jupiters. The color of each region denotes the irradiation temperature. Given the day-night temperature difference, CoRoT-2b lies in the low-recirculation efficiency region.}
    \label{Fig3: Energy Budget}
\end{figure}

We derive the day-to-night heat recirculation efficiency, $\epsilon$, and Bond albedo, $A_{\rm B}$, of CoRoT-2b using all existing transit and eclipse depths in the infrared along with our best fit phase amplitude and offset, shown in Figure \ref{Fig3: Energy Budget}. Given the young age of the system (100--300 Mya) and the inflated radius of CoRoT-2b, we expect the planet to experience internal heating from residual heat of formation or tidal heating, but this should be dwarfed by the external heating of the star. The $\sim$35\% Bond albedo of CoRoT-2b shown in Figure \ref{Fig3: Energy Budget} is greater than its low optical geometric albedo of $12 \pm 2$ \% \cite{2009Alonso, 2010A&A...513A..76S}, suggesting significant NIR albedo, as reported for other hot Jupiters \cite{2015MNRAS.449.4192S}. The day--night temperature contrast is greater than has been inferred for HD 209458b (a hot Jupiter with similar irradiation temperature), suggesting that CoRoT-2b is less effective at transporting heat to its nightside.

\begin{figure*}[!htpb]
	\centering
	\includegraphics[width=\linewidth]{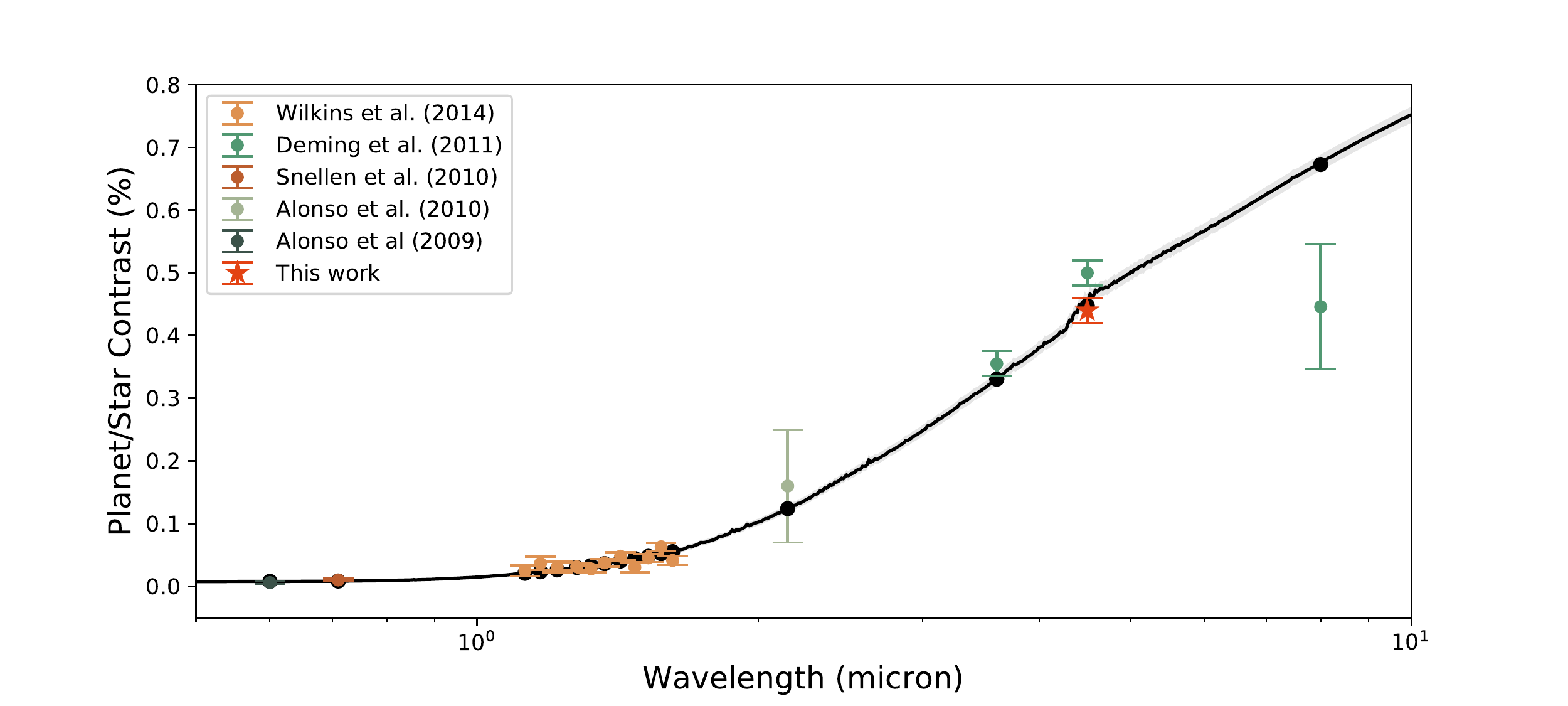}
	\caption{\textbf{Dayside emission spectrum of CoRoT-2b.} Our secondary eclipse depth and error, obtained from our fit using an MCMC, is shown along with previous measurements from \textit{CoRoT} observations \cite{2009Alonso,2010A&A...513A..76S}, \textit{Spitzer} data \cite{2011Deming}, \textit{HST}/WFC3 $\beta$ results \cite{2014ApJ...783..113W} and ground-based NIR measurement with their respective uncertainties \cite{2010Alonso}. The black line is the $1693 \pm 17$ K toy model with a geometric albedo of $0.12\pm 0.02$ and the black dots are the band-integrated eclipse depths (per-datum $\chi ^2 = 1.34$).}
    \label{Fig: Emission Spectrum}
\end{figure*}

The emission spectrum of CoRoT-2b has been difficult to understand since no spectral model could fit all the data within the uncertainties \cite{2014ApJ...783..113W}. Using our new measurement at 4.5 $\mu m$, along with published eclipse depth measurements at other wavelengths, we fit a toy model including thermal emission and reflected light shown in Figure \ref{Fig: Emission Spectrum}, described in Methods. The model with a geometric albedo of 0.12$\pm$0.02 and dayside effective temperature of 1693$\pm$17 K best fits the data with chi-squared per datum of 1.34.

Water vapor is expected in hot Jupiter' atmospheres and therefore we expect to see water absorption features in \textit{HST} data as well as at 4.5 $\mu$m. However, these features are not apparent in the emission spectrum of CoRoT-2b which could mean one of two things: 1) wavelengths in and outside of \ce{H2O} bands are probing the same pressure or 2) they are probing a vertically isothermal region of the atmosphere. For example, optically thick clouds would prevent deeper observations into the atmosphere and could be responsible for the absence of water absorption features \cite{2017arXiv170900349D}. Alternatively, it would mean that infrared emission originates from a vertically isothermal layer of the atmosphere.

Given that the irradiation temperature of CoRoT-2b is similar to that of HD~209458b, global circulation models predict an eastward hot spot shift, due to fast and broad equatorial jets at and near the photosphere \cite{2002A&A...385..166S}. We propose three possible explanations for the westward offset seen on CoRoT-2b: 1) westward winds due to sub-synchronous rotation\cite{2014ApJ...790...79R}, 2) westward winds due to magnetic effects \cite{2014ApJ...794..132R, 2017NatAs...1E.131R}, and/or 3) inhomogeneous clouds that are optically thick in the mid-IR \cite{2013ApJ...776L..25D, 2016ApJ...828...22P, 2016A&A...594A..48L, 2017arXiv170907459R}.  In practice, these scenarios could be causally related, since spin rate affects wind direction \cite{2014ApJ...790...79R}, and wind direction affects cloud patterns \cite{2016A&A...594A..48L, 2017arXiv170907459R}.

We note that the planetary photometry for CoRoT-2b exhibits a broad minimum rather than the distinct trough near or before transit seen in other phase curves. Sub-synchronous rotation not only produce westward atmospheric circulation, but the entire wind and temperature pattern is different than the standard eastward-jet pattern. Simulation of HD 209458 b with sub-synchronous rotation also show a long minimum in the phase curve \cite{2015ApJ...799..241R}.

The high temperature of CoRoT-2b allows for collisional ionization of alkali metals in the atmosphere, hence creating a partially ionized atmosphere. The presence of a deep-seated magnetic field could create temporary directional winds causing atmospheric variability as seen on HAT-P-7b \cite{2016NatAs...1E...4A, 2017NatAs...1E.131R}. We note that the host star is relatively young and spectrally active. CoRoT-2b is then subject to high X-ray and extreme ultraviolet flux (XUV) potentially leading to time-variable photo-ionization. While magnetism in the atmosphere of hot Jupiters is not yet understood, it is conceivable that the photo-ionization could also lead to time-variable coupling of the atmosphere with the magnetic dynamo of the planet. We estimate that the strength of the magnetic field needs to be $B\sim 230$ G to produce significant westward winds (see Supplementary Information). A recent study \cite{2017arXiv170905676Y} found that some hot Jupiters with energetic interiors can have magnetic fields up to 250 G. Since an inflated radius suggests a high entropy interior, it is plausible that CoRoT-2b could have such a strong deep-seated magnetic field. Additionally, the coupling effect of magnetic drag can slow down wind speed \cite{2012ApJ...745..138M} which could explain the low heat redistribution efficiency, as well as the broad minimum in the light curve.

It is well established that some hot Jupiters have inhomogeneous clouds which produce non-trivial reflected phase variations \cite{2013ApJ...776L..25D}. As previously mentioned, Kepler-7b's optical phase curve exhibits a westward offset. Such observations can be caused by reflective clouds located west of the substellar meridian \cite{2016A&A...594A..48L,2017arXiv170907459R} due to nightside clouds advected by the eastward jets. In contrast with Kepler-7b, CoRoT-2b whould have western cloud coverage which require different aerosol formation and transportation mechanisms.

The equilibrium temperature of CoRoT-2b of $1521\pm18$ K  allows for \ce{MnS}, \ce{Cr} and \ce{MgSiO3} clouds to form on the dayside hemisphere \cite{2016ApJ...828...22P}. Inhomogeneous clouds covering the East and night side of the planet with particles large enough to block thermal emission at 4.5 $\mu m$ could explain CoRoT-2b's unusual phase curve \cite{2017arXiv170900349D}. Based on our current understanding, clouds tend to form on the cooler nightside hemisphere. Therefore,  in the presence of westward winds, one would expect eastern cloud coverage \cite{2016A&A...594A..48L}. Alternatively, photochemical hazes (produced on the dayside) coupled with standard eastward jets could result in aerosols located east of the substellar point \cite{2017ApJ...845L..20K}. If inhomogeneous clouds are responsible for the shape of the phase curve, the dayside emission spectrum would be an average of a blackbody spectrum and a clear spectrum. This could explain why no spectral model, so far, could fit all the data within the errors \cite{2016ApJ...829...52F}. However, if this is the case, it is possible that phase curves of other planets are also sculpted by clouds.

All three options are attractive because they might also explain other features of the planet. Asynchronous rotation can lead to tidal heating which, if deposited deep enough, can prevent contraction, explaining the inflated radius of CoRoT-2b \cite{2010ApJ...714....1A}. Magnetic effects are also attractive as they can explain the large temperature difference between the dayside and the nightside of the planet \cite{2012ApJ...745..138M} and the inflated radius \cite{2017arXiv170905676Y}. Partial cloud coverage is appealing because it may explain the anomalous dayside emission spectrum of the gas giant. It should be possible to distinguish between these scenarios with phase curve observations with wider spectral coverage including 4.5 $\mu$m, either with both \textit{Spitzer} channels or \textit{JWST}. Non-synchronous rotation should have an impact on the phase curve at all wavelengths, while unusual cloud coverage may betray itself at short wavelengths dominated by reflected light or by spectral emission features. On the other hand, the circulation due to magnetic effects should be variable on an Alfv\'en timescale of $\tau _A \sim 23$ days \cite{2017NatAs...1E.131R, 2016NatAs...1E...4A} (see Methods for calculation), so a new phase curve at the same wavelength would show a different location for the peak.

The westward offset of CoRoT-2b is another example that hot Jupiters are not all cut from the same cloth and cannot be organized into a simple one-parameter family. More broadly, each scenario outlined above challenges our understanding of short period planets. If the westward hot spot offset is due to non-synchronous rotation, then our understanding of tidal interaction between planets and their host star is not fully understood. If magnetic effects are responsible for the unusual shape of the phase curve, our result would represent one of the few observable effects of a hot Jupiter's magnetic field, allowing further understanding of the magnetic fields of hot Jovians. Lastly, if it is caused by Eastern clouds, then our understanding of large-scale atmospheric circulation on tidally locked planets is incomplete. Hence, an exhaustive understanding of these phenomena is necessary for the characterization of short-period planets, including the potentially habitable variety.
\end{multicols}

\begin{multicols}{2}
\section*{Methods}

\subsection*{Data Source}
We acquired observations of CoRoT-2b \cite{Alonso2008} with the Infrared Array Camera (IRAC) \cite{2004ApJS..154...10F} on the Spitzer Space Telescope \cite{2004ApJS..154....1W} with the 4.5 $\mu m$ channel on January 3-5, 2016, during the \textit{Post-Cryogenic Mission}. The system was observed for approximately 49 hours from shortly before a secondary eclipse to slightly after the next secondary eclipse. We used the subarray mode with 2.0 s exposures (1.92 s effective exposure time) to minimize the data volume and to make the observations as uniform as possible. This generated data cubes of 64 images with $32 \times 32$ pixel ($39" \times 39"$) dimensions. Our observations were divided into 5 Astronomical Observation Requests (AORs) and includes a total of 1374 data cubes covering the full orbit of CoRoT-2. We elect to discard the first and last AOR containing 12 and 4 data cubes, respectively, since they are dithered, placing the target on different pixels than the rest of the data.

\subsection*{Data Reduction}
We convert the pixel intensity from MJy/str to electron counts and mask bad pixels, i.e., 4$\sigma$ outliers and \textit{NaN} pixels. We discard all frames with containing a bad pixel in the vicinity of the target. Observations of CoRoT-2 ($K=10.31$) \cite{2009A&A...506..501C} show the presence of a close-in visual companion, 2MASS J19270636+0122577 ($K=12.03$) \cite{2003yCat.2246....0C}. Due to the proximity of the companion, we experiment with various photometric extraction schemes to remove contamination from the companion and choose the strategy resulting the smallest RMS scatter. We then bin the data into bins of 64 frames before fitting the data (see the Supplementary Information for details about the data reduction and photometry extraction).

\subsection*{Astrophysical Model}
\label{Sec: Model}

We model the measured flux $F_{model}(t)$ as the product of the astrophysical signal $A(t)$ and the detector response $\tilde{D}$, 
\begin{equation}
F_{model}(t) = A(t) \times \tilde{D}.
\end{equation}

Our astrophysical model is the sum of the emitted flux from the host star, $F_*$, and from the planet, $F_p$, as seen by a distant observer
\begin{equation}
A(t) = F_*(t) + F_p(t). 
\end{equation}

To model the occultations, we use the Python package \texttt{batman} \cite{2015Kreidberg}. Using the quadratic limb-darkening model \cite{2002MandelAgol} supported by \texttt{batman}, we obtain the stellar intensity profile during transit $T(t)$.  The secondary eclipse $E(t)$ is modeled using a uniform disk.

\subsubsection*{Stellar Model}

CoRoT-2 is a young active star (100--300 Mya) \cite{Schroter2011} with a rotational period $P_*$ of 4.522$\pm 0.024$ days \cite{2009A&A...493..193L}. Unfortunately, the CoRoT-2 system was not visible from Earth at the time of the \textit{Spitzer} observation, so stellar activity could not be monitored in real time. Typically, stellar variation due to star spots should not have a large effect in the mid-infrared, but given that CoRoT-2b is an active star with a short rotational period, it would be unwise to ignore stellar variability. We use optical observations acquired by \textit{CoRoT} \cite{2009Alonso} to estimate the magnitude of the stellar variation at 4.5$\mu m$ on a 2 days time scale. We find that the stellar flux can vary by 1.1\% (see Supplementary Information). We experiment with and without the inclusion of stellar variability to test the robustness of our fit.

The apparent stellar brightness is modeled as 
\begin{equation}
F_* (t) = \Phi _* (t) + T(t)
\end{equation}
\noindent where $\Phi _* (t)$ is the stellar variability and $T(t)$, as mentioned before is the transit curve. This is modeled as a sinusoid with a period equal to the rotational period of the host star
\begin{equation}
\Phi_* (t) = S_1 \cos \left( \frac{2\pi(t)}{P_*} -S_2  \right) 
\end{equation} 
\noindent where $S_1$  and $S_2$ are the semi-amplitude and the offset of the phase stellar variation model included as free parameters, respectively.

By construction, modeling the stellar variability as a sinusoid is not ideal for optimization purposes since this can lead to degeneracy between stellar and planetary model. In other words, both the phase variation model and the stellar variation model can mimic the shape of the planetary phase variation which explains the large uncertainty obtained for the fits which include a varying stellar brightness (see Supplementary Table 4 and 5). Although a few fits including stellar variability yield eastward offsets, they can be ruled out based on their significantly lower Bayesian Evidence, as described below.

\subsubsection*{Planetary Model}

In the \texttt{batman} package, the time of secondary eclipse $t_e$ is not an explicit parameter, instead it is defined as the time when true anomaly equals $3\pi /2 - \omega$, where $\omega$ is the longitude of periastron. We did not account for the light travel time as it is only a matter on 28.04 seconds and does not affect our analysis.

The planet's flux is given by
\begin{equation}
F_p (t) = \Phi _p (t) \times E(t)
\end{equation}
\noindent where $\Phi _p(t)$ is the phase variation and $E(t)$ is the secondary eclipse. In the \texttt{batman} package, the eclipse $E(t)$ is scaled such that the flux is unity during eclipse and the eclipse depth is given in terms of stellar flux. We re-scaled it such that $E(t) = 0$ during complete occulation and $E(t)=\delta _e$ outside of eclipse.

Previous studies have reported that the orbit of CoRoT-2b is nearly circular \cite{Alonso2008, 2010Gillon} and therefore the phase variation of the planet's apparent brightness can be modeled \cite{2008ApJ...678L.129C} to first order as:
\begin{equation}
\Phi _p (t) = 1 + A \left[\cos\left(\frac{2\pi (t-t_e)}{P}\right) -1\right] + B \sin \left(\frac{2\pi (t-t_e)}{P}\right).
\end{equation}
\noindent and to second order as:
\begin{equation}
\begin{split}
\Phi _p (t) = 1 + A \left[\cos\left(\frac{2\pi (t-t_e)}{P}\right) -1\right] + B \sin \left(\frac{2\pi (t-t_e)}{P}\right) \\ + C \left[\cos\left(\frac{4\pi (t-t_e)}{P}\right) -1\right] + D \sin \left(\frac{4\pi (t-t_e)}{P}\right)
\end{split}
\end{equation}
\noindent where $t_e$ is the time of eclipse center. Note that $\Phi _p (t_e) =1$, which allows us to make the eclipse depth $\delta _e$ an explicit model variable. 

\subsection*{Detector Models}

Photometric data obtained using $Spitzer/IRAC$ exhibit a well-studied instrumental effect due to intrapixel sensitivity variations \cite{2005Charbonneau}. The total number of electron counts varies with small changes in the position of the PSF of the target on the detector. The measured flux variation is dependent on both the sensitivity variation across the detector and the shape and position of the PSF. We experiment with various methods to decorrelate the astrophysical signal from the detector sensitivity. Although the PSF spans many pixels, most of the flux falls in the core of the PSF. Ultimately, we ignore the effect of the PSF widths $\sigma _x$ and $\sigma _y$ on the photometry.

\subsubsection*{2D Polynomial}

Our first approach to correct the intrapixel sensitivity variation is to model the detector systematics as a $n^{th}$-degree polynomial in the centroid $x_0$ and $y_0$:
\begin{equation}
\tilde{D} (x_0, y_0) = D_0  + \sum ^{n} _{i=0} \sum ^{i} _{j=0} c_{ij}(x_0 - \langle x_0\rangle)^{j}(y_0 - \langle y_0\rangle)^{i-j}
\end{equation} 
\noindent where $n$ is the order of the polynomial. The model has $(n+1)(n+2)/2$ parameters and we experiment with polynomials of orders varying from 2 to 5. The shortcoming of this model is the requirement of accurate PSF location on the detector. As mentioned before, binning data improves the precision of centroid measurements. As we do not expect the location of the PSF to change significantly in $\sim$ 2 minutes, it is sensible to bin the centroids by datacube.

\subsubsection*{BLISS Mapping}
In recent years, many researchers have used BiLinear Interpolated Subpixel Sensitivity (BLISS) mapping \cite{2012ApJ...754..136S,2016AJ....152...44I, 2017PASP..129a4001S}. This non-parametric detector response model has the advantage of running quickly in a Markov-Chain-Monte-Carlo (MCMC) routine because the detector model has no explicit parameters. First, BLISS defines a set of locations on the pixel referred to as ``knots''. Then, it divides the astrophysical model from the light curve at each step of the MCMC and averages the residuals surrounding each knot to obtain the detector sensitivity at each location. Finally, it uses the sensitivity values at the knots to interpolate the detector sensitivity $\tilde{D}(x_0, y_0)$ at each centroid. 

Again, the drawback of this method is the necessity for accurate centroid measurements: a greater number of knots requires more precise centroid measurements. To mitigate the problem, we used binned centroids to obtain the detector sensitivity. Due to the relation between the inter-knot distance and the requirements for precised PSF location measurements, we chose a distance between the knots to be approximately the size of the centroid scatter within a datacube.

The shortcoming of such non-parametric models is that they do not properly marginalize over the detector uncertainty \cite{2017PASP..129a4001S} and they have an indeterminate number of parameters  which makes it difficult to assess the Bayesian evidence for the model as explained below.

\subsubsection*{Pixel-Level Decorrelation}
Finally, we experiment with Pixel Level Decorrelation (PLD) \cite{2015ApJ...805..132D,2016AJ....152...44I} using a modified version of the systematics model. As mentioned earlier, the PSF of the target spans many pixels. One can express the total flux measurements as a general function of the pixels level fluxes. Astrophysical variations are expected to affect all pixels equally. Therefore, variations in the fraction of the total measured by each pixel are caused by the detector systematics such as variations in the telescope pointing, intra-pixel sensitivity variation, pixel coupling, and oscillation due to heating. Hence, one can express the detector sensitivity as a general function of the fraction to total flux recorded by each pixel.

We define the detector model as

\begin{equation}
\tilde{D^t} = \sum _{i=1} ^N a_i \frac{P_i^t}{\sum _k ^N P_k^t}
\end{equation} 

\noindent where $N$ is the number of pixels used, $a_i$ is the linear PLD coefficient for the $i^{th}$ pixel and $P_i$ is the value of the $i^{th}$ pixel. In contrast with the original formulation \cite{2015ApJ...805..132D}, we elect to include $\tilde{D ^t}$ as a multiplicative factor rather than an additive factor since it describes the detector systematics more accurately. The difference between including the systematics as an additive term or a multiplicative factor is the $\delta A(t) \cdot \delta \tilde{D}^t$ cross-term which can be as large as $0.03 \cdot 0.005 = 0.000 15$ (150 parts per million) for a 3\% transit depth \cite{2017ApJ...834..187B}. 

Although our observations were acquired in staring mode, the telescope pointing can drift significantly in long time series observations. In our case, the image position on the detector varied by a third of a pixel, hence the first order PLD performed poorly compared to other decorrelation methods.

\subsection*{Model Fitting and Error Estimates}

To estimate model parameters and their uncertainties, we use the package \texttt{emcee} \cite{2013PASP..125..306F}, an Affine Invariant Markov Chain Monte Carlo (MCMC) implemented in Python. We use parameters from the literature \cite{Alonso2008, 2010Gillon} as an initial estimate of the astrophysical model. We use a Levenberg-Marquardt to estimate the detector coefficients from the residuals obtained after removing the initial astrophysical signal guess. We initialize 500 MCMC walkers with initial positions in parameter space distributed around the initial guess. We define the likelihood function as 

\begin{equation}
\ln L = -\frac{1}{2} \chi ^2 - N_{dat}\ln \sigma _F
\end{equation}

\noindent where $\sigma _F$ is the photometric uncertainty which we make a jump parameter, $N_{dat}$ is the number of data and $\chi ^2$ is the badness-of-fit which is defined as

\begin{equation}
\chi ^2 = \frac{\sum _i [F_{data} (t) - F_{model} (t)]^2}{\sigma _F ^2}
\end{equation}

\noindent where $F_{data}$ is the measured flux obtained from photometry. Since the measured flux varies by at most 4\%, we adopt the same photometric uncertainty $\sigma _F$ for the entire data set.

Each fit has a different burn in period for the MCMC. To ensure that our MCMC fit has converged, it has to satisfy the following criteria: 1) over the last 2000 MCMC steps of all the walkers, the likelihood of the best fit did not change and 2) over the last 2000 MCMC steps of all the walkers, the distribution the MCMC walker along each parameters was approximately constant. We find that depending on the complexity of model, the burn in period is about 4000 to 15000 steps for each MCMC walker.

Instead of using a covariance matrix to estimate the uncertainty on our parameter estimates, we marginalize over all the walkers over the last 2000 MCMC steps to get a posterior distribution for each jump parameter.

\subsection*{Priors}

Our observations only include one orbit of CoRoT-2b which does not allow us to constrain astrophysical parameters such as the period, $P$, the semi-major axis, $a$, and the inclination, $i$, as precisely as values available in the literature. We therefore adopt informative priors for these parameters in the MCMC. We use the values and uncertainties obtained from 152 days of continuous observations of the system  \cite{Alonso2008} to impose Gaussian priors on $a$ and $i$. Since the uncertainty on $P$ is merely 0.0001\% of the value, we choose to fix the period to reduce the number of jump parameters in our analysis.

The time of transit and secondary eclipse allows us to constrain $e \cos \omega$ while the relative duration of the transit and the secondary eclipse allows us to constrain $e \sin \omega$ \cite{2005Charbonneau}.  While this puts a strong constrain on $e \cos \omega$, the duration of the occultations is usually too short to strongly constrain $e \sin \omega$. Since eccentricity can only range between 0 and 1 and $\omega$ can be any value, we put a uniform prior on $e \cos \omega$ and $e \sin \omega$ ranging from -1 to 1.

Additionally, we specify priors on the limb-darkening coefficient. We consider quadratic limb-darkening where the stellar intensity $I(\mu)$ is described as

\begin{equation}
I(\mu)/I_0 = 1 - u_1(1-\mu) - u_2(1-\mu)^2
\end{equation}

\noindent where $I_0$ is the stellar intensity at the center of the disk, $u_1$ and $u_2$ are the limb-darkening coefficients and $\mu = \sqrt{1-r^2}$ with $r$ defined as the distance from the center of the disc. The common way to deal with limb darkening coefficients is to estimate the limb darkening coefficients prior to the fit and keep them fixed. The drawback is that the coefficients are dependent on the stellar atmosphere model adopted 
\cite{2015MNRAS.450.1879E}. Alternatively, one can make the coefficients jump parameters which is more statistically robust as it makes no assumption about the star; we chose the latter solution and to ensure that we make no assumption about the intensity profile of the host star while never exploring unphysical solutions, we used the following parametrization \cite{2013MNRAS.435.2152K}:

\begin{equation}
q_1 = (u_1 + u_2)^2,
\end{equation}

\begin{equation}
q_2 = \frac{1}{2(u_1 + u_2)},
\end{equation}

\noindent with uniform prior on $q_1$ and $q_2$ ranging from 0 to 1, which the author claims to yield both realistic and robust uncertainties.

Most importantly, the full orbit phase curve of the system allows us to obtain a longitudinal surface brightness map of the planet \cite{2008ApJ...678L.129C}. We use a physical prior rejecting models with phase variation coefficients that yield negative brightness at any longitude \cite{2017arXiv170903502K}. 

\subsection*{Model Comparison}

Generally, a fit to data improves as we increase the number of model parameters. To compare the various astrophysical and detector models, we estimate the Bayesian Evidence by analogy with the Bayesian Information Criterion (BIC)\cite{schwarz1978estimating, 2012BIC}:

\begin{equation}
E = \ln L - \frac{N_{par}}{2}\ln N_{dat} = -\frac{\text{BIC}}{2}
\end{equation}

\noindent where $N_{par}$ is the number of model parameters. By this definition, a greater Bayesian evidence is preferred. 

We experiment with various models: different detector models, planetary signature models, and the inclusion of stellar variabilty signatures. Although BLISS mapping is a non-parameteric model and therefore cannot be assigned a Bayesian Evidence, those fits yield lower log-likelihood than polynomials and therefore were ruled out as best fits. Comparing the Bayesian Evidence, we find 3 almost equivalently best models and 2 substantially good models \cite{Kass1995}: 
\begin{itemize}
\item no stellar variability, 2$^{nd}$ order polynomial detector model, 2$^{nd}$ order phase variation, $E = 7695.64$ ($\Delta \text{BIC} = 0$)
\item with stellar variability, 3$^{th}$ order polynomial detector model, 1$^{st}$ order phase variation, $E = 7695.14$ ($\Delta \text{BIC} = 1.00$)
\item with stellar variability, 4$^{th}$ order polynomial detector model, 1$^{st}$ order phase variation, $E = 7694.62$ ($\Delta \text{BIC} = 2.04$)
\item with stellar variability, 2$^{nd}$ order polynomial detector model, 2$^{st}$ order phase variation, $E = 7693.52$ ($\Delta \text{BIC} = 4.24$)
\item with stellar variability, 2$^{nd}$ order polynomial detector model, 1$^{st}$ order phase variation, $E = 7693.49$ ($\Delta \text{BIC} = 4.30$)
\end{itemize}
 
\noindent where $\Delta \text{BIC}$ is the difference in BIC compared to the fit with lowest BIC, and by analogy with greatest Bayesian Evidence. Fits with a $\Delta \text{BIC}>6$ can be strongly ruled out as best-fits \cite{Kass1995}. The most probable parameters, log-likelihood, and Bayesian Evidence values for each models are reported in Supplementary Table \ref{Tab:Astro-Params-Estimate-v1}, \ref{Tab:Astro-Params-Estimate-v2}, \ref{Tab:Astro-Params-Estimate-v1-stellar}, and \ref{Tab:Astro-Params-Estimate-v2-stellar}; the best fits are highlighted. All five models favor a westward hotspot offset on the planet, but with varying significance.

Since the Bayesian Evidence does not allow us to discern which of the 5 is the best model, we elect to look at the in-eclipse portion of the lightcurve to discriminate between the models. The in-eclipse portions of the phase curve should be unity once we remove the detector systematics and stellar variability. Fitting a linear function to the in-eclipse segments of the corrected lightcurves, we find the model in which we assumed no stellar variability to be the most consistent with the absence of a trend: the linear fits to the first and second eclipses have slopes of $-0.009 \pm 0.013$ and $-0.003 \pm 0.01$, respectively (see Supplementary Figures \ref{Fig: In-eclipse_poly2-v2}, \ref{Fig: In-eclipse_poly3-v1-stellar} and \ref{Fig: In-eclipse_poly4-v1-stellar} in Supplementary Figures). Since the in-eclipse flatness test favors the model with stable stellar flux over the models with the inclusion of stellar variability, it suggests that the models with stable stellar flux is a better representation of the underlying astrophysics.

As mentioned previously, the inclusion of a sinusoidal stellar variation model can introduce degeneracy as seen in Supplementary Tables \ref{Tab:Astro-Params-Estimate-v1-stellar} and \ref{Tab:Astro-Params-Estimate-v2-stellar}, which can lead to less consistent parameter estimates and larger uncertainties. As a sanity check, we include a fit performed using an entirely independent photometry and fitting pipeline \cite{2017arXiv171007642Z}. Using a higher-order PLD method coupled with a linear trend as model for the stellar variability, we find a westward offset of $25.6 \pm 1.9$ degrees, confirming the westward offset obtained with the first pipeline.

Note, moreover, that all of the models exhibit a slightly declining flux during eclipse which is consistent with the fit obtained using the higher-order PLD pipeline. This suggests that if anything, we are slightly underestimating the magnitude of the westward hotspot offset: if the eclipse bottom were flat, then the flux would be even greater after eclipse. 

\subsection*{Surface Brightness Map}

Due to the orbital motion and rotation of the planet, the region of the planet facing us changes over time. One can translate the phase variation of the planet as seen from a distant observer into a longitudinal brightness map of the planet. We map the surface brightness of CoRoT-2b \cite{2008ApJ...678L.129C} (see Supplementary Information) and see a clear westward offset of $23 \pm 4$ degrees.

In the case of a non-synchronously rotating planet, the same formalism can be used to obtained the longitudinal brightness of the planet, but the longitude would correspond to stellar zenith angle \cite{2012ApJ...757...80C}.

\subsection*{Energy Budget}

Thermal phase variations of short-period planet constrain the day-to-night heat recirculation efficiency of the atmosphere \cite{2011ApJ...729...54C, 2013ApJ...776..134P, 2015MNRAS.449.4192S}.

Using published eclipse and transit depths, including ours, along with the phase curve amplitude and offset, we obtain the dayside and nightside brightness temperatures with the inverse Planck function. We constrain the Bond albedo and recirculation effiency from the derived effective dayside and nightside temperature shown in Figure \ref{Fig3: Energy Budget} \cite{2011ApJ...729...54C, 2017arXiv170705790S}. 

\subsection*{Emission Spectroscopy}
We used a toy emission spectrum model accounting for the contribution from reflected light, $F_{reflected}$, and thermal emission, $F_{thermal}$ to describe the planet's emission spectrum. We model the emission spectrum of the host star, $F_*(\lambda)$, using the Kurucz Atlas from the \texttt{pysynphot} package \cite{2013ascl.soft03023S} with an effective temperature of $T_*=$ 5625 K, a metallicity [Fe/H]=0, and a surface gravity $\log g=4.71$. The emission spectrum of the planet as the sum of thermal emission and reflected light \cite{2017arXiv170903502K}:

\begin{equation}
\begin{split}
F_p(\lambda) = \left(\frac{R_p}{R_*}\right)^2B_\lambda(T_{day}, \lambda) + \frac{A_g}{(a/R_*)^2}\left(\frac{R_p}{R_*}\right)^2F_*(\lambda)
\end{split}
\end{equation}

\noindent where $T_{day}$ is the effective dayside emission and $A_g$ is the geometric albedo of the planet. Using \texttt{emcee}, we fit the presented model to CoRoT-2b's emission spectrum and we find that the model with a geometric albedo, $A_g$, of 0.12$\pm$0.02 and dayside effective temperature of 1693$\pm$17 K best fits the data with chi-squared per datum, $\chi _{dat}$, of 1.34. We note that our eclipse depth measurement at 4.5 $\mu$m is shallower than that reported using observation taken during the cryogenic \textit{Spitzer} era \cite{2010Gillon,2011Deming}. This discrepancy may be due to improvements observational in strategies and self-calibration techniques over time \cite{2014Hansen, 2016AJ....152...44I}.

\subsection*{Constraint on Magnetic Field Strength of CoRoT-2b}

To estimate the lower limit on the magnetic strength of CoRoT-2b's dynamo required to explain westward winds, we calculate the ionization fraction, $\chi _e$. The importance of the effect of magnetism on zonal winds can be approximated as the ratio of magnetic to wave timescales $\tau_{mag}/\tau_{wave}$. The magnetic timescale if defined as

\begin{equation}
\tau_{mag} = \frac{4\pi \rho \eta}{B^2}
\end{equation}

\noindent where $\rho$ is the density, $\eta$ is the magnetic diffusivity, $B$ is the magnetic field strength, $g$ is the gravity. Hence, we can define the lower limit of $B\sim \sqrt{4\pi \rho \eta/ \tau_{wave}}$, i.e. when $\tau_{mag}/\tau_{wave}\sim 1$. The magnetic diffusivity is defined as \cite{2017NatAs...1E.131R}

\begin{equation}
\eta = 230 \sqrt{T}/\chi _e
\end{equation}

\noindent where $\chi _e$ is the ionization fraction given which we evaluate using a simplified Saha equation that only account for potassium \cite{2010ApJ...719.1421P}:
\begin{equation}
\begin{split}
\chi _e = 6.47 \times 10^{-13} \left( \frac{a_K}{10^{-7}} \right)^{1/2}\left( \frac{T}{10^{3}} \right)^{3/4} \\ \left( \frac{2.4 \times 10^{15}}{n_n} \right)^{1/2}  \frac{\exp(-25188/T)}{1.15 \times 10^{-11}}
\end{split}
\end{equation}

\noindent where $a_K$ is the abundance of potassium, $T$ is the dayside temperature of the planet, and $n_n$ is the number density of neutrals and is defined as $n_n = \rho/\bar{m}$ where $\bar{m}$ is the molecular mass of hydrogen.

So, assuming that the planet is mainly made of molecular hydrogen with a gas constant of $R=3523J/kg \cdot K$, we calculate a density of $\rho = P/RT= 1.635 \times 10^{-5}$ g/cm$^3$ at a pressure of $P=1$ bar and a temperature $T=1736$ K. Therefore, we get a number density of neutrals of $n_n = \rho / \bar{m} = 4.89 \times 10^{18}$. Approximating a potassium abundance of $a_K = 10^{-7}$, we find an ionization fraction of $9.41 \times 10^{-10}$ and magnetic diffusivity of $\eta = 1.02 \times 10^{13}$. The wave timescale is defined as

\begin{equation}
\tau_{wave} = \frac{L}{\sqrt{gH}}
\end{equation}

\noindent where $L$ is the characteristic length scale of the horizontal flow, $g = 4185$ cm/s$^{2}$ is the gravity, and $H$ is the depth of the atmosphere. Approximating the characteristic length scale as $L\sim r_p$, where $r_p = 1.06 \times 10^8$ m, and calculating $H=kT/\bar{m}g$, where $k$ is the Boltzmann constant, we calculate a wave timescale of $\tau_{wave}= 3.83 \times 10^4$ seconds.
 
Using the derived values of $\rho$, $\eta$, and $\tau _{wave}$ above, we find a lower limit on the magnetic field strength of $B \sim 230$ G to cause atmospheric variability. We note that this is a rough estimation of the planet's magnetic field, better constraints would require magnetohydrodynamic simulations.

Additionally, the circulation should be variable on an Alfv\'en timescale \cite{2017NatAs...1E.131R}$, \tau _A = a/v_A$, where $a$ is a characteristic scale of the system and $v_A$ is the Alfv\'en velocity. We calculate the Alfv\'en velocity defined as $v_A = B/\sqrt{\mu _0 \rho}$, where $\mu _0$ is the permeability of vacuum. Approximating as $a \sim \pi r_p$ and using the density $\rho$ derived above, we obtain a timescale of $\tau _A \sim 23$ days.

\subsection*{Data Availability Statement}
Individual fit parameters and uncertainties using various models are provided in the Supplementary Information.  Any other data that support the plots within this paper and other findings of this study are available from the corresponding author upon reasonable request.
\end{multicols}

\section*{Correspondence and Requests}
\textbf{Correspondence and requests for materials} should be addressed to L.D.

\section*{Acknowledgements}

L.D. thanks Sean Carey, James Ingalls, and William Glaccum from the Spitzer IRAC team for the helpful discussions that contributed to the reduction of the data. Funding for this work was provided in part by the Natural Sciences and Engineering Research Council of Canada (NSERC) discovery grant and the California Institute of Technology's Infrared Processing and Analysis Center (Caltech/IPAC) Visiting Graduate Research Fellowship. Work by S. S. was funded by the Google Summer of Code program. This work is based on observations made with the Spitzer Space Telescope, which is operated by the Jet Propulsion Laboratory, California Institute of Technology under a contract with NASA. Support for this work was provided by NASA through an award issued by JPL/Caltech.

\section*{Author contributions statement}
L.D. extracted the photometric measurements from the data, detrended the data, developed and fit the phase curve models,led the analysis and wrote the manuscript. N.B.C. is the P.I. of the successful Spitzer proposal from which we obtained the observations and contributed to the writing of the manuscript. J.C.S contributed materials to the main text. E.R. contributed to the interpretation for the results and to the discussion. M.Z. and H.K. verified the robustness of the analysis and contributed to the interpretation of results. S.S. contributed to the photometric measurements pipeline. J.C.S., E.R, H.K., I.D., M.L., D.D., J.J.F., and M.Z. are co-I.'s of the successful Spitzer proposal from which we obtained the observations. All authors commented on the manuscript.\\
\textit{\textbf{Facilities}}: Spitzer Space Telescope \cite{2004ApJS..154....1W}, Infrared Array Camera \cite{2004ApJS..154...10F}\\
\textit{\textbf{Software}}: Numpy \cite{Numpy}, Scipy \cite{Numpy}, Astropy \cite{2013A&A...558A..33A}, Matplotlib \cite{Matplotlib}, Emcee \cite{2013PASP..125..306F}, Batman \cite{2015Kreidberg}, Corner \cite{corner}, Jupyter \cite{IPython}, pysynphot \cite{2013ascl.soft03023S}

\section*{Additional information}
\textbf{Supplementary information} is availalbe for this paper.

\section*{Competing Interests}
The authors declare no competing financial interests.

\renewcommand{\refname}{Further References}

\newpage

\section*{Supplementary Tables}
\renewcommand{\arraystretch}{1.3}
\renewcommand{\tablename}{Supplementary Table}
\setcounter{table}{0}
\renewcommand*{\thefootnote}{\fnsymbol{footnote}}

\begin{table}[!htpb]
	\centering
	\scalebox{0.85}{
	\begin{tabular}{lccl} \hline
		\textbf{Name}                          & \textbf{Symbol} & \textbf{Constraint} & \textbf{Reference} \\ \hline
		
		\multicolumn{2}{c}{\textit{Fitted}}                  \\
		Time of transit (days from start of observations \footnote{Time of start of observation (BMJD): 57390.7636202 days})      & $t_0$       & -- & --\\
		Radius of planet                       & $R_p/R_*$   & -- & --\\
		Semi-major axis                        & $a/R_*$     & $6.70\pm 0.03$ & Alonso et al. (2008)\\
		Orbital inclination (degrees)          & $i$         & $87.84\pm 0.1$ & Alonso et al. (2008)\\
		Orbital eccentricity                   & $e$         & $[0,1]$ & --\\
		Longitude of periastron                & $\omega$    & $[0, 2 \pi]$& --\\
		Limb darkening coefficient             & $q_1$       &  $[0,1]$ & Kipping (2013)\\
		Limb darkening coefficient             & $q_2$       &  $[0,1]$ & Kipping (2013)\\
		Eclipse depth ($F_p/F_*$)              & $\delta _e$ & $[0,1]$ & --\\
		Phase variation even coefficient ($1^{st}$ order)      & $A$     &  $F_p \geq 0$ & Keating \& Cowan (2017)  \\
		Phase variation odd coefficient ($1^{st}$ order)      & $B$      & $F_p \geq 0$ & Keating \& Cowan (2017)  \\
		Phase variation even coefficient ($2^{nd}$ order)      & $C$      & $F_p \geq 0$ & Keating \& Cowan (2017) \\
		Phase variation odd coefficient ($2^{nd}$ order)       & $D$         & $F_p \geq 0$ & Keating \& Cowan (2017) \\ 
		Stellar variation even coefficient ($1^{st}$ order)      & $S_1$      & -- &  -- \\
		Stellar variation odd coefficient ($1^{st}$ order)       & $S_2$     & -- & --     \\ \hline
		\multicolumn{2}{c}{\textit{Fixed}}                   \\
		Orbital period   (days)                  & $P$     & 1.7429964 &   Alonso et al. (2008)  \\ 
		Rotational period of the host star (days)  & $P_*$     & 4.522 &   Lanza et al. (2009)  \\ \hline
				\multicolumn{2}{c}{\textit{Derived}}                   \\
		Phase Amplitude   (units of Stellar Flux)                  & $A_p$     & -- &   --  \\ 
		Phase Offset (in rad) & $\Phi _p$     & -- &   --  \\ \hline
	\end{tabular}
	}
    
	\caption{\textbf{Astrophysical Model Parameters.} Note that positive values of $\Phi _p$ corresponds to peak occurring after eclipse.}
	\label{Tab:Astro-Params}
\end{table}

\clearpage\clearpage
\begin{table}[!htpb]
	\centering
	\scalebox{0.88}{
	\begin{tabular}{ccccccc} \hline

		\textbf{Parameter} & \textbf{Poly2}&\textbf{Poly3} &\textbf{Poly4} & \textbf{Poly5} & \textbf{PLD1} & \textbf{BLISS}\\ \hline
$t_0$ & $1.0744^{+20.1353}_{-20.1353}$ & $1.0744^{+0.0002}_{-0.0002}$ & $1.0744^{+0.0002}_{-0.0002}$ & $1.0745^{+0.0002}_{-0.0002}$ & $1.0746^{+0.0002}_{-0.0002}$ & $1.0745^{+0.0002}_{-0.0002}$ \\
$R_p/R_*$ & $0.1698^{+0.0008}_{-0.0009}$ & $0.1692^{+0.0009}_{-0.001}$ & $0.1689^{+0.001}_{-0.001}$ & $0.1689^{+0.0009}_{-0.001}$ & $0.1694^{+0.0012}_{-0.0013}$ & $0.1685^{+0.001}_{-0.001}$ \\
$a/R_*$ & $6.6843^{+0.0237}_{-0.0238}$ & $6.6848^{+0.0248}_{-0.0247}$ & $6.6868^{+0.0246}_{-0.0265}$ & $6.6891^{+0.0243}_{-0.0244}$ & $6.6991^{+0.0252}_{-0.0253}$ & $6.6858^{+0.0241}_{-0.0246}$ \\
$i$ & $87.8598^{+0.0926}_{-0.0909}$ & $87.8639^{+0.0919}_{-0.0897}$ & $87.8603^{+0.0969}_{-0.0949}$ & $87.8592^{+0.0944}_{-0.0917}$ & $87.8411^{+0.0922}_{-0.0954}$ & $87.8626^{+0.0928}_{-0.0922}$ \\
$e \cos{\omega}$ & $-1.6e-05^{+0.000373}_{-0.000388}$ & $1e-05^{+0.000365}_{-0.000355}$ & $2.3e-05^{+0.000409}_{-0.000393}$ & $9e-06^{+0.000376}_{-0.000398}$ & $-4e-06^{+0.000356}_{-0.000361}$ & $-1e-06^{+0.000483}_{-0.000489}$ \\
$e \sin{\omega}$ & $-0.0^{+0.000384}_{-0.000377}$ & $-5e-06^{+0.000334}_{-0.000334}$ & $7e-06^{+0.000394}_{-0.000386}$ & $1e-06^{+0.000375}_{-0.000398}$ & $-8e-06^{+0.000359}_{-0.00036}$ & $-1e-05^{+0.000493}_{-0.000497}$ \\
$q_1$ & $0.0112^{+0.0162}_{-0.0074}$ & $0.0129^{+0.0189}_{-0.0087}$ & $0.0147^{+0.0229}_{-0.0099}$ & $0.0151^{+0.0228}_{-0.0101}$ & $0.0227^{+0.0314}_{-0.0146}$ & $0.0161^{+0.0216}_{-0.0103}$ \\
$q_2$ & $0.3211^{+0.3796}_{-0.2341}$ & $0.3071^{+0.3943}_{-0.228}$ & $0.254^{+0.3774}_{-0.1892}$ & $0.2686^{+0.3938}_{-0.2023}$ & $0.3494^{+0.3923}_{-0.2506}$ & $0.2913^{+0.3692}_{-0.2149}$ \\
$\delta _e$ & $0.0045^{+0.0002}_{-0.0002}$ & $0.0046^{+0.0002}_{-0.0002}$ & $0.0044^{+0.0002}_{-0.0002}$ & $0.0044^{+0.0002}_{-0.0002}$ & $0.0042^{+0.0002}_{-0.0002}$ & $0.0044^{+0.0002}_{-0.0002}$ \\
$A$ & $0.3984^{+0.008}_{-0.009}$ & $0.3921^{+0.0097}_{-0.0104}$ & $0.3908^{+0.0101}_{-0.0115}$ & $0.3907^{+0.0117}_{-0.0135}$ & $0.3972^{+0.0113}_{-0.0125}$ & $0.3923^{+0.0124}_{-0.0141}$ \\
$B$ & $0.2415^{+0.0246}_{-0.0245}$ & $0.2498^{+0.0251}_{-0.0267}$ & $0.2567^{+0.0281}_{-0.0286}$ & $0.2474^{+0.0319}_{-0.0317}$ & $0.2368^{+0.0348}_{-0.0359}$ & $0.2382^{+0.0341}_{-0.0354}$ \\ \hline
$\sigma _F$ & $0.00149^{+3e-05}_{-2.9e-05}$ & $0.001474^{+3.1e-05}_{-2.9e-05}$ & $0.001468^{+3.1e-05}_{-3e-05}$ & $0.001461^{+3e-05}_{-2.9e-05}$ & $0.0021^{+0.0}_{-0.0}$ & $0.0016^{+0.0}_{-0.0}$ \\ \hline
$\log(L)$ & $7741.73$ & $7758.25$ & $7765.04$ & $\textcolor{blue}{\textbf{7775.53}}$ & $7296.34$ & $7684.27$ \\
$E$ & $7677.28$ & $\textcolor{blue}{\textbf{7679.48}}$ & $7668.37$ & $7657.37$ & $7221.15$ & $--$ \\ \hline
$\log(L)$\footnote{\label{Unbinned}Calculated for unbinned data} & $339938.37$ & $340400.45$ & $\textcolor{blue}{\textbf{340611.54}}$ & $340528.36$ & $314954.3$ & $337867.36$ \\
$E$\footnoteref{Unbinned} & $339836.54$ & $340276.0$ & $\textcolor{blue}{\textbf{340458.8}}$ & $340341.68$ & $314835.69$ & $--$ \\ \hline
$A_p$ & $0.0042^{+0.0002}_{-0.0002}$ & $0.0042^{+0.0002}_{-0.0002}$ & $0.0041^{+0.0002}_{-0.0002}$ & $0.0041^{+0.0002}_{-0.0002}$ & $0.0039^{+0.0002}_{-0.0002}$ & $0.004^{+0.0002}_{-0.0002}$ \\
$\Phi _p$ & $0.544^{+0.0503}_{-0.0566}$ & $0.5692^{+0.0629}_{-0.0503}$ & $0.5818^{+0.0629}_{-0.0629}$ & $0.5629^{+0.0692}_{-0.0692}$ & $0.5378^{+0.0818}_{-0.0755}$ & $0.544^{+0.0755}_{-0.0755}$ \\ \hline
	\end{tabular}
	}
	\caption{\textbf{Fit parameters using a first order Fourier series to model the phase variation without stellar variation for different detector models.} Most probable astrophysical parameter estimates obtained from posterior probability distribution from the MCMC routine using various systematics models. The planetary brightness phase variation is modeled as a first order Fourier series and the model assumes no stellar variability. The largest value of $\log L$ and $E$ for both binned and unbinned data are in bold blue. The errors on the parameters are the 68\% confidence region bounds of the posterior distribution obtained from \texttt{emcee}.}
\label{Tab:Astro-Params-Estimate-v1}
\end{table} 


\begin{table}[!htpb]
	\centering
	\scalebox{0.88}{
	\begin{tabular}{caccccc} \hline
		\textbf{Parameter} & \textbf{Poly2}&\textbf{Poly3} &\textbf{Poly4} & \textbf{Poly5} & \textbf{PLD1} &  \textbf{BLISS}\\ \hline
$t_0$ & $1.0744^{+0.0002}_{-0.0002}$ & $1.0744^{+0.0002}_{-0.0002}$ & $1.0744^{+0.0002}_{-0.0002}$ & $1.0745^{+0.0002}_{-0.0002}$ & $1.0746^{+0.0002}_{-0.0002}$ & $1.0744^{+0.0002}_{-0.0002}$ \\
$R_p/R_*$ & $0.1696^{+0.0009}_{-0.0009}$ & $0.1691^{+0.0009}_{-0.0009}$ & $0.1689^{+0.0009}_{-0.0009}$ & $0.1688^{+0.0009}_{-0.0009}$ & $0.1702^{+0.0011}_{-0.0012}$ & $0.1686^{+0.001}_{-0.001}$ \\
$a/R_*$ & $6.6792^{+0.0249}_{-0.024}$ & $6.6821^{+0.0238}_{-0.0246}$ & $6.6867^{+0.0237}_{-0.0244}$ & $6.6856^{+0.0243}_{-0.0247}$ & $6.6946^{+0.0251}_{-0.0257}$ & $6.6837^{+0.0244}_{-0.025}$ \\
$i$ & $87.8684^{+0.0929}_{-0.0923}$ & $87.8671^{+0.0937}_{-0.09}$ & $87.8686^{+0.0915}_{-0.0945}$ & $87.8645^{+0.0966}_{-0.093}$ & $87.8405^{+0.092}_{-0.0922}$ & $87.8694^{+0.092}_{-0.0926}$ \\
$e \cos{\omega}$ & $-0.0^{+0.0004}_{-0.0004}$ & $-0.0^{+0.0004}_{-0.0003}$ & $-0.0^{+0.0003}_{-0.0003}$ & $0.0^{+0.0004}_{-0.0004}$ & $0.0^{+0.0003}_{-0.0003}$ & $-0.0^{+0.0005}_{-0.0005}$ \\
$e \sin{\omega}$ & $0.0^{+0.0004}_{-0.0004}$ & $-0.0^{+0.0004}_{-0.0004}$ & $0.0^{+0.0004}_{-0.0004}$ & $-0.0^{+0.0004}_{-0.0004}$ & $0.0^{+0.0003}_{-0.0003}$ & $-0.0^{+0.0005}_{-0.0005}$ \\
$q_1$ & $0.0133^{+0.0178}_{-0.0087}$ & $0.0129^{+0.0196}_{-0.0085}$ & $0.0133^{+0.0182}_{-0.0089}$ & $0.015^{+0.0207}_{-0.01}$ & $0.0205^{+0.0271}_{-0.0131}$ & $0.0158^{+0.0231}_{-0.0103}$ \\
$q_2$ & $0.2957^{+0.3659}_{-0.2192}$ & $0.2951^{+0.3934}_{-0.2149}$ & $0.2889^{+0.3688}_{-0.2103}$ & $0.2842^{+0.3695}_{-0.2112}$ & $0.38^{+0.385}_{-0.2688}$ & $0.2874^{+0.3852}_{-0.2132}$ \\
$\delta _e$ & $0.0044^{+0.0002}_{-0.0002}$ & $0.0046^{+0.0002}_{-0.0002}$ & $0.0045^{+0.0002}_{-0.0002}$ & $0.0044^{+0.0002}_{-0.0002}$ & $0.0048^{+0.0002}_{-0.0002}$ & $0.0045^{+0.0002}_{-0.0002}$ \\
$A$ & $0.4443^{+0.0133}_{-0.0148}$ & $0.4257^{+0.0187}_{-0.0195}$ & $0.4325^{+0.0171}_{-0.0196}$ & $0.427^{+0.0194}_{-0.0215}$ & $0.3745^{+0.0212}_{-0.0213}$ & $0.422^{+0.0198}_{-0.0223}$ \\
$B$ & $0.1934^{+0.0341}_{-0.0324}$ & $0.1647^{+0.0355}_{-0.0346}$ & $0.1754^{+0.0411}_{-0.039}$ & $0.1775^{+0.0474}_{-0.0446}$ & $0.1322^{+0.0319}_{-0.0298}$ & $0.1748^{+0.0459}_{-0.0443}$ \\
$C$ & $0.0669^{+0.0132}_{-0.013}$ & $0.0802^{+0.015}_{-0.0163}$ & $0.0756^{+0.0161}_{-0.016}$ & $0.0746^{+0.0182}_{-0.0189}$ & $0.1096^{+0.0135}_{-0.014}$ & $0.0784^{+0.018}_{-0.0187}$ \\
$D$ & $0.0681^{+0.0117}_{-0.012}$ & $0.0627^{+0.0134}_{-0.0152}$ & $0.0628^{+0.0141}_{-0.0154}$ & $0.0638^{+0.0174}_{-0.0177}$ & $0.1096^{+0.0135}_{-0.014}$ & $0.0715^{+0.0163}_{-0.0177}$ \\ \hline
$\sigma _F$ & $0.001461^{+2.9e-05}_{-2.9e-05}$ & $0.001454^{+3e-05}_{-2.9e-05}$ & $0.001448^{+2.9e-05}_{-2.9e-05}$ & $0.001446^{+2.9e-05}_{-2.9e-05}$ & $0.002013^{+4e-05}_{-3.9e-05}$ & $0.001536^{+3.1e-05}_{-3e-05}$ \\ \hline
$\log(L)$ & $7767.25$ & $7776.56$ & $7784.37$ & $\textcolor{blue}{\textbf{7785.60}}$ & $7350.24$ & $7700.07$ \\
$E$ & $\textcolor{blue}{\textbf{7695.64}}$ & $7690.63$ & $7680.54$ & $7660.28$ & $7267.89$ & $--$ \\ \hline
$\log(L)$\footnoteref{Unbinned} & $340798.6$ & $340917.09$ & $\textcolor{blue}{\textbf{341131.14}}$ & $340976.52$ & $317233.61$ & $338357.01$ \\
$E$\footnoteref{Unbinned} & $340685.46$ & $340781.32$ & $\textcolor{blue}{\textbf{340967.09}}$ & $340778.52$ & $317103.7$ & $--$ \\ \hline
$A_p$ & $0.0043^{+0.0002}_{-0.0002}$ & $0.0042^{+0.0002}_{-0.0002}$ & $0.0042^{+0.0002}_{-0.0002}$ & $0.0041^{+0.0002}_{-0.0002}$ & $0.004^{+0.0002}_{-0.0002}$ & $0.0041^{+0.0003}_{-0.0003}$ \\
$\Phi _p$ & $0.412^{+0.0629}_{-0.0566}$ & $0.3491^{+0.0692}_{-0.0692}$ & $0.3679^{+0.0755}_{-0.0755}$ & $0.3742^{+0.0881}_{-0.0943}$ & $0.3365^{+0.0629}_{-0.0629}$ & $0.3805^{+0.0881}_{-0.0881}$  \\ \hline
	\end{tabular}
    }
	\caption{\textbf{Fit parameters using a second order Fourier series to model the phase variation without stellar variation for different detector models.} Most probable astrophysical parameters estimates obtained from posterior probability distribution from the MCMC routine using various systematics models. The planetary brightness phase variation is modeled as a second order Fourier series and the model assumes no stellar variability. The highlighted fit yields the greatest Bayesian Evidence. The largest value of $\log L$ and $E$ for both binned and unbinned data are in bold blue. The errors on the parameters are the 68\% confidence region bounds of the posterior distribution obtained from \texttt{emcee}.}
	\label{Tab:Astro-Params-Estimate-v2}
\end{table} 

\begin{table}[!htpb]
	\centering
	
	\scalebox{0.88}{
		\begin{tabular}{caaaccc} \hline
			\textbf{Parameter} & \textbf{Poly2}&\textbf{Poly3} &\textbf{Poly4} & \textbf{Poly5} & \textbf{PLD1} & \textbf{BLISS}\\ \hline
$t_0$ & $1.0744^{+0.0002}_{-0.0002}$ & $1.0744^{+0.0002}_{-0.0002}$ & $1.0745^{+0.0002}_{-0.0002}$ & $1.0745^{+0.0002}_{-0.0002}$ & $1.0746^{+0.0002}_{-0.0002}$ & $1.0745^{+0.0002}_{-0.0002}$ \\
$R_p/R_*$ & $0.1693^{+0.0009}_{-0.0009}$ & $0.1687^{+0.0009}_{-0.001}$ & $0.1683^{+0.0009}_{-0.0009}$ & $0.1684^{+0.0009}_{-0.001}$ & $0.169^{+0.0012}_{-0.0012}$ & $0.1678^{+0.001}_{-0.001}$ \\
$a/R_*$ & $6.6805^{+0.0242}_{-0.0235}$ & $6.6824^{+0.0231}_{-0.0248}$ & $6.6883^{+0.0249}_{-0.0242}$ & $6.6881^{+0.024}_{-0.0237}$ & $6.6929^{+0.0249}_{-0.0257}$ & $6.6842^{+0.0247}_{-0.0242}$ \\
$i$ & $87.8783^{+0.0916}_{-0.0935}$ & $87.8717^{+0.0908}_{-0.0918}$ & $87.8616^{+0.0893}_{-0.0906}$ & $87.8593^{+0.0943}_{-0.0929}$ & $87.8499^{+0.0921}_{-0.0904}$ & $87.8652^{+0.0918}_{-0.0929}$ \\
$e \cos{\omega}$ & $-3e-06^{+0.000398}_{-0.000393}$ & $2.3e-05^{+0.000388}_{-0.000382}$ & $-1.4e-05^{+0.000479}_{-0.000451}$ & $-2.4e-05^{+0.000467}_{-0.000492}$ & $-5e-06^{+0.000398}_{-0.000401}$ & $-5e-06^{+0.000554}_{-0.000555}$ \\
$e \sin{\omega}$ & $7e-06^{+0.000405}_{-0.000399}$ & $1.8e-05^{+0.000382}_{-0.000383}$ & $-1e-06^{+0.000442}_{-0.000446}$ & $2e-06^{+0.000461}_{-0.000451}$ & $-1e-05^{+0.000407}_{-0.000397}$ & $-6e-06^{+0.000558}_{-0.000561}$ \\
$q_1$ & $0.0137^{+0.0202}_{-0.0091}$ & $0.0146^{+0.0202}_{-0.0098}$ & $0.0145^{+0.0194}_{-0.0094}$ & $0.0144^{+0.0197}_{-0.0094}$ & $0.0231^{+0.0284}_{-0.0147}$ & $0.0163^{+0.0223}_{-0.0106}$ \\
$q_2$ & $0.2917^{+0.3798}_{-0.2135}$ & $0.2688^{+0.3909}_{-0.1969}$ & $0.2643^{+0.3814}_{-0.1963}$ & $0.2748^{+0.39}_{-0.2033}$ & $0.3483^{+0.3772}_{-0.248}$ & $0.2998^{+0.3779}_{-0.2219}$ \\
$\delta _e$ & $0.0041^{+0.0002}_{-0.0002}$ & $0.0043^{+0.0002}_{-0.0002}$ & $0.004^{+0.0002}_{-0.0002}$ & $0.0041^{+0.0002}_{-0.0002}$ & $0.0038^{+0.0002}_{-0.0003}$ & $0.004^{+0.0002}_{-0.0003}$ \\
$A$ & $0.2966^{+0.0676}_{-0.0745}$ & $0.3543^{+0.0495}_{-0.0817}$ & $0.3744^{+0.0413}_{-0.0715}$ & $0.3682^{+0.0466}_{-0.0846}$ & $-0.2809^{+0.1462}_{-0.1489}$ & $0.2414^{+0.1175}_{-0.1459}$ \\
$B$ & $0.1721^{+0.0424}_{-0.0468}$ & $0.118^{+0.0416}_{-0.0457}$ & $0.0685^{+0.0506}_{-0.0539}$ & $0.0779^{+0.051}_{-0.0562}$ & $-0.0097^{+0.0741}_{-0.0746}$ & $-0.1545^{+0.0829}_{-0.0866}$ \\ \hline
$S_1$ & $-0.0028^{+0.0008}_{-0.001}$ & $-0.003^{+0.0006}_{-0.0009}$ & $-0.0043^{+0.0007}_{-0.0008}$ & $-0.0044^{+0.0008}_{-0.0009}$ & $-0.2809^{+0.1462}_{-0.1489}$ & $-0.0103^{+0.0016}_{-0.0017}$ \\
$S_2$ & $0.1947^{+0.1199}_{-0.0779}$ & $0.4273^{+0.1414}_{-0.1493}$ & $0.4732^{+0.0856}_{-0.1018}$ & $0.4775^{+0.0938}_{-0.1155}$ & $-0.0097^{+0.0741}_{-0.0746}$ & $0.4637^{+0.0807}_{-0.0806}$ \\
$\sigma _F$ & $0.001467^{+2.9e-05}_{-2.8e-05}$ & $0.001454^{+2.9e-05}_{-2.9e-05}$ & $0.001436^{+2.8e-05}_{-2.7e-05}$ & $0.001438^{+2.8e-05}_{-2.8e-05}$ & $0.002^{+0.0}_{-0.0}$ & $0.0015^{+0.0}_{-0.0}$ \\ \hline
$\log(L)$ & $7765.1$ & $7780.55$ & $7798.97$ & $\textcolor{blue}{\textbf{7799.72}}$ & $7331.22$ & $7728.31$ \\
$E$ & $7693.49$ & $7694.62$ & $\textcolor{blue}{\textbf{7695.14}}$ & $7674.41$ & $7248.87$ & $--$ \\ \hline
$\log(L)$\footnoteref{Unbinned} & $340661.32$ & $340980.0$ & $\textcolor{blue}{\textbf{341190.60}}$ & $341021.73$ & $316405.51$ & $338845.64$ \\
$E$\footnoteref{Unbinned} & $340548.18$ & $340844.23$ & $\textcolor{blue}{\textbf{341026.55}}$ & $340823.74$ & $316275.61$ & $--$ \\ \hline
$A_p$ & $0.0028^{+0.0007}_{-0.0006}$ & $0.0033^{+0.0008}_{-0.0004}$ & $0.0031^{+0.0006}_{-0.0004}$ & $0.0031^{+0.0007}_{-0.0004}$ & $0.0021^{+0.001}_{-0.0012}$ & $0.0024^{+0.0006}_{-0.0006}$ \\
$\Phi _p$ & $0.5315^{+0.1006}_{-0.1006}$ & $0.3365^{+0.1132}_{-0.1006}$ & $0.1855^{+0.1447}_{-0.1258}$ & $0.217^{+0.1509}_{-0.1321}$ & $-2.7768^{+0.2704}_{-5.7674}$ & $-0.5629^{+0.5157}_{-0.3271}$ \\ \hline
		\end{tabular}
	}

	\caption{\textbf{Fit parameters using a first order Fourier series to model the phase variation with stellar variation for different detector models.} Most probable astrophysical parameters estimates obtained from posterior probability distribution from the MCMC routine using various systematics models. The planetary brightness phase variation is modeled as a first order Fourier series and the model includes stellar variability. The highlighted fits yield the greatest Bayesian Evidences. We note that the fit presented in the two last columns are inconsistent with a westward offset but their $\log(L)$ are significantly lower than the $\log(L)$ of Poly2 presented in Supplementary Table 3. The largest value of $\log L$ and $E$ for both binned and unbinned data are in bold blue. The errors on the parameters are the 68\% confidence region bounds of the posterior distribution obtained from \texttt{emcee}.}
	\label{Tab:Astro-Params-Estimate-v1-stellar}
\end{table} 


\begin{table}[!ht]
	\centering
	
	\scalebox{0.88}{
	\begin{tabular}{caccccc} \hline
		\textbf{Parameter} & \textbf{Poly2}&\textbf{Poly3} &\textbf{Poly4} & \textbf{Poly5} & \textbf{PLD1} &  \textbf{BLISS}\\ \hline
$t_0$ & $1.0744^{+0.0002}_{-0.0002}$ & $1.0744^{+0.0002}_{-0.0002}$ & $1.0745^{+0.0002}_{-0.0002}$ & $1.0745^{+0.0002}_{-0.0002}$ & $1.0747^{+0.0002}_{-0.0002}$ & $1.0745^{+0.0002}_{-0.0002}$ \\
$R_p/R_*$ & $0.1695^{+0.0009}_{-0.001}$ & $0.1689^{+0.0009}_{-0.001}$ & $0.1686^{+0.0009}_{-0.001}$ & $0.1686^{+0.001}_{-0.001}$ & $0.1699^{+0.0012}_{-0.0012}$ & $0.1681^{+0.001}_{-0.001}$ \\
$a/R_*$ & $6.6803^{+0.0228}_{-0.0239}$ & $6.6785^{+0.0245}_{-0.0238}$ & $6.686^{+0.0237}_{-0.024}$ & $6.6849^{+0.0234}_{-0.0224}$ & $6.6888^{+0.0246}_{-0.0251}$ & $6.6823^{+0.0245}_{-0.025}$ \\
$i$ & $87.8756^{+0.0888}_{-0.0886}$ & $87.8709^{+0.0911}_{-0.0896}$ & $87.8662^{+0.0937}_{-0.0926}$ & $87.8592^{+0.0968}_{-0.0943}$ & $87.8581^{+0.0923}_{-0.0909}$ & $87.8693^{+0.0954}_{-0.0927}$ \\
$e \cos{\omega}$ & $0.0^{+0.0004}_{-0.0004}$ & $-0.0^{+0.0004}_{-0.0004}$ & $-0.0^{+0.0005}_{-0.0005}$ & $-0.0^{+0.0004}_{-0.0004}$ & $0.0^{+0.0003}_{-0.0003}$ & $-0.0^{+0.0006}_{-0.0005}$ \\
$e \sin{\omega}$ & $0.0^{+0.0004}_{-0.0004}$ & $-0.0^{+0.0004}_{-0.0004}$ & $-0.0^{+0.0005}_{-0.0005}$ & $-0.0^{+0.0005}_{-0.0004}$ & $0.0^{+0.0003}_{-0.0003}$ & $0.0^{+0.0006}_{-0.0006}$ \\
$q_1$ & $0.0126^{+0.019}_{-0.0084}$ & $0.0142^{+0.0205}_{-0.0095}$ & $0.0147^{+0.0202}_{-0.0097}$ & $0.0139^{+0.0198}_{-0.0093}$ & $0.0206^{+0.0263}_{-0.0132}$ & $0.0164^{+0.0225}_{-0.0106}$ \\
$q_2$ & $0.2912^{+0.3934}_{-0.2163}$ & $0.2931^{+0.3882}_{-0.2177}$ & $0.2969^{+0.3635}_{-0.2218}$ & $0.2899^{+0.3927}_{-0.2192}$ & $0.3475^{+0.3813}_{-0.2509}$ & $0.2986^{+0.3773}_{-0.2215}$ \\
$\delta _e$ & $0.0042^{+0.0002}_{-0.0002}$ & $0.0044^{+0.0002}_{-0.0002}$ & $0.0041^{+0.0002}_{-0.0002}$ & $0.0042^{+0.0002}_{-0.0002}$ & $0.0045^{+0.0002}_{-0.0002}$ & $0.004^{+0.0003}_{-0.0003}$ \\
$A$ & $0.3662^{+0.0423}_{-0.0554}$ & $0.3879^{+0.0422}_{-0.0615}$ & $0.3294^{+0.0619}_{-0.0837}$ & $0.3565^{+0.0605}_{-0.0793}$ & $0.0209^{+0.0307}_{-0.0154}$ & $0.2474^{+0.0925}_{-0.1138}$ \\
$B$ & $0.1452^{+0.0427}_{-0.0435}$ & $0.1058^{+0.0441}_{-0.0431}$ & $0.0047^{+0.0567}_{-0.0571}$ & $0.0238^{+0.0601}_{-0.0608}$ & $0.0151^{+0.0265}_{-0.0203}$ & $-0.1335^{+0.0822}_{-0.086}$ \\
$C$ & $0.0624^{+0.0258}_{-0.0292}$ & $0.0618^{+0.0324}_{-0.0351}$ & $0.0955^{+0.0342}_{-0.04}$ & $0.092^{+0.0318}_{-0.0366}$ & $0.1955^{+0.0165}_{-0.0195}$ & $0.064^{+0.0379}_{-0.0446}$ \\
$D$ & $0.0795^{+0.0233}_{-0.0247}$ & $0.044^{+0.0236}_{-0.0253}$ & $-0.0084^{+0.0367}_{-0.0384}$ & $0.0022^{+0.0332}_{-0.0377}$ & $0.143^{+0.029}_{-0.0289}$ & $0.0359^{+0.0422}_{-0.0425}$ \\ \hline
$S_1$ & $-0.0018^{+0.0006}_{-0.0008}$ & $-0.0019^{+0.0006}_{-0.0008}$ & $-0.0038^{+0.0008}_{-0.0009}$ & $-0.0037^{+0.0008}_{-0.0009}$ & $-0.0045^{+0.0004}_{-0.0004}$ & $-0.0089^{+0.0017}_{-0.0018}$ \\
$S_2$ & $0.137^{+0.1387}_{-0.1027}$ & $0.4255^{+0.2129}_{-0.1939}$ & $0.3925^{+0.1183}_{-0.12}$ & $0.4565^{+0.1249}_{-0.1388}$ & $-0.3939^{+0.085}_{-0.0901}$ & $0.4327^{+0.0806}_{-0.0777}$ \\
$\sigma _F$ & $0.00146^{+3e-05}_{-2.9e-05}$ & $0.00145^{+2.8e-05}_{-2.8e-05}$ & $0.001434^{+2.9e-05}_{-2.8e-05}$ & $0.001435^{+2.9e-05}_{-2.8e-05}$ & $0.001961^{+4e-05}_{-3.8e-05}$ & $0.001516^{+3.1e-05}_{-2.9e-05}$ \\ \hline
$\log(L)$ & $7772.29$ & $7782.91$ & $7800.13$ & $\textcolor{blue}{\textbf{7801.06}}$ & $7383.08$ & $7728.82$ \\
$E$ & $\textcolor{blue}{\textbf{7693.52}}$ & $7689.82$ & $7689.13$ & $7668.59$ & $7293.57$ & $--$ \\ \hline
$\log(L)$\footnoteref{Unbinned} & $340908.31$ & $304922.02$ & $\textcolor{blue}{\textbf{341265.45}}$ & $341080.61$ & $318363.35$ & $338882.05$ \\
$E$\footnoteref{Unbinned} & $340783.86$ & $304774.93$ & $\textcolor{blue}{\textbf{341090.08}}$ & $340871.3$ & $318222.15$ & $--$ \\ \hline
$A_p$ & $0.0034^{+0.0004}_{-0.0004}$ & $0.0036^{+0.0005}_{-0.0004}$ & $0.0027^{+0.0006}_{-0.0004}$ & $0.0031^{+0.0005}_{-0.0005}$ & $0.0024^{+0.0002}_{-0.0002}$ & $0.0027^{+0.0005}_{-0.0005}$ \\
$\Phi _p$ & $0.4245^{+0.0943}_{-0.1006}$ & $0.2925^{+0.1132}_{-0.1195}$ & $-0.022^{+0.1824}_{-0.1761}$ & $0.0409^{+0.1698}_{-0.1635}$ & $0.3176^{+0.0755}_{-0.0755}$ & $-0.1415^{+0.4088}_{-0.2704}$ \\ \hline
	\end{tabular}
}
	\caption{\textbf{Fit parameters using a second order Fourier series to model the phase variation with stellar variation for different detector models.} Most probable astrophysical parameters estimates obtained from posterior probability distribution from the MCMC routine using various systematics models. The planetary brightness phase variation is modeled as a second order Fourier series and the model includes stellar variability. The highlighted fits yield the greatest Bayesian Evidences. The highlighted fits yield the greatest Bayesian Evidences. The largest value of $\log L$ and $E$ for both binned and unbinned data are in bold blue. The errors on the parameters are the 68\% confidence region bounds of the posterior distribution obtained from \texttt{emcee}. We note that the fit presented in last column shows an eastward offset but the $\log(L)$ is significantly lower than the $\log(L)$ of Poly2 presented in Supplementary Table 3. Despite having a high $\log(L)$, Poly 4 and Poly 5, which are consistent with a null planetary offset, have a $\Delta E > 6$ (or $\Delta \text{BIC}>12$) when compared to the fit with highest $E$. Therefore, these fits are significantly worse than Poly2 presented in Supplementary Table 3 \cite{Kass1995}.}
	\label{Tab:Astro-Params-Estimate-v2-stellar}
\end{table}

\clearpage\clearpage

\begin{table}[!ht]
	\centering
	
	\scalebox{0.88}{
	\begin{tabular}{ccacc} \hline
		\textbf{Parameter} & \textbf{Scheme 1}&\textbf{Scheme 2} &\textbf{Scheme 3} & \textbf{Scheme 4} \\ \hline
$t_0$ & $1.0745^{+0.0002}_{-0.0002}$ & $1.0744^{+0.0002}_{-0.0002}$ & $1.0742^{+0.0002}_{-0.0002}$ & $1.0744^{+0.0002}_{-0.0002}$ \\
$R_p/R_*$ & $0.1672^{+0.0012}_{-0.0012}$ & $0.1697^{+0.0009}_{-0.0009}$ & $0.1709^{+0.0009}_{-0.001}$ & $0.1717^{+0.001}_{-0.001}$ \\
$a/R_*$ & $6.7048^{+0.0254}_{-0.0264}$ & $6.6818^{+0.0228}_{-0.0229}$ & $6.6732^{+0.0241}_{-0.0244}$ & $6.67^{+0.0248}_{-0.0243}$ \\
$i$ & $87.8299^{+0.094}_{-0.0962}$ & $87.8771^{+0.0895}_{-0.092}$ & $87.8992^{+0.0906}_{-0.0947}$ & $87.8904^{+0.0943}_{-0.0882}$ \\
$e\cos{\omega}$ & $-0.0^{+0.0003}_{-0.0003}$ & $0.0^{+0.0004}_{-0.0004}$ & $-0.0^{+0.0005}_{-0.0005}$ & $0.0^{+0.0004}_{-0.0004}$ \\
$e\sin{\omega}$ & $0.0^{+0.0003}_{-0.0003}$ & $0.0^{+0.0004}_{-0.0004}$ & $-0.0^{+0.0005}_{-0.0005}$ & $-0.0^{+0.0004}_{-0.0004}$ \\
$q_1$ & $0.0218^{+0.0305}_{-0.0145}$ & $0.0118^{+0.0174}_{-0.0079}$ & $0.0085^{+0.0125}_{-0.0059}$ & $0.0125^{+0.017}_{-0.0082}$ \\
$q_2$ & $0.3157^{+0.3819}_{-0.2321}$ & $0.3198^{+0.3903}_{-0.2364}$ & $0.4278^{+0.3581}_{-0.302}$ & $0.4092^{+0.368}_{-0.2808}$ \\
$\delta _e$ & $0.0043^{+0.0002}_{-0.0002}$ & $0.0044^{+0.0002}_{-0.0002}$ & $0.0047^{+0.0002}_{-0.0002}$ & $0.0047^{+0.0002}_{-0.0002}$ \\
$A$ & $0.4^{+0.0196}_{-0.0216}$ & $0.445^{+0.0134}_{-0.0155}$ & $0.4633^{+0.0084}_{-0.0113}$ & $0.4424^{+0.0125}_{-0.0129}$ \\
$B$ & $0.1684^{+0.0364}_{-0.0346}$ & $0.1966^{+0.0329}_{-0.0326}$ & $0.1758^{+0.0318}_{-0.0316}$ & $0.2345^{+0.0348}_{-0.0349}$ \\
$C$ & $0.0936^{+0.0143}_{-0.0149}$ & $0.066^{+0.0136}_{-0.0129}$ & $0.0627^{+0.0111}_{-0.0115}$ & $0.0243^{+0.0153}_{-0.0153}$ \\
$D$ & $0.0782^{+0.0128}_{-0.0132}$ & $0.0686^{+0.0118}_{-0.0118}$ & $0.0554^{+0.0108}_{-0.0101}$ & $0.07^{+0.0139}_{-0.0142}$ \\
$\sigma _F$ & $0.002^{+0.0}_{-0.0}$ & $0.0015^{+0.0}_{-0.0}$ & $0.0017^{+0.0}_{-0.0}$ & $0.0018^{+0.0}_{-0.0}$ \\
$\log(L)$ & $7352.42$ & $7767.37$ & $7566.86$ & $7519.72$ \\
$E$ & $7280.82$ & $7695.76$ & $7495.25$ & $7448.11$ \\ \hline
$\log(L)$\footnoteref{Unbinned} & $283683.77$ & $340759.43$ & $309315.81$ & $297155.75$ \\
$E$\footnoteref{Unbinned} & $283570.62$ & $340646.28$ & $309428.95$ & $297268.89$ \\ \hline
$A_p$ & $0.0043^{+0.0002}_{-0.0002}$ & $0.0046^{+0.0003}_{-0.0003}$ & $0.0047^{+0.0002}_{-0.0002}$ & $0.0047^{+0.0002}_{-0.0002}$ \\
$\Phi _p$ & $0.412^{+0.0629}_{-0.0566}$ & $0.3994^{+0.0692}_{-0.0692}$ & $0.3616^{+0.0566}_{-0.0629}$ & $0.4874^{+0.0629}_{-0.0629}$ \\ \hline
	\end{tabular}
}
	\caption{\textbf{Fit parameters using a second order Fourier series to model the phase variation without stellar variation for different photometric extraction schemes.} Most probable astrophysical parameters estimates obtained from posterior probability distribution from the MCMC routine using various photometric schemes (See Photometry Extraction in Methods). The planetary brightness phase variation is modeled as a second order Fourier series and the detector model is second order polynomial. The highlighted fits yield the greatest Bayesian Evidences. The largest value of $\log L$ and $E$ for both binned and unbinned data are in bold blue. The errors on the parameters are the 68\% confidence region bounds of the posterior distribution obtained from \texttt{emcee}.}
	\label{Tab:Astro-Params-Estimate-v2-Photo}
\end{table}

\newpage
\section*{Supplementary Figures}

\renewcommand{\figurename}{Supplementary Figure}
\setcounter{figure}{0}

\begin{figure}[h!]
	\centering
	\includegraphics[width=\linewidth]{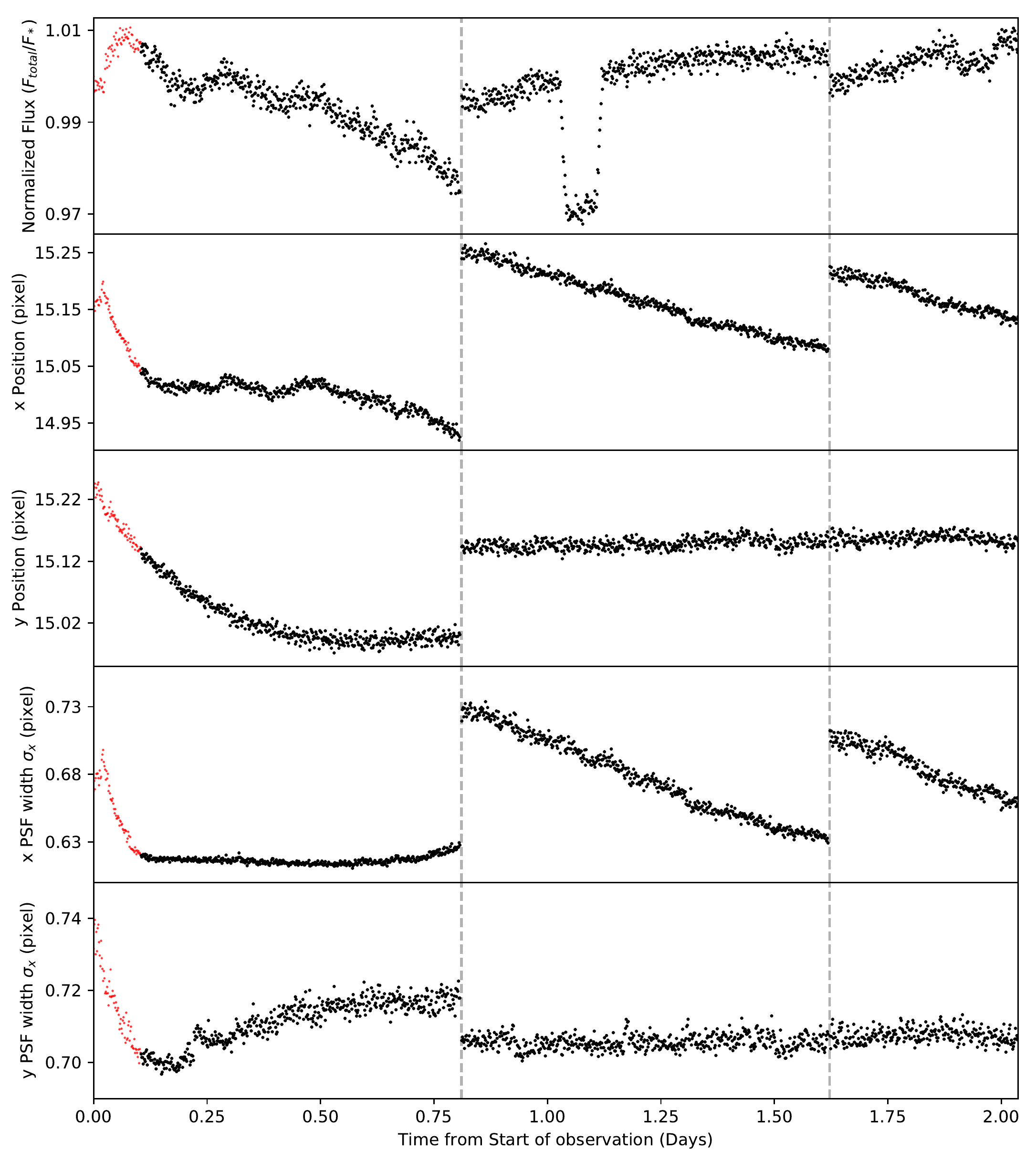}
	\caption{\textbf{Raw normalized photometry and PSF diagnostics.} \textit{Spitzer} 4.5 $\mu$m photometry and PSF diagnostics after median binning by data cube excluding the discarded AORs. The vertical gray dashed lines denotes the start and end of the different AORs. The red dots represent the data excluded from our analysis due to the rapid change in telescope pointing.}
	\label{Fig: Observation}
\end{figure}

\newpage

\begin{figure*}[!htpb]
	\centering
	\includegraphics[width=\textwidth]{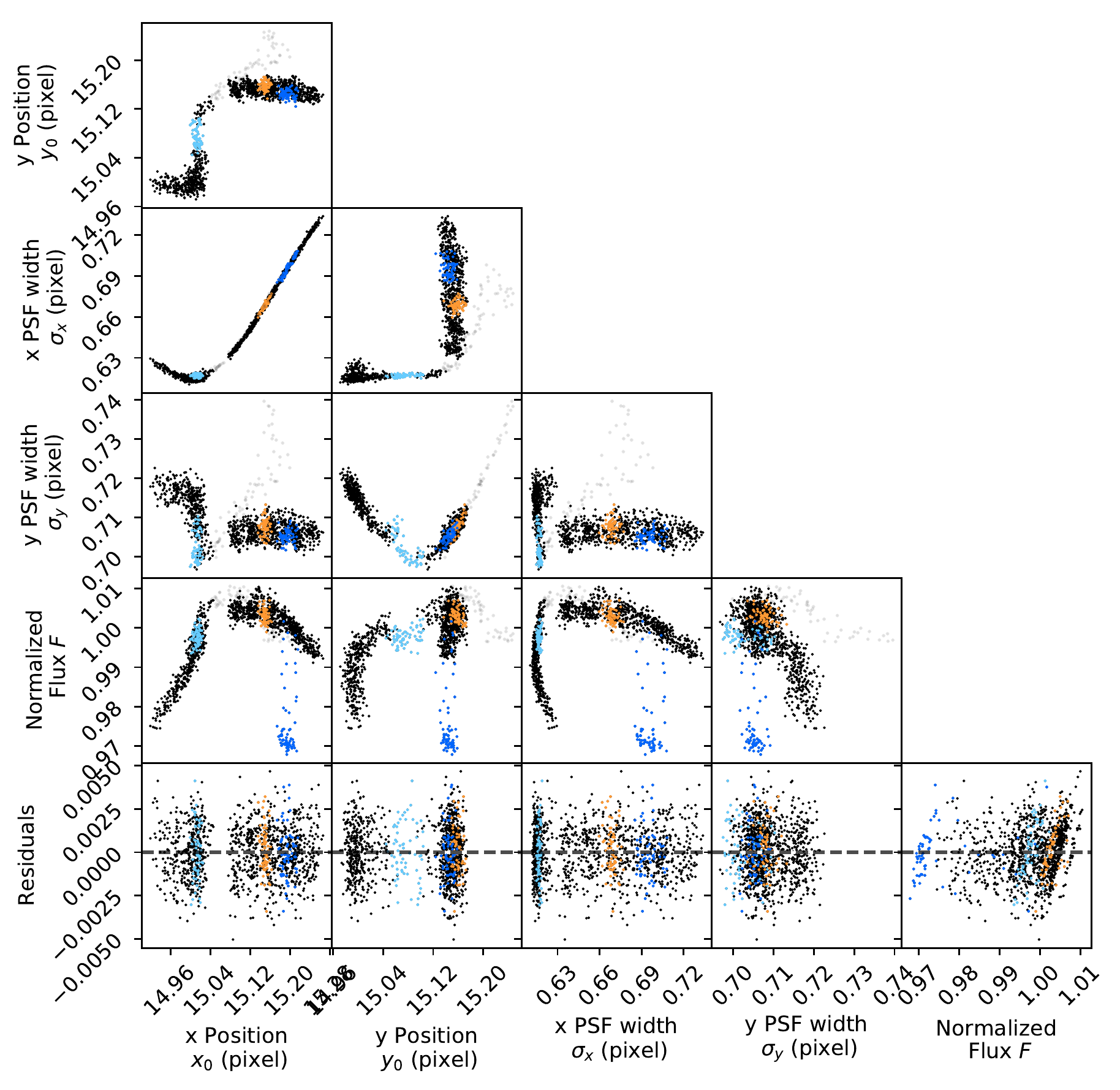}
	\caption{\textbf{PSF diagnostics for our observations of CoRoT-2b.} Each point represent the median of a data cube. The light and dark blue dots denote the first and second secondary eclipse, respectively. The orange dots represent the transit and the gray dots are the data cubes at the start of the observations that were discarded from our analysis.}
	\label{Fig: PSF Metric}
\end{figure*}

\newpage

\begin{figure}[!htpb]
	\centering
	\includegraphics[width=0.7\linewidth]{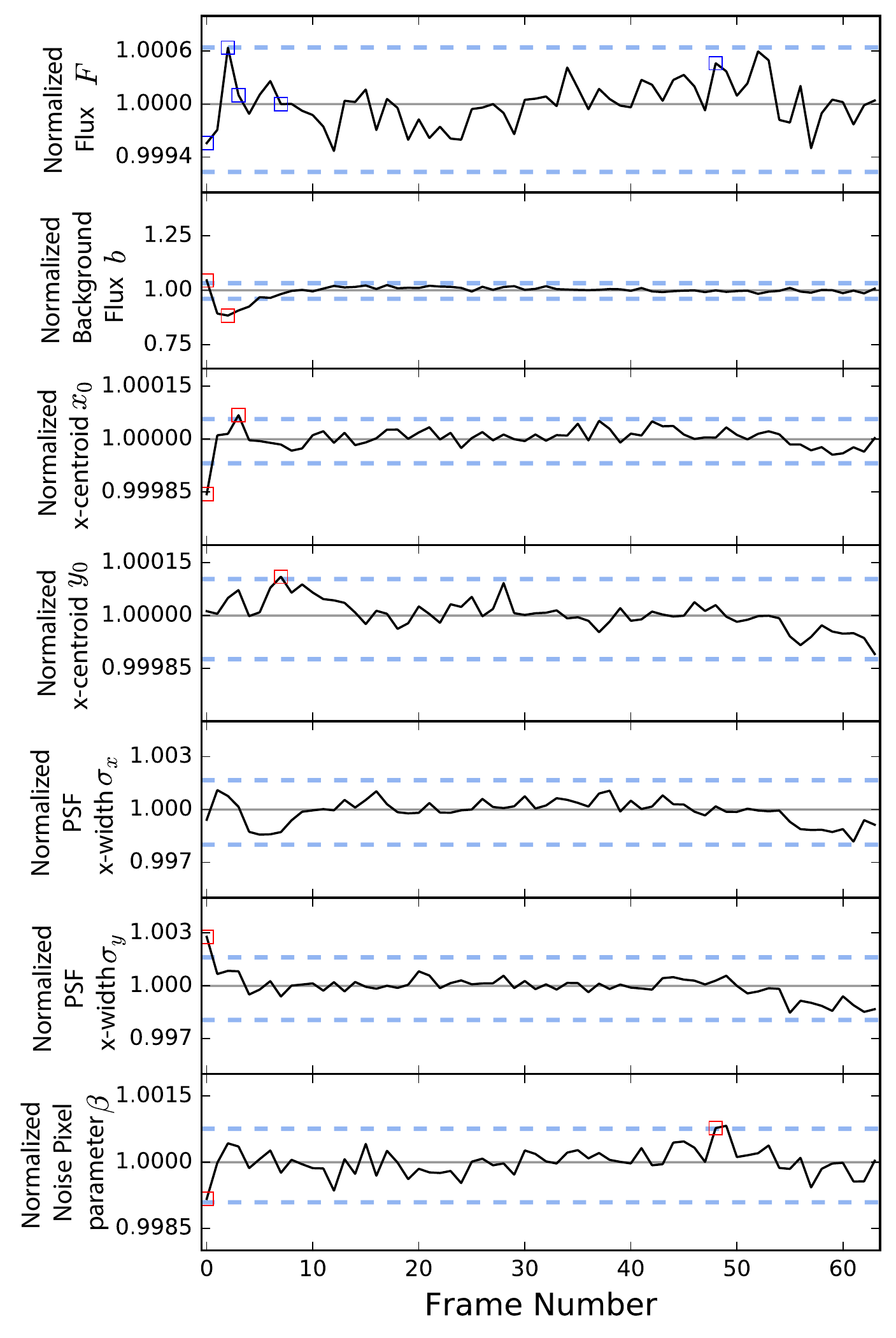}
	\caption{\textbf{Systematic changes in photometry within a 64-frame data cube.} All the values presented in this figure are normalized to their respective data cube median. The top panel shows the background-subtracted photometry using a 2.5 pixel radius hard circular aperture centered on the pixel (15, 15). The second panel shows systematic changes in the background flux and the panels below are the systematic changes in the PSF diagnostics. The last panel shows the systematic changes in the noise pixel parameter. The gray lines represent the mean parameter values and the blue dashed lines are the 3 $\sigma$ boundaries. The blue squares highlight to unusual frames with usable photometry despite their unusual PSF metrics identified with red squares.}
	\label{Fig: Frame Diagnotics}
\end{figure}

\begin{figure}[!htpb]
		\includegraphics[width=0.333\linewidth]{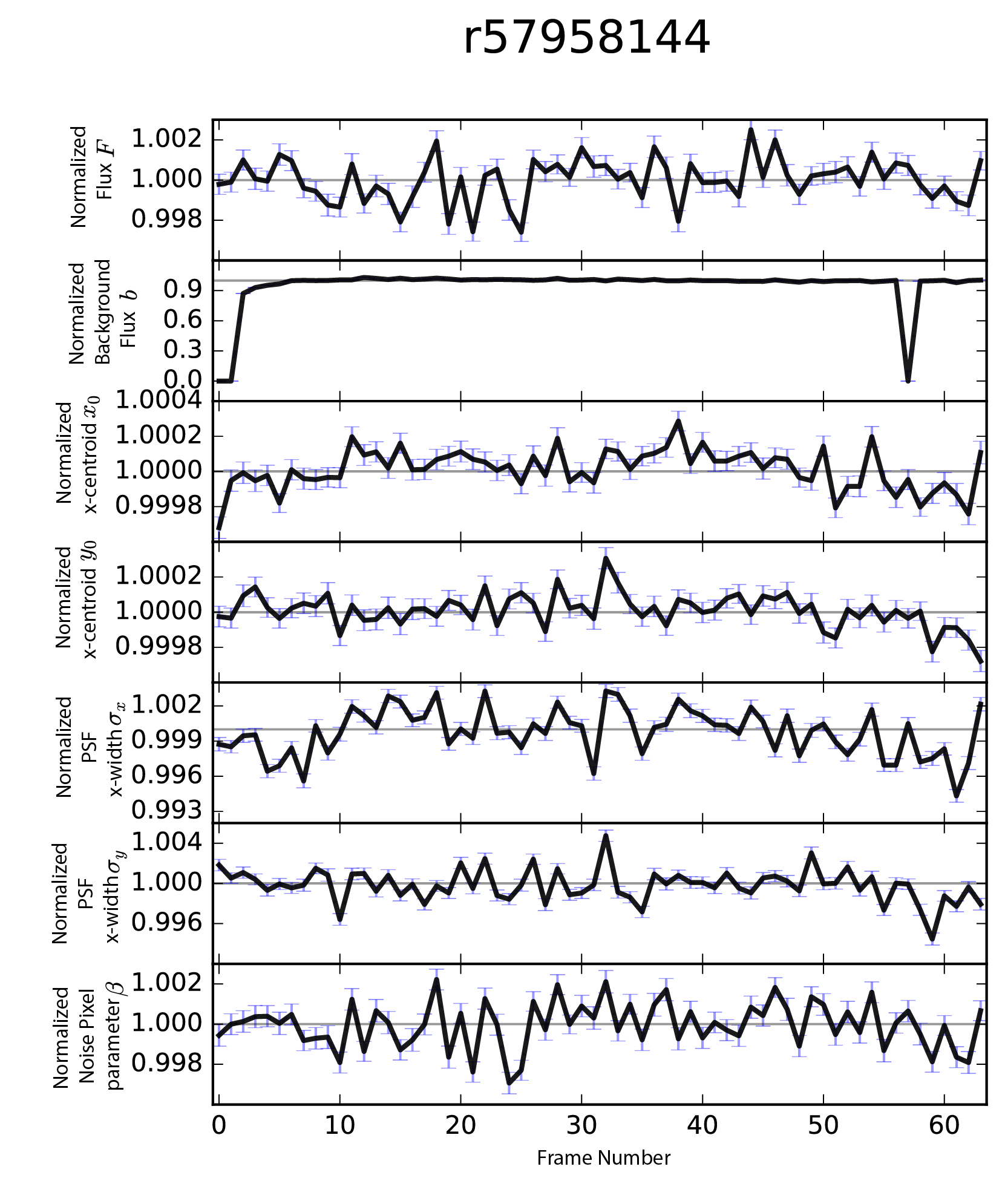}
		\includegraphics[width=0.333\linewidth]{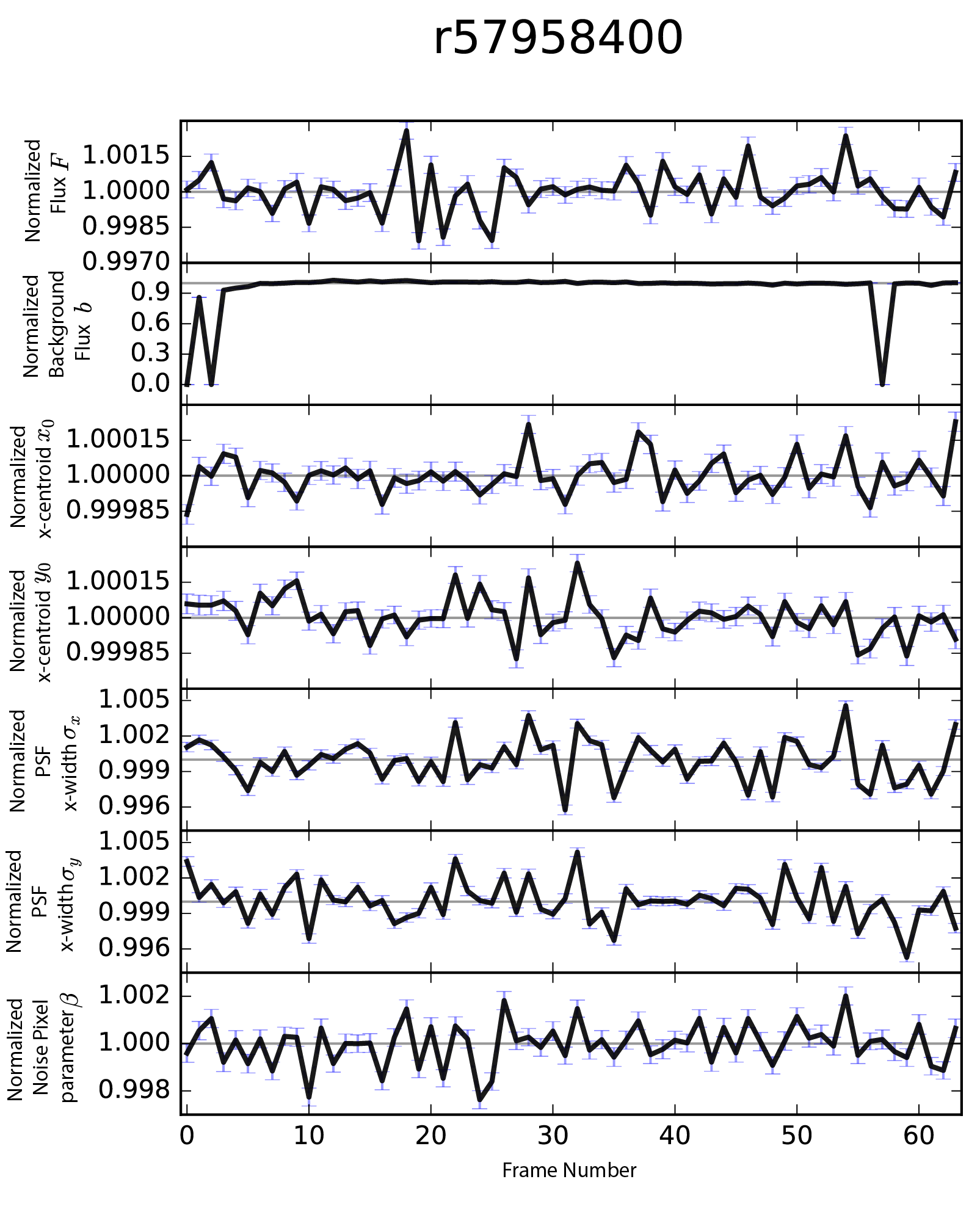}
		\includegraphics[width=0.333\linewidth]{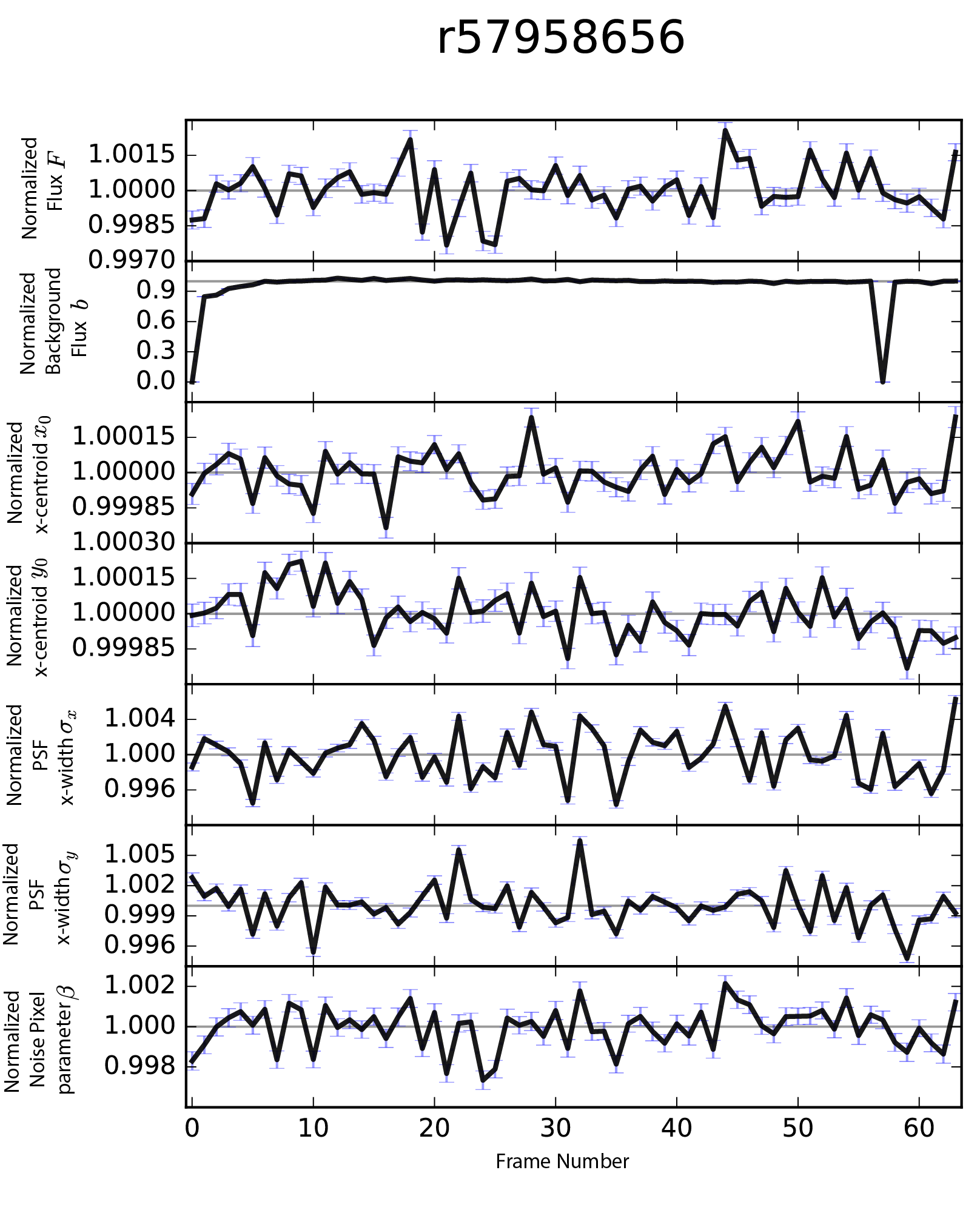}
	\caption{\textbf{Frame diagnostics for AORs r57958144, r57958400, and r57958656 respectively}. The background subtracted flux exhibit a repeating zigzag pattern between the 18$^{th}$ and 26$^{th}$ frames. This effect was introduced at the Sky Dark subtraction stage, the only frame-dependent process that affects IRAC data. We used an image stack provided by the IRAC team to remove this effect and also correct for the known low 58$^{th}$ frame background level error. Note that the first few frames will still have low backgrounds, which is due to the \textit{first frame effect} that impacts every IRAC observation and depends on the delay time since last exposure. This was not corrected by the image stack, but since it does not affect our photometry significantly, we chose to keep the first frame photometry.}
	\label{Fig: Frame-Dependent Systematics}
\end{figure}

\begin{figure}[!htpb]
	\centering
	\includegraphics[width=0.48\linewidth]{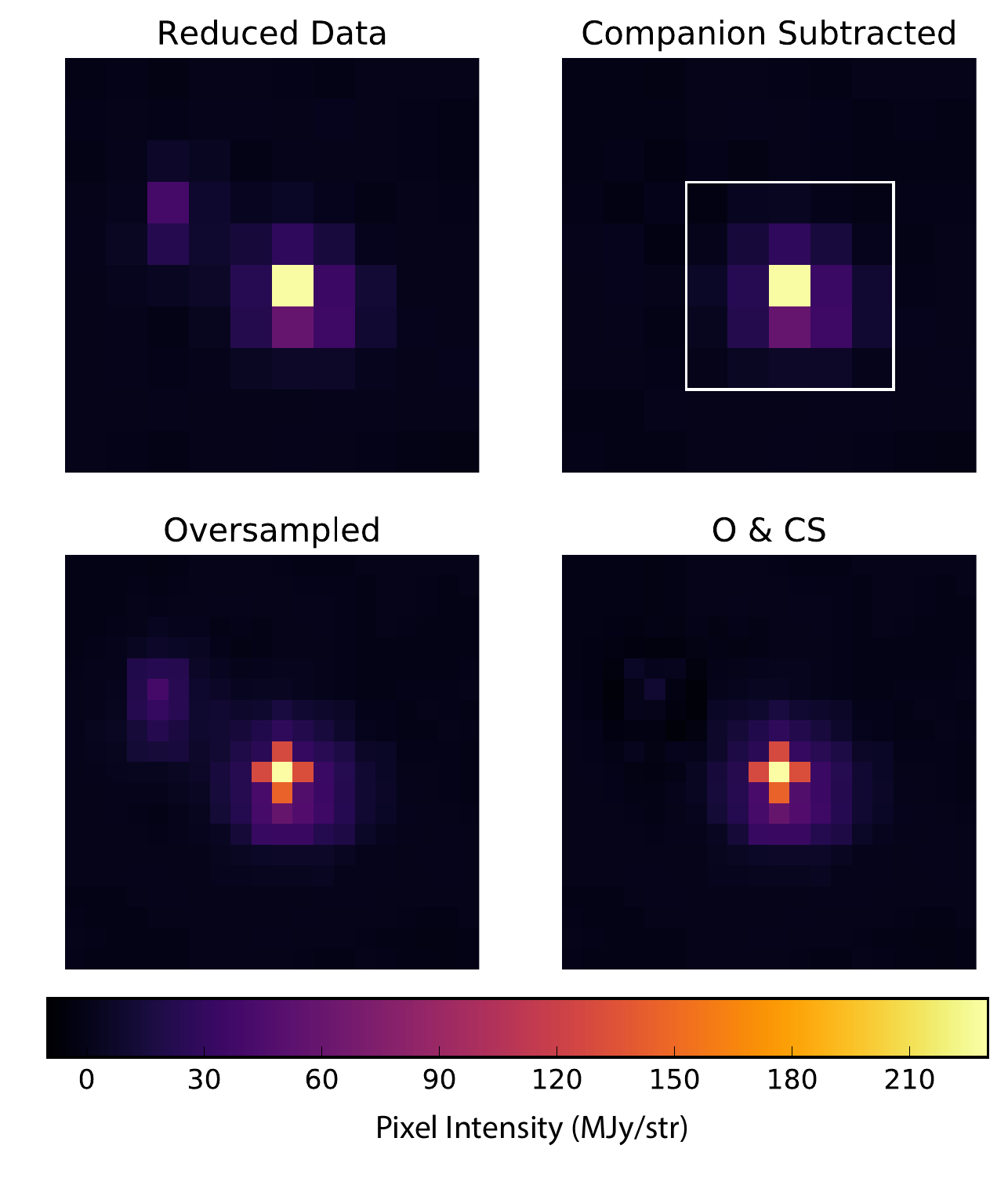}
	\caption{\textbf{Modified images for different photometric schemes.} \emph{Top Left:} Background-subtracted image of CoRoT-2b and its visual companion, 2MASS J19270636+0122577, on the left. \emph{Top Right:} Same image after the subtraction of the companion. The white rectangle encapsulates the pixels used to estimate the centroid and widths of the PSF. \emph{Bottom Left:} Oversampled image of the background subtracted image (top-left). \emph{Bottom Right:} Same image (bottom left) after the subtraction of the PSF of the companion.}
	\label{Fig: Companion Subtraction}
\end{figure}

\begin{figure}[!htpb]
	\centering
	\includegraphics[width=\linewidth]{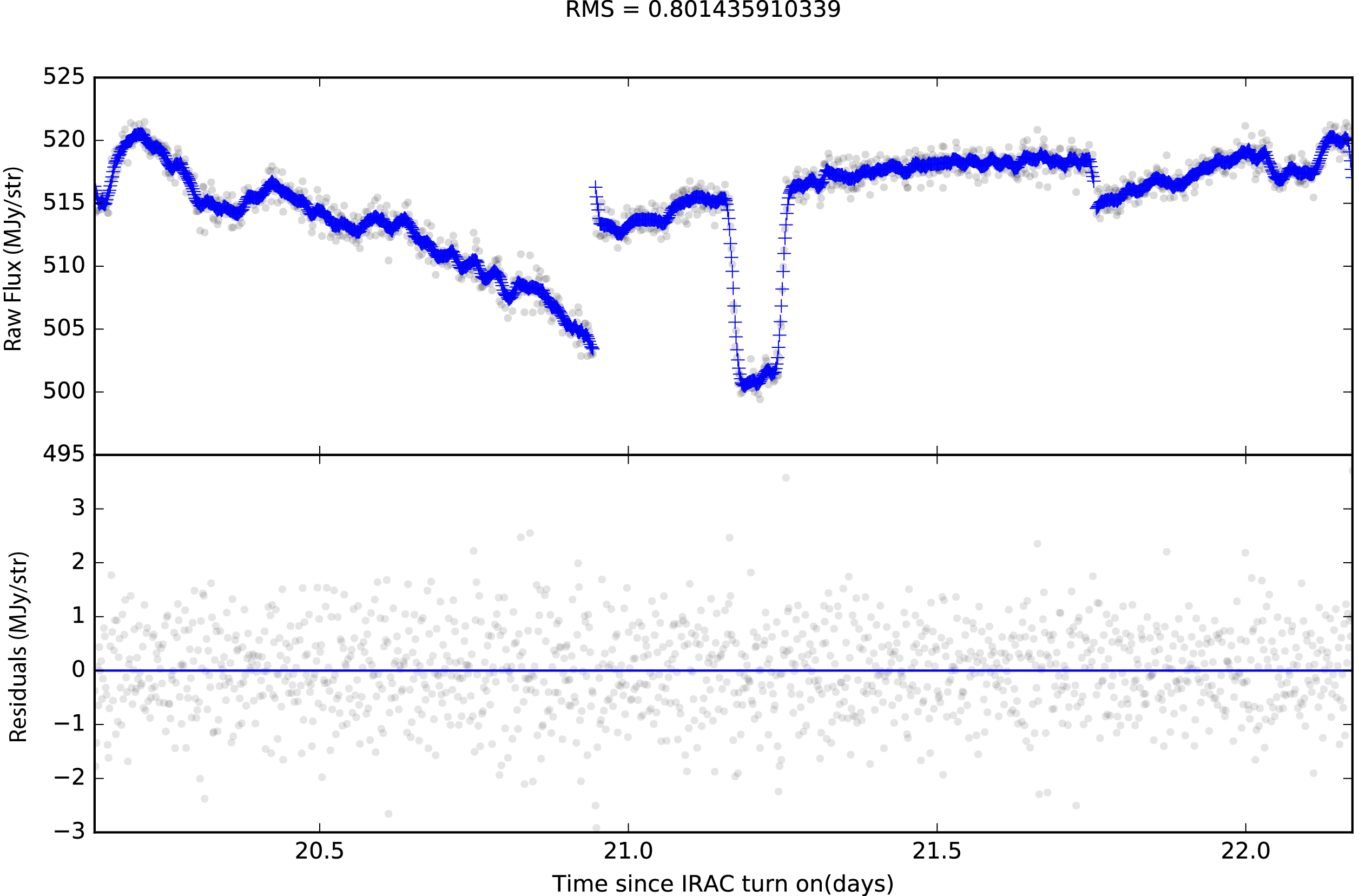}
	\caption{\textbf{Root mean square calculation example.} In the top panel, the grey points are raw data and the blue light curve is the smoothed lightcurve obtained by boxcar averaging with a length of 50 using the \texttt{astropy.convolution} module in Python. The lower panel show the difference between the raw and smoothed lightcurve used to estimate the RMS scatter.}
	\label{Fig: RMS Scatter Ex}
\end{figure}

\begin{figure*}[!htpb]
	\centering
	\includegraphics[width=\textwidth]{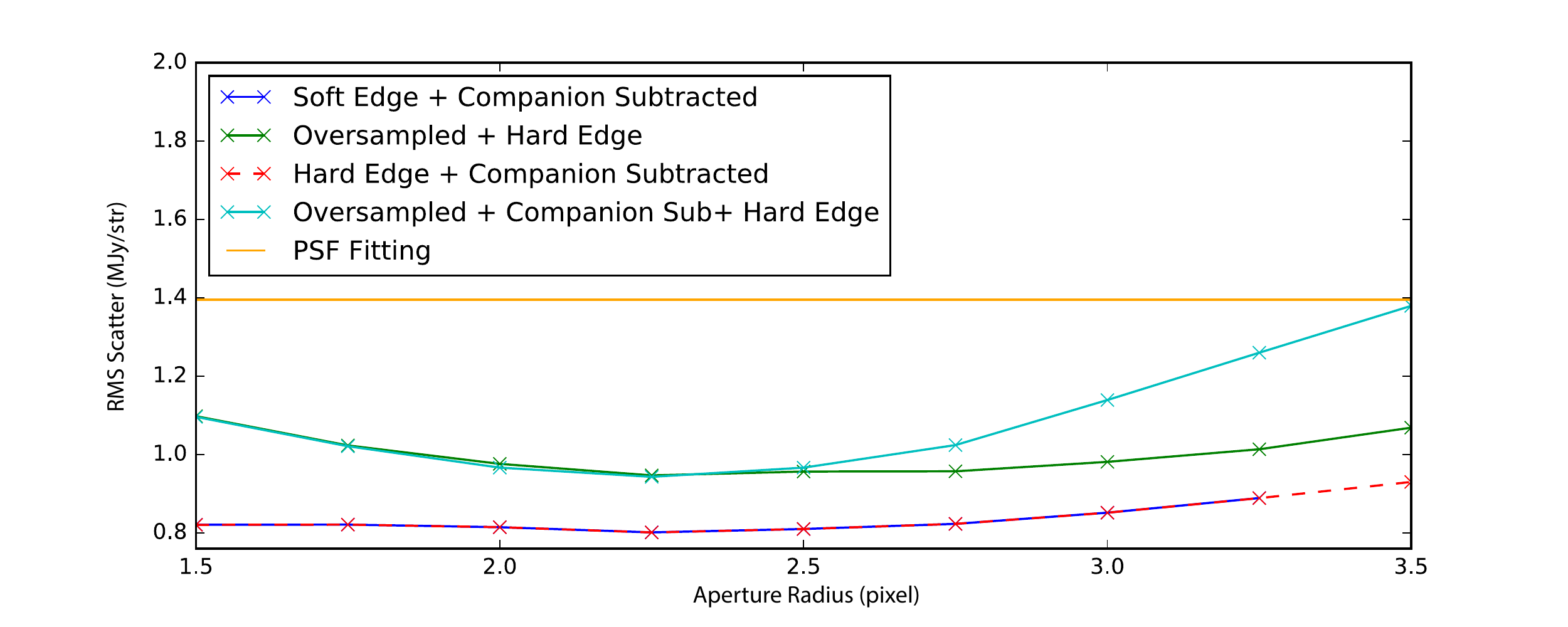}
	\caption{\textbf{RMS scatter for different photometric schemes.} In all cases, an aperture with a radius of $r=2.25$ is optimal as it minimizes the RMS scattering. The non-oversampled, soft-edge and companion subtracted photometric scheme yield the smallest RMS scatter. Note that the RMS scatter for PSF fitting is constant since there is no aperture involved.}
	\label{Fig: Photometry Comparison}
\end{figure*}

\begin{figure*}[!htpb]
	\centering
	\includegraphics[width=0.9\linewidth]{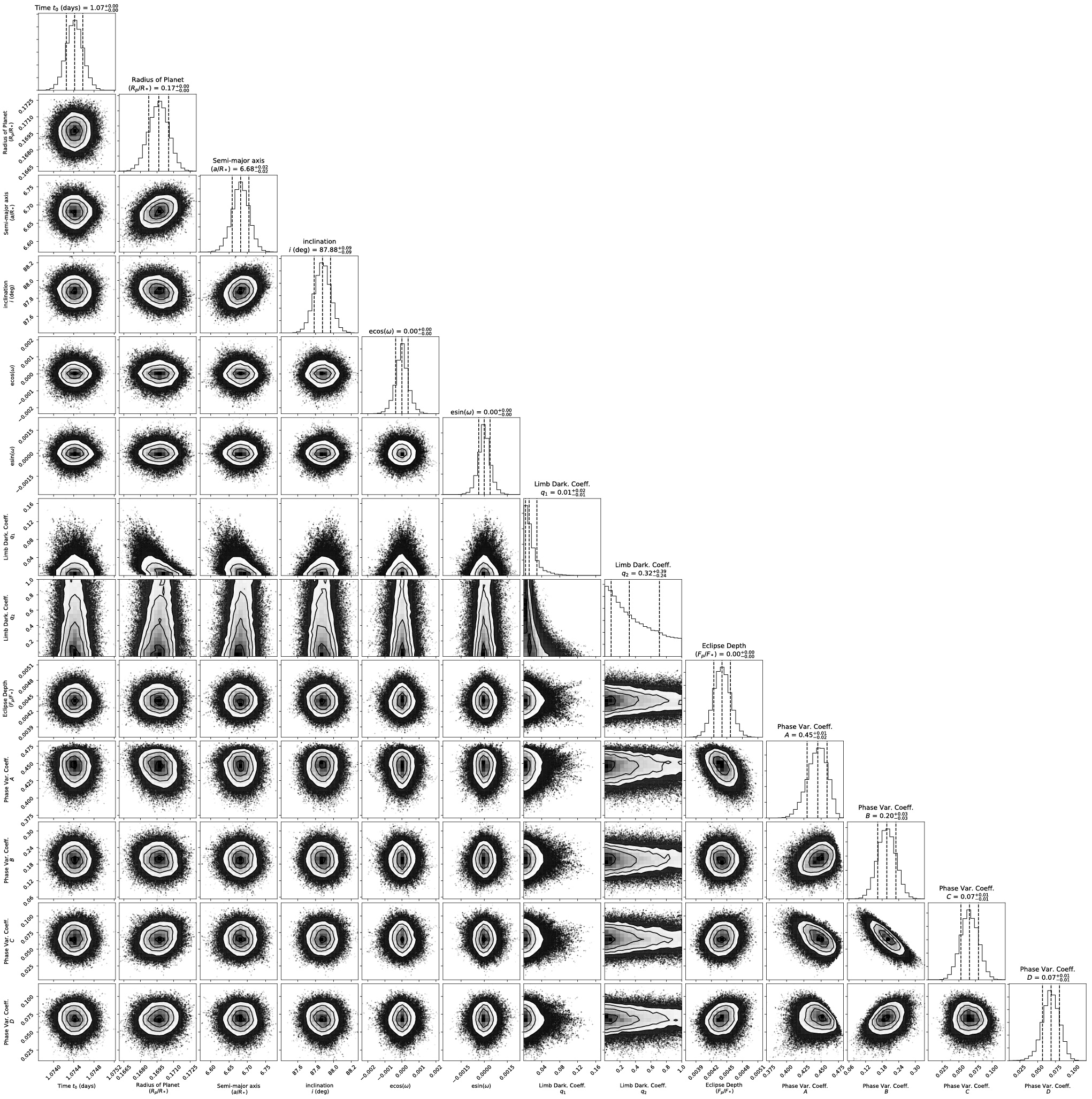}
	\caption{\textbf{Corner plot of the fit parameters' posterior distribution.} Pairs plot showing the posterior distribution of the astrophysical fitting parameters from MCMC. The panels on the diagonal show the marginalized
posterior distribution for each fitting parameter. The 68\% credible confidence region is marked by vertical dashed lines and quantified above the panel. The off-diagonal panels show the two-dimensional marginalized distribution for pairs of parameters, with the gray shading corresponding to the probability density and black contours indicating the 68\%, 95\%, and 98\% confidence regions. The outer black points are individual MCMC walkers positions outside of the 98\% confidence region.
This plot is made using the \texttt{corner} Python package \cite{corner}.}
    \label{Fig: Corner-Plot}
\end{figure*}

\newpage

\begin{figure*}[!htpb]
	\centering
	\includegraphics[width=0.9\linewidth]{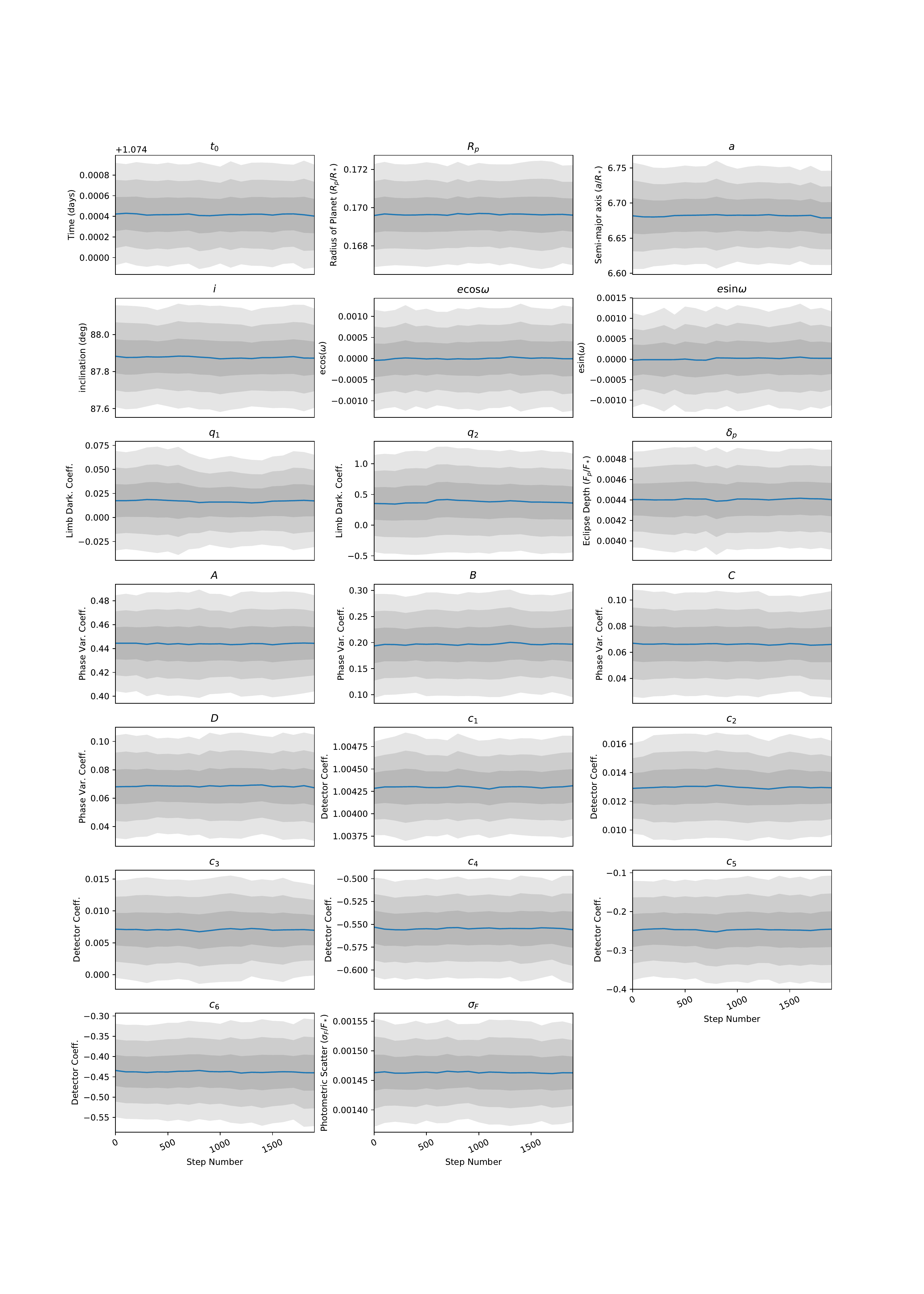}
	\caption{\textbf{Distribution of walkers positions for the last 2000 steps of the MCMC for our best fit model.} The blue line denoted the best-fit parameter value at each step and the gray areas are the 68\%, 95\% and 98\% confidence regions obtained from the distribution of the walkers at each step.}
    \label{Fig: Convergence-Plot}
\end{figure*}

\begin{figure}[!htpb]
	\centering
	\includegraphics[width=0.9\linewidth]{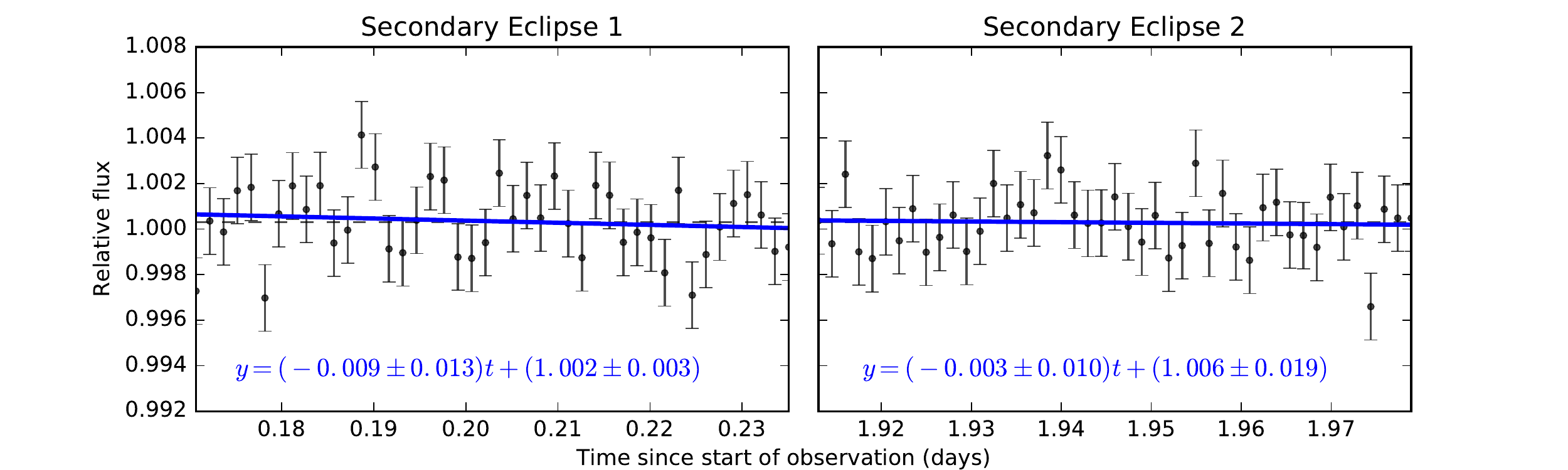}
	\caption{\textbf{In-eclipse diagnostics for the model with the greatest $E$.} The left and right panels show the first and second in-eclipse portions of the lightcurve respectively. The black points are the photometry after the removal of detector systematics (see Table \ref{Tab:Astro-Params-Estimate-v2}; Poly 2). We fit a linear function to the in-eclipse portion and find that the fit is consistent within 1 $\sigma$ with the absence of trend. The error bars are the photometric scatter estimated with a Markov Chain Monte Carlo (MCMC) in the fitting routine.}
    \label{Fig: In-eclipse_poly2-v2}
\end{figure}

\begin{figure}[!htpb]
	\centering
	\includegraphics[width=0.9\linewidth]{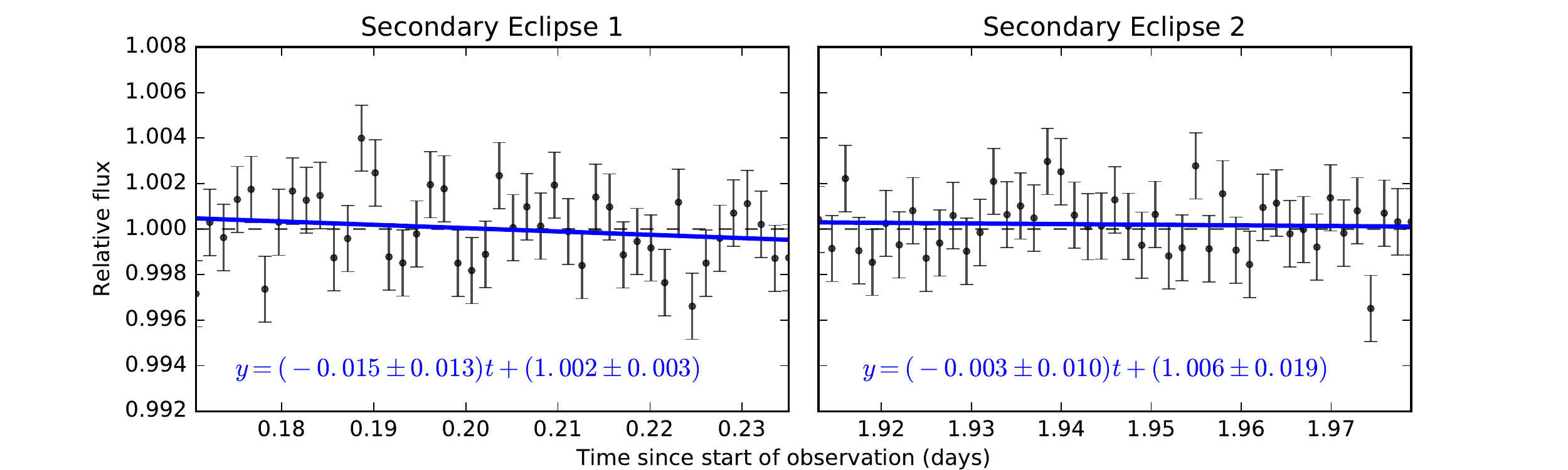}
	\caption{\textbf{In-eclipse diagnostics for the model with the second greatest $E$.} The left and right panels show the first and second in-eclipse portions of the lightcurve respectively. The black points are the photometry after the removal of detector systematics and stellar variability (see Table \ref{Tab:Astro-Params-Estimate-v1-stellar}; Poly3). We fit a linear function to the in-eclipse portion. We find that the fit for the second eclipse is consistent within 1 $\sigma$ with the absence of trend, but the first in-eclipse portion exhibit a trend with a slope of $-0.015 \pm 0.013$. The error bars are the photometric scatter estimated with a Markov Chain Monte Carlo (MCMC) in the fitting routine.}
    \label{Fig: In-eclipse_poly3-v1-stellar}
\end{figure}

\begin{figure}[!htpb]
	\centering
	\includegraphics[width=0.9\linewidth]{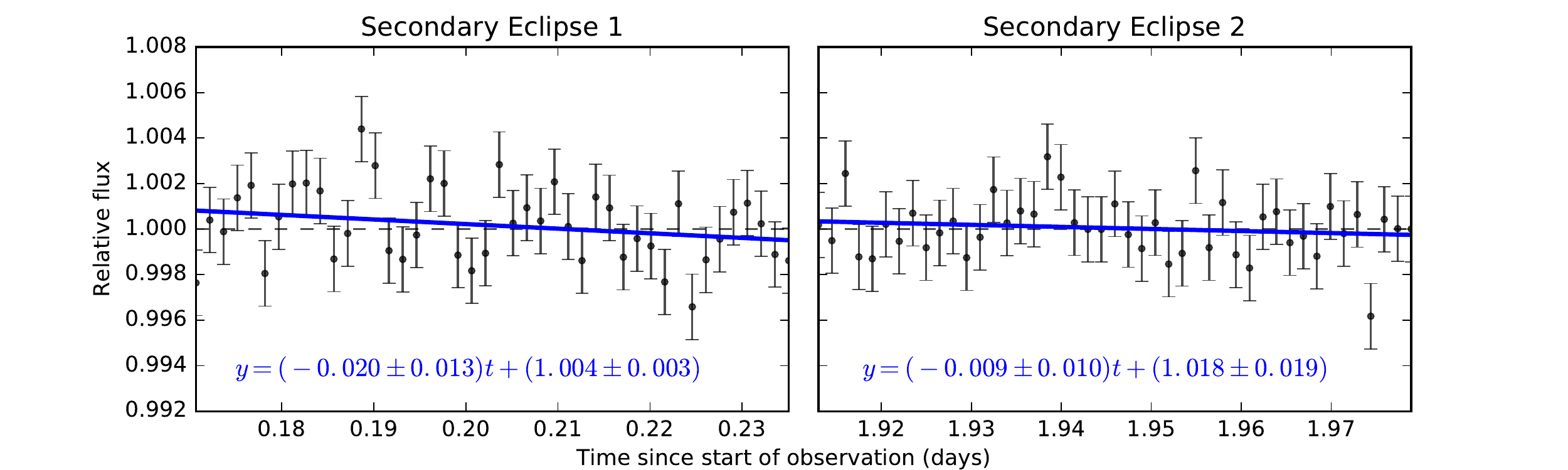}
	\caption{\textbf{In-eclipse diagnostics for the model with the third greatest $E$.} The left and right panels show the first and second in-eclipse portions of the lightcurve respectively. The black points are the photometry after the removal of detector systematics and stellar variability (see Table \ref{Tab:Astro-Params-Estimate-v1-stellar}; Poly4). We fit a linear function to the in-eclipse portion. We find that the fit for the second eclipse is consistent within 1 $\sigma$ with the absence of trend, but the first in-eclipse portion exhibit a trend with a slope of $-0.020 \pm 0.013$. The error bars are the photometric scatter estimated with a Markov Chain Monte Carlo (MCMC) in the fitting routine.}
	\label{Fig: In-eclipse_poly4-v1-stellar}
\end{figure}

\begin{figure}[!htpb]
	\centering
	\includegraphics[width=0.7\linewidth]{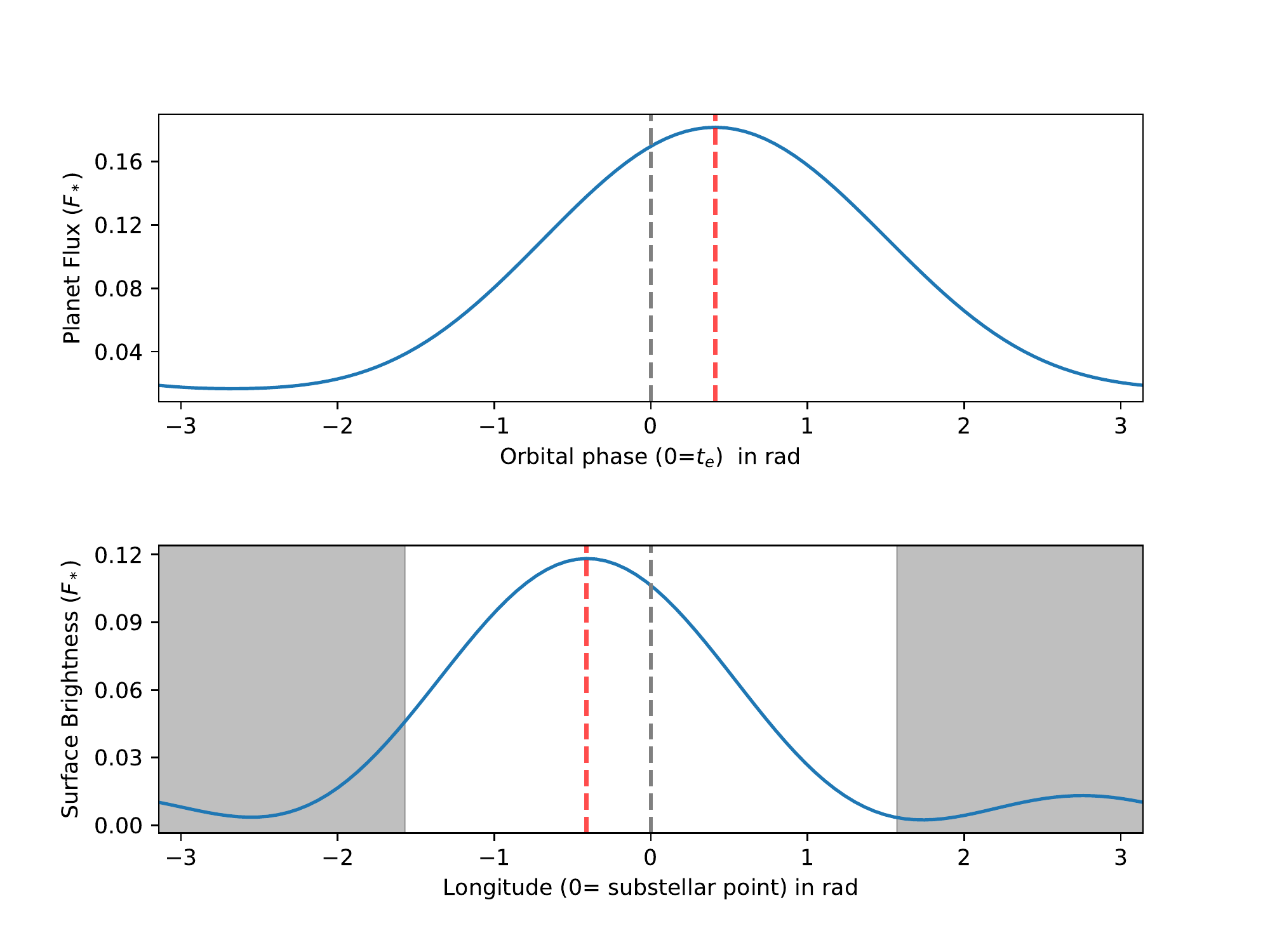}
	\caption{\textbf{1D brightness map of CoRoT-2b from inverting the orbital phase function}. \emph{Top panel}: The brightness variation of the planet as a function of orbital phase. The gray dashed line denote the orbital phase of secondary eclipse and the red dashed line denoted the orbital phase of the peak of the phase variation which is 0.41$\pm$0.06 rad after the secondary eclipse. \emph{Bottom panel} The surface brightness of CoRoT-2b as a function of longitude shown in the bottom panel \cite{2008ApJ...678L.129C}. The gray dashed line denotes the substellar meridian of the planet and the red dashed line denotes the brightest longitude on the planet located west of the substellar meridian. The gray shaded area represents the night hemisphere of the planet.}
	\label{Fig: Longitudinal Brightness}
\end{figure}

\begin{figure*}[!htpb]
	\centering
	\includegraphics[width=\linewidth]{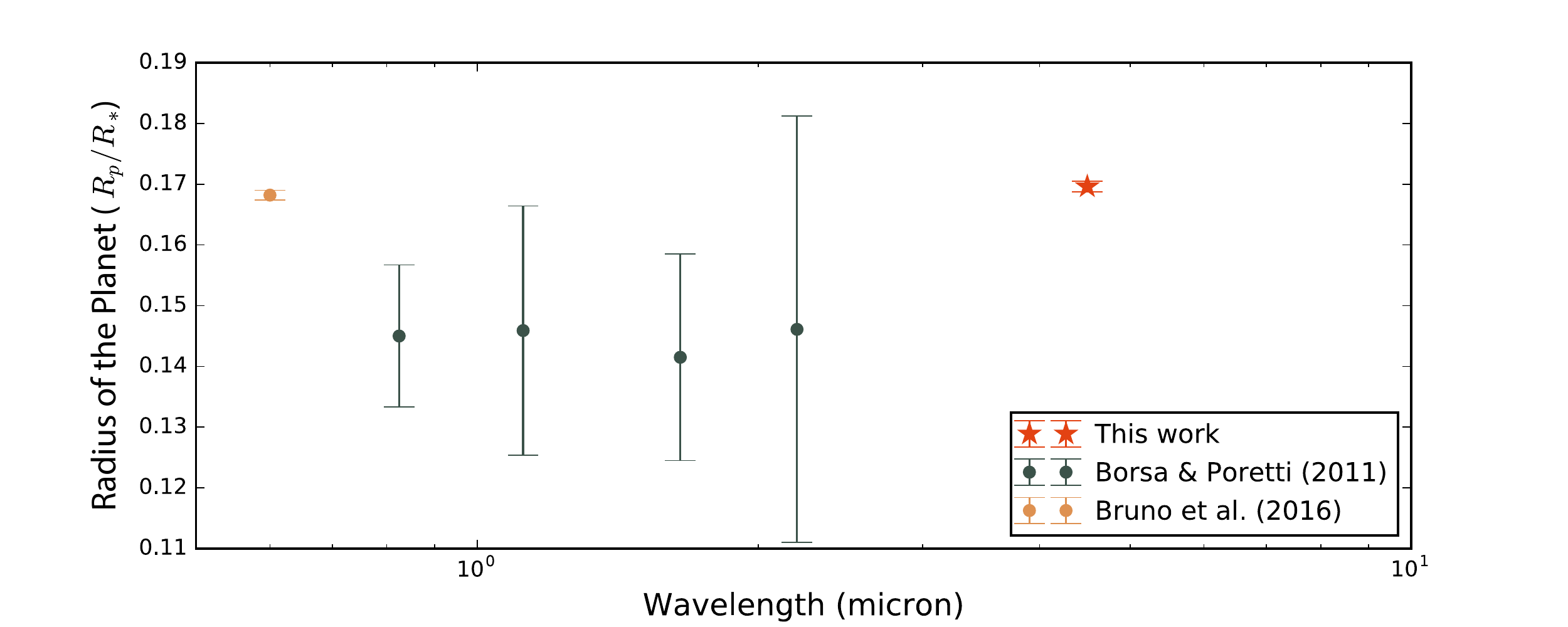}
	\caption{\textbf{Transmission spectrum of CoRoT-2b.} Our transit depth measurement and error estimates obtained with a Markov Chain Monte Carlo (MCMC) is shown along with ground-based measurements and their respective uncertainties \cite{2011MSAIS..16...80B} and the re-analysis of the \textit{CoRoT} observations \cite{2016A&A...595A..89B}. Given the large errors on the ground-based measurements and the sparsity of the measurements, we did not attempt to fit the transmission spectrum. Observations from future space mission such as JWST would be required to obtain a meaningful transmission spectrum.}
	\label{Fig: Transmission Spectrum}
\end{figure*}

\begin{figure*}[!htpb]
	\centering
	\includegraphics[width=\linewidth]{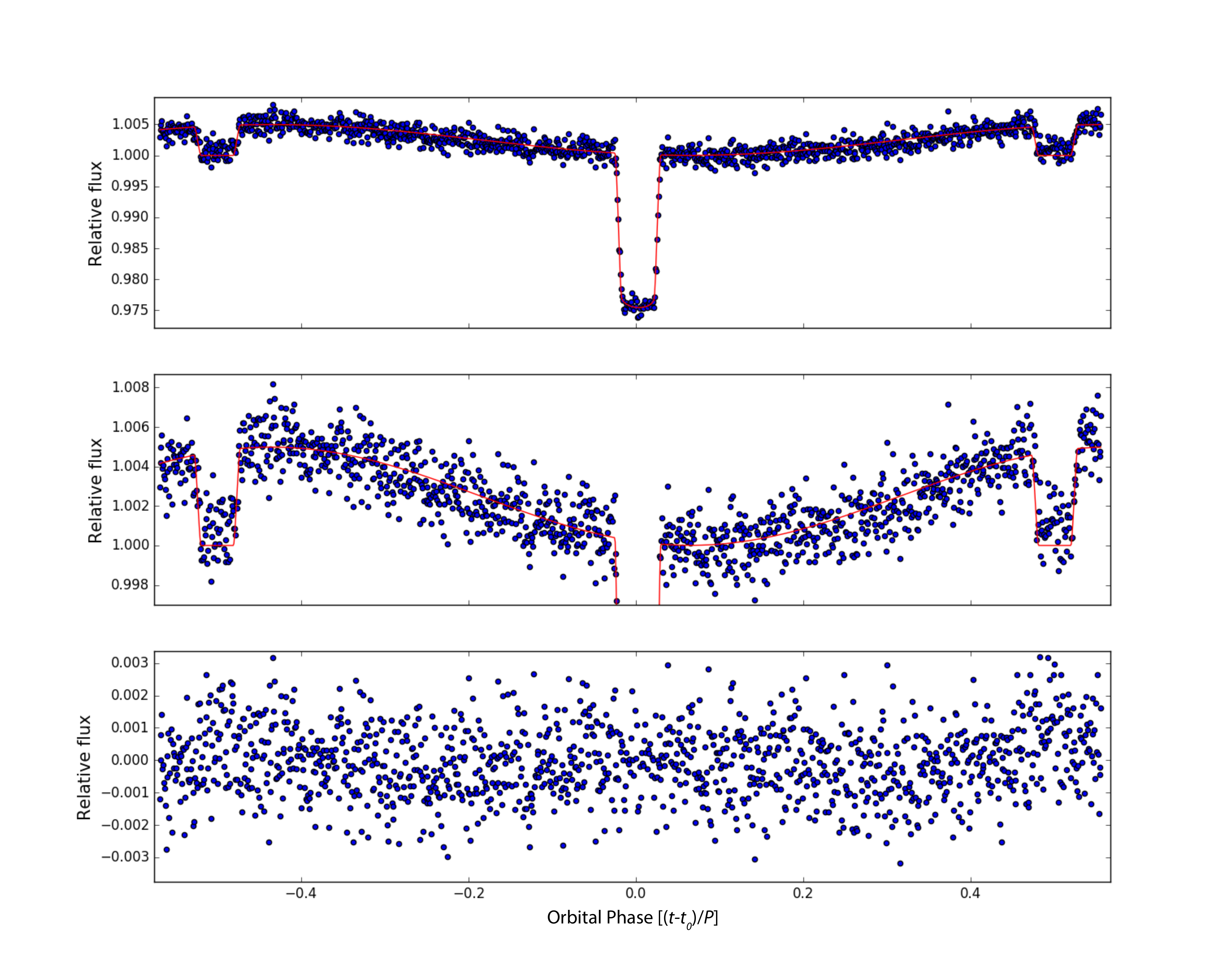}
	\caption{\textbf{Fit using an independent photometry and fitting pipeline \cite{2017arXiv171007642Z}.} The top panel show the astrophysical fit (line) and the data after the removal of the systematics (dots). The second panel is a zoomed-in version of the first panel to better see the planetary phase variation. The bottom panel shows the residuals between the corrected data and the astrophysical fit. Using an independent fitting pipeline, the result show a westward offset of $25.6 \pm 1.9$ degrees, which is consistent with the result obtained using the method described in this paper.}
	\label{Fig: Zhang Fit}
\end{figure*}

\newpage
\clearpage\clearpage

\section*{Supplementary Information}
\renewcommand\theequation{S\arabic{equation}}
\setcounter{equation}{0}

\subsection*{Data Reduction}
For our analysis, we use basic calibrated data which are corrected (dark subtracted, flat-fielded, linearized and flux calibrated) using the S19.2.0 IRAC pipeline. After the data cube diagnostics described in Supplementary Information, we find a frame respective flux modulation introduced by inaccurate dark subtraction. We then correct for the flux modulation using an image stack provided by the IRAC team, which also fixes the known 58th frame error in \textit{Spitzer} sub-array data \cite{2011Deming}.

We convert the pixel intensity from MJy/str to electron counts by multiplying the pixel values by GAIN $\times$ EXPTIME/FLUXCONV. We use the parameter values AINTBEG and ATIMEEND to obtain the middle of each exposure assuming uniform temporal spacing between each frame. We mask the pixels with $NaN$ values which are a result of energetic particle hits or pixel defects. Masking is preferred over replacing them with average values to minimize the correction and manipulation of the data. 

We perform pixel-level sigma clipping by comparing each pixel with the median of the same pixel of all the frames in its respective data cube and masking $4\sigma$ outliers. Frames containing a sigma-clipped pixel located in a $5 \times 5$ pixel box centered on the central pixel of the target are discarded entirely. A total of 191 images were tossed out, representing 0.22\% of the total data. 

We perform frame-by-frame background subtraction where sky background level is estimated as the median pixel value of the frame excluding a $7 \times 7$ pixel box centered on the pixel (15, 15) containing both the target and the companion.
Additionally, the retained data exhibit a 2.5 hours ramp-like behavior at the beginning with rapidly changing PSF metrics. This effect may be related to the settling of the telescope at a new pointing \cite{2012ApJ...754...22K}. Experimenting with and without removing the ramp-like behavior, we find that it is difficult to constrain the detector model during the 2.5 hours as the the PSF properties are notably different from the rest of the data, as described below. Since trimming is standard practice for \textit{Spitzer} phase curves \cite{2012ApJ...754...22K, 2015ApJ...811..122W}, we elect to discard the first 2.5 hours of data. After data removal, the remaining data we use for our analysis contains 1288 data cubes.

\subsection*{Photometry Extraction}

Observations of CoRoT-2 ($K=10.31$) \cite{2009A&A...506..501C} show the presence of a close-in visual companion, 2MASS J19270636+0122577 ($K=12.03$) \cite{2003yCat.2246....0C}. Due to the proximity of the companion, naively performing aperture photometry could lead to inaccurate estimation of the transit and secondary eclipse depths. We experiment with different strategies to retrieve our target's flux while reducing the contamination from the second source.

First, we fit for both sources simultaneously using two 2D Gaussians for each frame and retrieve the photometry from the fit. The second strategy is to fit for both sources, then subtract the fit for the companion from each frame and perform aperture photometry on the companion-subtracted image. The third strategy is to oversample the images by a factor of 2 and use aperture photometry. The fourth scheme combines the second and third strategy: we fit for both sources simultaneously using the oversampled images and then subtract the fit for the companion and use aperture photometry.

To retrieve the target's flux, we experiment with various apertures: hard-edged and soft-edged circular apertures of various radii. While the PSF metrics vary from one frame to another, we chose to keep the position of the aperture fixed. As the centroids only moves over a tenth of the area of a pixel throughout the observation, an aperture of radius 2-3 pixels is large enough to collect all the flux despite the changes in centroid. In principle, an aperture varying in shape and size should improve the photometry, but in practice, we find that a fixed aperture performs better. This suggests that the  uncertainties on measurements of the PSF's position and shape for each frame introduce noise into the time-varying aperture photometry.

To determine the best photometric schemes, we calculate the root-mean-squared (RMS) scatter for each light curve as shown in Supplementary Fig. \ref{Fig: Photometry Comparison} and choose the one exhibiting the smallest RMS scatter. In general, we find that the light curves obtained from PSF fitting and aperture photometry on oversampled images exhibit larger RMS scatter. Ultimately, we use photometry on non-oversampled images using a soft-edged circular aperture with a radius of 2.25 pixel after the subtraction of the PSF of the companion as it yields the smallest RMS scatter. 

Moreover, the residual flux from the companion subtraction is less than 0.05\% of the target flux, which does not significantly impact centroid measurements. We estimate the residual flux by placing a circular aperture ($r=2.25$ pixels) at a symmetric location on the other side of CoRoT-2's companion on the companion-subtracted images. Therefore, the counts in this aperture are the residuals from companion subtraction.

\subsection*{Centroids}

Due to drift and jitter of the telescope pointing, the position of the target point spread function (PSF) on the detector varies with time. After the frame by frame removal of the companion, we determine the centroid ($x_0, y_0$) of CoRoT-2 in each frame by calculating the flux-weighted mean of a $5 \times 5$ box centered on the brightest pixel located at (15, 15):

\begin{equation}
\centering
x_0 = \frac{\sum_i F_i x_i}{\sum_i F_i},
\end{equation} 

\begin{equation}
\centering
y_0 = \frac{\sum_i F_i y_i}{\sum_i F_i}.
\end{equation} 

The shape of the PSF also changes from one frame to another. We first calculate the target's noise pixel parameter, $\tilde{\beta}$ \cite{2005Mighell}:

\begin{equation}
\centering
\tilde{\beta} = \frac{(\sum _i I _i)^2}{\sum _i I _i ^2}.
\end{equation}.

The noise pixel parameter is commonly used as an estimate of the PSF width assuming an isotropic PSF. We instead opt to estimate the $x$ and $y$ extent of the point spread function of the target separately by computing the standard deviation along each direction for each frame:

\begin{equation}
\centering
\sigma _x = \sqrt{\frac{\sum _i F_i (x_i - x_0)^2}{\sum _i F_i}},
\end{equation}

\begin{equation}
\centering
\sigma _y = \sqrt{\frac{\sum _i F_i (y_i - y_0)^2}{\sum _i F_i}}.
\end{equation}

As shown in Figure \ref{Fig: PSF Metric}, the size of the PSF shape is a non-monotonic function of the centroid position on the pixel. As previously mentioned, the data collected during the first 2.5 hours of the observation exhibit a ramp-like behavior which coincide with a brief change in the position and shape of the PSF \cite{2014A&A...572A..73L}.

\subsection*{Noise in \textit{Spitzer} IRAC Data due to Bias Dark Subtraction}


Unlike cryogenic \textit{Spitzer} data, the \textit{Warm Spitzer} sub-array data exhibit a frame dependent background flux systematics \cite{2011Deming}. Such known systematics include the \textit{58$^{th}$ frame error} and the \textit{first frame effect}. The 58$^{th}$ frame error is due to a problem in the skydark subtraction stage which leaves the background level in that frame different from the rest. On the other hand, the first few frames have low backgrounds, due the "first frame effect" which impacts every IRAC observation and depends on the delay time since last exposure. In principle, the background-subtracted flux of the target should be immune to such variations, but this has not been borne out in practice, leading researchers to remove certain frames from their analysis \cite{2011Deming}. 

Once we obtained the sky background level, centroid position, and PSF shape for each frame, we perform aperture photometry on the background and companion subtracted images for each frame using a soft-edged 2.25 pixel radius circular aperture. We normalize each value to its respective data cube median. We then find the median value for each frame number presented in Figure \ref{Fig: Frame Diagnotics}. While performing this analysis, we notice that the background subtracted flux exhibits a repeating zigzag pattern between the $18^{th}$ and  $26^{th}$ from one AOR to the other as shown in Figure \ref{Fig: Frame-Dependent Systematics}. This modulation was introduced at the Sky Dark subtraction stage of the S19.2.0 IRAC pipeline, the only frame-dependent process that affect IRAC data. Indeed, when we perform aperture photometry on the central pixels of the dark calibration cube, we see the same zigzag pattern in reverse. The IRAC team provided us with an image stack to remove this effect which also fixed the 58$^{th}$ frame error. After the correction, our analysis shows no significant flux variation, therefore all frames within a data cube provide usable photometry.

\subsection*{Binning}
Although the recalibration cube provided by the IRAC team corrected the obvious frame dependent systematics (see Supplementary Information), subarray data are still subject to effects such as the \textit{first frame effect}, which in principle should not affect the photometry. Nonetheless, we choose to play it safe and elect to median bin the data by datacube. Given our 2 seconds exposures, the binned data have a temporal resolution of 128 seconds. Since the duration of ingress and egress of the system is over 1400 seconds, such resolution is still short enough to resolve the shape of occultations.

Binning data before fitting a model has many advantages \cite{2015ApJ...805..132D}. First, binning data filters out high frequency noise, including the datacube systematics. Secondly, it increases the accuracy of our measurement of the PSF metrics. Our instrumental models are a function of PSF metrics, hence more accurate measurements improve our ability to decorrelate the detector systematics from the astrophysical signal. Finally, reducing the number of data ultimately makes model fitting significantly faster. 

\subsection*{Upper Limit on Stellar Variability at 4.5 $\mu m$}

The Spitzer Space Telescope is on a heliocentric Earth-trailing orbit and is drifting away from Earth at about 0.1 AU per year. Consequently, at the time of our CoRoT-2 system observations, \textit{Spitzer} was approximately 1.5 AU away from Earth and therefore had a significantly different field of regard than the Earth. For this reason, we could not obtain ground-based optical observations of the system around the time of the \textit{Spitzer} observations to monitor stellar variability. 

Instead, we estimate the upper limit of the magnitude of stellar variability at 4.5 $\mu$m using the observations acquired by the \textit{CoRoT} mission. The \textit{CoRoT} observations show that the optical stellar flux varies by at most 5\% in 2-day intervals \cite{Alonso2008}, due to the inhomogeneous star spot area coverage of CoRoT-2.  Using the reported mean star spots temperature \cite{2010A&A...510A..25S}, $T_{\circ}$, of 4700 $\pm$ 300 K, so assuming an isophotal wavelength, $\lambda$, of the CoRoT passband of 700 nm, one can approximate an out of transit stellar flux as:

\begin{equation}
\centering
F_s(T_*, T_{\circ}, f, \lambda) = (1-f) \; B_{\lambda}(T_*, \lambda) \; + \; f \; B_{\lambda}(T_{\circ}, \lambda)
\end{equation}

\noindent where $T_* = 5625$ K is the effective photospheric temperature of star, $f$ is the fraction of total spot area and $B_{\lambda}(T, \lambda)$ is Planck's law. Assuming a 4.0\% spot coverage on one of the hemispheres, we calculate that a 13\% spot coverage on the other hemisphere corresponds to the maximal 5\% flux variation. Extrapolating this to 4.5 $\mu m$, we find a stellar variability upper limit of 2.0\%,


\subsection*{Surface Brightness}

As described in the methods section, the planetary phase variation can be described more generally as a Fourier series of order $N$:

\begin{equation}
\centering
\label{Eq: flux phase}
F_p(\xi) = F_0 + \sum ^N _{j=1} C_j \cos (j\xi) + D_j \sin (j\xi)
\end{equation}

\noindent where $\xi$ is the orbital phase. Note that sinusoidal modes with odd $j$ other than $j=1$ are not expected to have a phase function signature for an edge-on orbit \cite{2008ApJ...678L.129C, 2013MNRAS.434.2465C}. If the rotation period of the planet is known, then phase variations allow us to constrain the longitudinal brightness of a planet. For tidally locked planets, the above phase variation corresponds to a longitudinal surface brightness map $J(\phi)$ given by:

\begin{equation}
\label{Eq: longitude}
J(\phi) = A_0 + \sum ^N _{j=1} A_j \cos (j\xi) + B_j \sin (j\xi)
\end{equation}

\noindent where $\phi$ is the longitude from the substellar point. One can directly relate the coefficient from equations \ref{Eq: flux phase} and \ref{Eq: longitude} \cite{2008ApJ...678L.129C}:
\begin{equation}
\centering
\begin{split}
A_0 & = \tfrac{1}{2}F_0\\
A_1 & = \tfrac{2}{\pi}C_1\\
B_1 & = \tfrac{-2}{\pi}D_1\\
& \vdots\\
A_j & = (-1)^{j/2}\left[\tfrac{-(j^2-1)}{2}\right]C_j\\
B_j & = (-1)^{j/2}\left[\tfrac{(j^2-1)}{2}\right]D_j
\end{split}
\end{equation}

\noindent where $j$ is even. Phase variations do not provide any latitudinal brightness constraints for an edge-on orbit. One can therefore express the flux contribution of an infinitesimal longitudinal slice as:

\begin{equation}
\centering
J(\phi) = \int ^\pi _0 I(\phi, \theta) \sin^2 \theta d\theta
\end{equation}

\noindent where $\theta$ is the latitude from the substellar point and $I(\phi, \theta)$ is the brightness at the coordinates $(\phi, \theta)$. Assuming that  the brightness drops off away from the equator as the sine of co-latitude, one can express intensity at a infinitesimal surface area at longitude $\phi$ and latitude $\theta$ as:

\begin{equation}
I(\phi, \theta) = \frac{3}{4} J(\phi) \sin (\theta).
\end{equation}

\renewcommand{\refname}{Supplementary References}


\begin{thebibliography}{10}
	\expandafter\ifx\csname url\endcsname\relax
	\def\url#1{\texttt{#1}}\fi
	\expandafter\ifx\csname urlprefix\endcsname\relax\def\urlprefix{URL }\fi
	\expandafter\ifx\csname doiprefix\endcsname\relax\def\doiprefix{DOI }\fi
	\providecommand{\bibinfo}[2]{#2}
	\providecommand{\eprint}[2][]{\url{#2}}
	
	\bibitem{2007Natur.447..183K}
	\bibinfo{author}{{Knutson}, H.~A.} \emph{et~al.}
	\newblock \bibinfo{journal}{\bibinfo{title}{{A map of the day-night contrast of
				the extrasolar planet HD 189733b}}}.
	\newblock {\emph{\JournalTitle{Nature}}} \textbf{\bibinfo{volume}{447}},
	\bibinfo{pages}{183--186} (\bibinfo{year}{2007}).
	\newblock \doiprefix 10.1038/nature05782.
	\newblock \eprint{0705.0993}.
	
	\bibitem{2002A&A...385..166S}
	\bibinfo{author}{{Showman}, A.~P.} \& \bibinfo{author}{{Guillot}, T.}
	\newblock \bibinfo{journal}{\bibinfo{title}{{Atmospheric circulation and tides
				of ``51 Pegasus b-like'' planets}}}.
	\newblock {\emph{\JournalTitle{Astronomy and Astrophysics}}}
	\textbf{\bibinfo{volume}{385}}, \bibinfo{pages}{166--180}
	(\bibinfo{year}{2002}).
	\newblock \doiprefix 10.1051/0004-6361:20020101.
	\newblock \eprint{astro-ph/0202236}.
	
	\bibitem{2012ApJ...747...82C}
	\bibinfo{author}{{Cowan}, N.~B.} \emph{et~al.}
	\newblock \bibinfo{journal}{\bibinfo{title}{{Thermal Phase Variations of
				WASP-12b: Defying Predictions}}}.
	\newblock {\emph{\JournalTitle{Astrophysical Journal}}}
	\textbf{\bibinfo{volume}{747}}, \bibinfo{pages}{82} (\bibinfo{year}{2012}).
	\newblock \doiprefix 10.1088/0004-637X/747/1/82.
	\newblock \eprint{1112.0574}.
	
	\bibitem{2012ApJ...754...22K}
	\bibinfo{author}{{Knutson}, H.~A.} \emph{et~al.}
	\newblock \bibinfo{journal}{\bibinfo{title}{{3.6 and 4.5 {$\mu$}m Phase Curves
				and Evidence for Non-equilibrium Chemistry in the Atmosphere of Extrasolar
				Planet HD 189733b}}}.
	\newblock {\emph{\JournalTitle{Astrophysical Journal}}}
	\textbf{\bibinfo{volume}{754}}, \bibinfo{pages}{22} (\bibinfo{year}{2012}).
	\newblock \doiprefix 10.1088/0004-637X/754/1/22.
	\newblock \eprint{1206.6887}.
	
	\bibitem{2013MNRAS.428.2645M}
	\bibinfo{author}{{Maxted}, P.~F.~L.} \emph{et~al.}
	\newblock \bibinfo{journal}{\bibinfo{title}{{Spitzer 3.6 and 4.5 {$\mu$}m
				full-orbit light curves of WASP-18}}}.
	\newblock {\emph{\JournalTitle{Monthly Notices of the Royal Astronomical
				Society}}} \textbf{\bibinfo{volume}{428}}, \bibinfo{pages}{2645--2660}
	(\bibinfo{year}{2013}).
	\newblock \doiprefix 10.1093/mnras/sts231.
	\newblock \eprint{1210.5585}.
	
	\bibitem{2014ApJ...790...53Z}
	\bibinfo{author}{{Zellem}, R.~T.} \emph{et~al.}
	\newblock \bibinfo{journal}{\bibinfo{title}{{The 4.5 {$\mu$}m Full-orbit Phase
				Curve of the Hot Jupiter HD 209458b}}}.
	\newblock {\emph{\JournalTitle{Astrophysical Journal}}}
	\textbf{\bibinfo{volume}{790}}, \bibinfo{pages}{53} (\bibinfo{year}{2014}).
	\newblock \doiprefix 10.1088/0004-637X/790/1/53.
	\newblock \eprint{1405.5923}.
	
	\bibitem{2015ApJ...811..122W}
	\bibinfo{author}{{Wong}, I.} \emph{et~al.}
	\newblock \bibinfo{journal}{\bibinfo{title}{{3.6 and 4.5 {$\mu$}m Phase Curves
				of the Highly Irradiated Eccentric Hot Jupiter WASP-14b}}}.
	\newblock {\emph{\JournalTitle{Astrophysical Journal}}}
	\textbf{\bibinfo{volume}{811}}, \bibinfo{pages}{122} (\bibinfo{year}{2015}).
	\newblock \doiprefix 10.1088/0004-637X/811/2/122.
	\newblock \eprint{1505.03158}.
	
	\bibitem{2016ApJ...823..122W}
	\bibinfo{author}{{Wong}, I.} \emph{et~al.}
	\newblock \bibinfo{journal}{\bibinfo{title}{{3.6 and 4.5 {$\mu$}m Spitzer Phase
				Curves of the Highly Irradiated Hot Jupiters WASP-19b and HAT-P-7b}}}.
	\newblock {\emph{\JournalTitle{Astrophysical Journal}}}
	\textbf{\bibinfo{volume}{823}}, \bibinfo{pages}{122} (\bibinfo{year}{2016}).
	\newblock \doiprefix 10.3847/0004-637X/823/2/122.
	\newblock \eprint{1512.09342}.
	
	\bibitem{2016MNRAS.455.2018D}
	\bibinfo{author}{{Demory}, B.-O.}, \bibinfo{author}{{Gillon}, M.},
	\bibinfo{author}{{Madhusudhan}, N.} \& \bibinfo{author}{{Queloz}, D.}
	\newblock \bibinfo{journal}{\bibinfo{title}{{Variability in the super-Earth 55
				Cnc e}}}.
	\newblock {\emph{\JournalTitle{Monthly Notices of the Royal Astronomical
				Society}}} \textbf{\bibinfo{volume}{455}}, \bibinfo{pages}{2018--2027}
	(\bibinfo{year}{2016}).
	\newblock \doiprefix 10.1093/mnras/stv2239.
	\newblock \eprint{1505.00269}.
	
	\bibitem{2017AJ....153...68S}
	\bibinfo{author}{{Stevenson}, K.~B.} \emph{et~al.}
	\newblock \bibinfo{journal}{\bibinfo{title}{{Spitzer Phase Curve Constraints
				for WASP-43b at 3.6 and 4.5 {$\mu$}m}}}.
	\newblock {\emph{\JournalTitle{Astronomical Journal}}}
	\textbf{\bibinfo{volume}{153}}, \bibinfo{pages}{68} (\bibinfo{year}{2017}).
	\newblock \doiprefix 10.3847/1538-3881/153/2/68.
	\newblock \eprint{1608.00056}.
	
	\bibitem{2014ApJ...790...79R}
	\bibinfo{author}{{Rauscher}, E.} \& \bibinfo{author}{{Kempton}, E.~M.~R.}
	\newblock \bibinfo{journal}{\bibinfo{title}{{The Atmospheric Circulation and
				Observable Properties of Non-synchronously Rotating Hot Jupiters}}}.
	\newblock {\emph{\JournalTitle{Astrophysical Journal}}}
	\textbf{\bibinfo{volume}{790}}, \bibinfo{pages}{79} (\bibinfo{year}{2014}).
	\newblock \doiprefix 10.1088/0004-637X/790/1/79.
	\newblock \eprint{1402.4833}.
	
	\bibitem{2014ApJ...794..132R}
	\bibinfo{author}{{Rogers}, T.~M.} \& \bibinfo{author}{{Komacek}, T.~D.}
	\newblock \bibinfo{journal}{\bibinfo{title}{{Magnetic Effects in Hot Jupiter
				Atmospheres}}}.
	\newblock {\emph{\JournalTitle{Astrophysical Journal}}}
	\textbf{\bibinfo{volume}{794}}, \bibinfo{pages}{132} (\bibinfo{year}{2014}).
	\newblock \doiprefix 10.1088/0004-637X/794/2/132.
	\newblock \eprint{1409.0519}.
	
	\bibitem{2017NatAs...1E.131R}
	\bibinfo{author}{{Rogers}, T.~M.}
	\newblock \bibinfo{journal}{\bibinfo{title}{{Constraints on the magnetic field
				strength of HAT-P-7 b and other hot giant exoplanets}}}.
	\newblock {\emph{\JournalTitle{Nature Astronomy}}}
	\textbf{\bibinfo{volume}{1}}, \bibinfo{pages}{0131} (\bibinfo{year}{2017}).
	\newblock \doiprefix 10.1038/s41550-017-0131.
	\newblock \eprint{1704.06271}.
	
	\bibitem{2013ApJ...776L..25D}
	\bibinfo{author}{{Demory}, B.-O.} \emph{et~al.}
	\newblock \bibinfo{journal}{\bibinfo{title}{{Inference of Inhomogeneous Clouds
				in an Exoplanet Atmosphere}}}.
	\newblock {\emph{\JournalTitle{Astrophysical Journal Letters}}}
	\textbf{\bibinfo{volume}{776}}, \bibinfo{pages}{L25} (\bibinfo{year}{2013}).
	\newblock \doiprefix 10.1088/2041-8205/776/2/L25.
	\newblock \eprint{1309.7894}.
	
	\bibitem{2016ApJ...828...22P}
	\bibinfo{author}{{Parmentier}, V.}, \bibinfo{author}{{Fortney}, J.~J.},
	\bibinfo{author}{{Showman}, A.~P.}, \bibinfo{author}{{Morley}, C.} \&
	\bibinfo{author}{{Marley}, M.~S.}
	\newblock \bibinfo{journal}{\bibinfo{title}{{Transitions in the Cloud
				Composition of Hot Jupiters}}}.
	\newblock {\emph{\JournalTitle{Astrophysical Journal}}}
	\textbf{\bibinfo{volume}{828}}, \bibinfo{pages}{22} (\bibinfo{year}{2016}).
	\newblock \doiprefix 10.3847/0004-637X/828/1/22.
	\newblock \eprint{1602.03088}.
	
	\bibitem{2016A&A...594A..48L}
	\bibinfo{author}{{Lee}, G.}, \bibinfo{author}{{Dobbs-Dixon}, I.},
	\bibinfo{author}{{Helling}, C.}, \bibinfo{author}{{Bognar}, K.} \&
	\bibinfo{author}{{Woitke}, P.}
	\newblock \bibinfo{journal}{\bibinfo{title}{{Dynamic mineral clouds on HD
				189733b. I. 3D RHD with kinetic, non-equilibrium cloud formation}}}.
	\newblock {\emph{\JournalTitle{Astronomy and Astrophysics}}}
	\textbf{\bibinfo{volume}{594}}, \bibinfo{pages}{A48} (\bibinfo{year}{2016}).
	\newblock \doiprefix 10.1051/0004-6361/201628606.
	\newblock \eprint{1603.09098}.
	
	\bibitem{2017arXiv170907459R}
	\bibinfo{author}{{Roman}, M.} \& \bibinfo{author}{{Rauscher}, E.}
	\newblock \bibinfo{journal}{\bibinfo{title}{{Modeling the effects of
				inhomogeneous aerosols on the hot Jupiter Kepler-7b's atmospheric
				circulation}}}.
	\newblock {\emph{\JournalTitle{Astrophysical Journal}}}
	\textbf{\bibinfo{volume}{850}}, \bibinfo{pages}{17} (\bibinfo{year}{2017}).
	\newblock \doiprefix 10.3847/1538-4357/aa8ee4.
	
	\bibitem{2011Guillot}
	\bibinfo{author}{{Guillot}, T.} \& \bibinfo{author}{{Havel}, M.}
	\newblock \bibinfo{journal}{\bibinfo{title}{{An analysis of the CoRoT-2 system:
				a young spotted star and its inflated giant planet}}}.
	\newblock {\emph{\JournalTitle{Astronomy and Astrophysics}}}
	\textbf{\bibinfo{volume}{527}}, \bibinfo{pages}{A20} (\bibinfo{year}{2011}).
	\newblock \doiprefix 10.1051/0004-6361/201015051.
	\newblock \eprint{1010.1078}.
	
	\bibitem{2013ApJ...763...25M}
	\bibinfo{author}{{Moses}, J.~I.}, \bibinfo{author}{{Madhusudhan}, N.},
	\bibinfo{author}{{Visscher}, C.} \& \bibinfo{author}{{Freedman}, R.~S.}
	\newblock \bibinfo{journal}{\bibinfo{title}{{Chemical Consequences of the C/O
				Ratio on Hot Jupiters: Examples from WASP-12b, CoRoT-2b, XO-1b, and HD
				189733b}}}.
	\newblock {\emph{\JournalTitle{Astrophysical Journal}}}
	\textbf{\bibinfo{volume}{763}}, \bibinfo{pages}{25} (\bibinfo{year}{2013}).
	\newblock \doiprefix 10.1088/0004-637X/763/1/25.
	\newblock \eprint{1211.2996}.
	
	\bibitem{2014ApJ...783..113W}
	\bibinfo{author}{{Wilkins}, A.~N.} \emph{et~al.}
	\newblock \bibinfo{journal}{\bibinfo{title}{{The Emergent 1.1-1.7 {$\mu$}m
				Spectrum of the Exoplanet CoRoT-2b as Measured Using the Hubble Space
				Telescope}}}.
	\newblock {\emph{\JournalTitle{Astrophysical Journal}}}
	\textbf{\bibinfo{volume}{783}}, \bibinfo{pages}{113} (\bibinfo{year}{2014}).
	\newblock \doiprefix 10.1088/0004-637X/783/2/113.
	\newblock \eprint{1401.4464}.
	
	\bibitem{2009Alonso}
	\bibinfo{author}{{Alonso}, R.} \emph{et~al.}
	\newblock \bibinfo{journal}{\bibinfo{title}{{The secondary eclipse of the
				transiting exoplanet CoRoT-2b}}}.
	\newblock {\emph{\JournalTitle{Astronomy and Astrophysics}}}
	\textbf{\bibinfo{volume}{501}}, \bibinfo{pages}{L23--L26}
	(\bibinfo{year}{2009}).
	\newblock \doiprefix 10.1051/0004-6361/200912505.
	\newblock \eprint{0906.2814}.
	
	\bibitem{2010A&A...513A..76S}
	\bibinfo{author}{{Snellen}, I.~A.~G.}, \bibinfo{author}{{de Mooij}, E.~J.~W.}
	\& \bibinfo{author}{{Burrows}, A.}
	\newblock \bibinfo{journal}{\bibinfo{title}{{Bright optical day-side emission
				from extrasolar planet CoRoT-2b}}}.
	\newblock {\emph{\JournalTitle{Astronomy and Astrophysics}}}
	\textbf{\bibinfo{volume}{513}}, \bibinfo{pages}{A76} (\bibinfo{year}{2010}).
	\newblock \doiprefix 10.1051/0004-6361/200913338.
	\newblock \eprint{0909.4080}.
	
	\bibitem{2010Alonso}
	\bibinfo{author}{{Alonso}, R.}, \bibinfo{author}{{Deeg}, H.~J.},
	\bibinfo{author}{{Kabath}, P.} \& \bibinfo{author}{{Rabus}, M.}
	\newblock \bibinfo{journal}{\bibinfo{title}{{Ground-based Near-infrared
				Observations of the Secondary Eclipse of CoRoT-2b}}}.
	\newblock {\emph{\JournalTitle{Astronomical Journal}}}
	\textbf{\bibinfo{volume}{139}}, \bibinfo{pages}{1481--1485}
	(\bibinfo{year}{2010}).
	\newblock \doiprefix 10.1088/0004-6256/139/4/1481.
	\newblock \eprint{1001.3060}.
	
	\bibitem{2010Gillon}
	\bibinfo{author}{{Gillon}, M.} \emph{et~al.}
	\newblock \bibinfo{journal}{\bibinfo{title}{{The thermal emission of the young
				and massive planet CoRoT-2b at 4.5 and 8 {$\mu$}m}}}.
	\newblock {\emph{\JournalTitle{Astronomy and Astrophysics}}}
	\textbf{\bibinfo{volume}{511}}, \bibinfo{pages}{A3} (\bibinfo{year}{2010}).
	\newblock \doiprefix 10.1051/0004-6361/200913507.
	\newblock \eprint{0911.5087}.
	
	\bibitem{2011Deming}
	\bibinfo{author}{{Deming}, D.} \emph{et~al.}
	\newblock \bibinfo{journal}{\bibinfo{title}{{Warm Spitzer Photometry of the
				Transiting Exoplanets CoRoT-1 and CoRoT-2 at Secondary Eclipse}}}.
	\newblock {\emph{\JournalTitle{Astrophysical Journal}}}
	\textbf{\bibinfo{volume}{726}}, \bibinfo{pages}{95} (\bibinfo{year}{2011}).
	\newblock \doiprefix 10.1088/0004-637X/726/2/95.
	\newblock \eprint{1011.1019}.
	
	\bibitem{2015MNRAS.449.4192S}
	\bibinfo{author}{{Schwartz}, J.~C.} \& \bibinfo{author}{{Cowan}, N.~B.}
	\newblock \bibinfo{journal}{\bibinfo{title}{{Balancing the energy budget of
				short-period giant planets: evidence for reflective clouds and optical
				absorbers}}}.
	\newblock {\emph{\JournalTitle{Monthly Notices of the Royal Astronomical
				Society}}} \textbf{\bibinfo{volume}{449}}, \bibinfo{pages}{4192--4203}
	(\bibinfo{year}{2015}).
	\newblock \doiprefix 10.1093/mnras/stv470.
	\newblock \eprint{1502.06970}.
	
    \bibitem{2017arXiv170900349D}
	\bibinfo{author}{{Delorme}, P.} \emph{et~al.}
	\newblock \bibinfo{journal}{\bibinfo{title}{{In-depth study of moderately young
				but extremely red, very dusty substellar companion HD206893B}}}.
	\newblock {\emph{\JournalTitle{ArXiv e-prints}}}  (\bibinfo{year}{2017}).
    \newblock \doiprefix 10.1051/0004-6361/201731145
	\newblock \eprint{1709.00349}.
    
	\bibitem{2015ApJ...799..241R}
	\bibinfo{author}{{Rauscher}, E.} \& \bibinfo{author}{{Kempton}, E.~M.~R.}
	\newblock \bibinfo{journal}{\bibinfo{title}{{Erratum: ``The Atmospheric
				Circulation and Observable Properties of Non-synchronously Rotating Hot
				Jupiters''}}}.
	\newblock {\emph{\JournalTitle{Astrophysical Journal}}}
	\textbf{\bibinfo{volume}{799}}, \bibinfo{pages}{241} (\bibinfo{year}{2015}).
	\newblock \doiprefix 10.1088/0004-637X/799/2/241.
	
	\bibitem{2016NatAs...1E...4A}
	\bibinfo{author}{{Armstrong}, D.~J.} \emph{et~al.}
	\newblock \bibinfo{journal}{\bibinfo{title}{{Variability in the atmosphere of
				the hot giant planet HAT-P-7 b}}}.
	\newblock {\emph{\JournalTitle{Nature Astronomy}}}
	\textbf{\bibinfo{volume}{1}}, \bibinfo{pages}{0004} (\bibinfo{year}{2016}).
	\newblock \doiprefix 10.1038/s41550-016-0004.
	\newblock \eprint{1612.04225}.
	
	\bibitem{2017arXiv170905676Y}
	\bibinfo{author}{{Yadav}, R.~K.} \& \bibinfo{author}{{Thorngren}, D.~P.}
	\newblock \bibinfo{journal}{\bibinfo{title}{{Estimating the Magnetic Field
				Strength in Hot Jupiters}}}.
	\newblock {\emph{\JournalTitle{Astrophysical Journal Letters}}}
	\textbf{\bibinfo{volume}{849}}, \bibinfo{pages}{L12} (\bibinfo{year}{2017}).
	\newblock \doiprefix 10.3847/2041-8213/aa93fd
	\newblock \eprint{1709.05676}.
	
	\bibitem{2012ApJ...745..138M}
	\bibinfo{author}{{Menou}, K.}
	\newblock \bibinfo{journal}{\bibinfo{title}{{Magnetic Scaling Laws for the
				Atmospheres of Hot Giant Exoplanets}}}.
	\newblock {\emph{\JournalTitle{Astrophysical Journal}}}
	\textbf{\bibinfo{volume}{745}}, \bibinfo{pages}{138} (\bibinfo{year}{2012}).
	\newblock \doiprefix 10.1088/0004-637X/745/2/138.
	\newblock \eprint{1108.3592}.
	
	\bibitem{2017ApJ...845L..20K}
	\bibinfo{author}{{Kempton}, E.~M.-R.}, \bibinfo{author}{{Bean}, J.~L.} \&
	\bibinfo{author}{{Parmentier}, V.}
	\newblock \bibinfo{journal}{\bibinfo{title}{{An Observational Diagnostic for
				Distinguishing between Clouds and Haze in Hot Exoplanet Atmospheres}}}.
	\newblock {\emph{\JournalTitle{Astrophysical Journal Letters}}}
	\textbf{\bibinfo{volume}{845}}, \bibinfo{pages}{L20} (\bibinfo{year}{2017}).
	\newblock \doiprefix 10.3847/2041-8213/aa84ac.
	\newblock \eprint{1705.05847}.
	
	\bibitem{2016ApJ...829...52F}
	\bibinfo{author}{{Feng}, Y.~K.} \emph{et~al.}
	\newblock \bibinfo{journal}{\bibinfo{title}{{The Impact of Non-uniform Thermal
				Structure on the Interpretation of Exoplanet Emission Spectra}}}.
	\newblock {\emph{\JournalTitle{Astrophysical Journal}}}
	\textbf{\bibinfo{volume}{829}}, \bibinfo{pages}{52} (\bibinfo{year}{2016}).
	\newblock \doiprefix 10.3847/0004-637X/829/1/52.
	\newblock \eprint{1607.03230}.
	
	\bibitem{2010ApJ...714....1A}
	\bibinfo{author}{{Arras}, P.} \& \bibinfo{author}{{Socrates}, A.}
	\newblock \bibinfo{journal}{\bibinfo{title}{{Thermal Tides in Fluid Extrasolar
				Planets}}}.
	\newblock {\emph{\JournalTitle{Astrophysical Journal}}}
	\textbf{\bibinfo{volume}{714}}, \bibinfo{pages}{1--12}
	(\bibinfo{year}{2010}).
	\newblock \doiprefix 10.1088/0004-637X/714/1/1.
	\newblock \eprint{0912.2313}.
	
\setcounter{firstbib}{\value{enumiv}}

\end{thebibliography}

\begin{thebibliography}{10}
		\expandafter\ifx\csname url\endcsname\relax
		\def\url#1{\texttt{#1}}\fi
		\expandafter\ifx\csname urlprefix\endcsname\relax\def\urlprefix{URL }\fi
		\expandafter\ifx\csname doiprefix\endcsname\relax\def\doiprefix{DOI }\fi
		\providecommand{\bibinfo}[2]{#2}
		\providecommand{\eprint}[2][]{\url{#2}}
	\setcounter{enumiv}{\value{firstbib}}

\bibitem{Alonso2008}
\bibinfo{author}{{Alonso}, R.} \emph{et~al.}
\newblock \bibinfo{journal}{\bibinfo{title}{{Transiting exoplanets from the
			CoRoT space mission. II. CoRoT-Exo-2b: a transiting planet around an active G
			star}}}.
\newblock {\emph{\JournalTitle{Astronomy and Astrophysics}}}
\textbf{\bibinfo{volume}{482}}, \bibinfo{pages}{L21--L24}
(\bibinfo{year}{2008}).
\newblock \doiprefix 10.1051/0004-6361:200809431.
\newblock \eprint{0803.3207}.

\bibitem{2004ApJS..154...10F}
\bibinfo{author}{{Fazio}, G.~G.} \emph{et~al.}
\newblock \bibinfo{journal}{\bibinfo{title}{{The Infrared Array Camera (IRAC)
			for the Spitzer Space Telescope}}}.
\newblock {\emph{\JournalTitle{Astrophysical Journal Supplement Series}}}
\textbf{\bibinfo{volume}{154}}, \bibinfo{pages}{10--17}
(\bibinfo{year}{2004}).
\newblock \doiprefix 10.1086/422843.
\newblock \eprint{astro-ph/0405616}.

\bibitem{2004ApJS..154....1W}
\bibinfo{author}{{Werner}, M.~W.} \emph{et~al.}
\newblock \bibinfo{journal}{\bibinfo{title}{{The Spitzer Space Telescope
			Mission}}}.
\newblock {\emph{\JournalTitle{Astrophysical Journal Supplement Series}}}
\textbf{\bibinfo{volume}{154}}, \bibinfo{pages}{1--9} (\bibinfo{year}{2004}).
\newblock \doiprefix 10.1086/422992.
\newblock \eprint{astro-ph/0406223}.

\bibitem{2009A&A...506..501C}
\bibinfo{author}{{Cabrera}, J.} \emph{et~al.}
\newblock \bibinfo{journal}{\bibinfo{title}{{Planetary transit candidates in
			CoRoT-LRc01 field}}}.
\newblock {\emph{\JournalTitle{Astronomy and Astrophysics}}}
\textbf{\bibinfo{volume}{506}}, \bibinfo{pages}{501--517}
(\bibinfo{year}{2009}).
\newblock \doiprefix 10.1051/0004-6361/200912684.

\bibitem{2003yCat.2246....0C}
\bibinfo{author}{{Cutri}, R.~M.} \emph{et~al.}
\newblock \bibinfo{journal}{\bibinfo{title}{{VizieR Online Data Catalog: 2MASS
			All-Sky Catalog of Point Sources (Cutri+ 2003)}}}.
\newblock {\emph{\JournalTitle{VizieR Online Data Catalog}}}
\textbf{\bibinfo{volume}{2246}} (\bibinfo{year}{2003}).

\bibitem{2015Kreidberg}
\bibinfo{author}{{Kreidberg}, L.}
\newblock \bibinfo{journal}{\bibinfo{title}{{batman: BAsic Transit Model
			cAlculatioN in Python}}}.
\newblock {\emph{\JournalTitle{Publications of the Astronomical Society of the
			Pacific}}} \textbf{\bibinfo{volume}{127}}, \bibinfo{pages}{1161--1165}
(\bibinfo{year}{2015}).
\newblock \doiprefix 10.1086/683602.
\newblock \eprint{1507.08285}.

\bibitem{2002MandelAgol}
\bibinfo{author}{{Mandel}, K.} \& \bibinfo{author}{{Agol}, E.}
\newblock \bibinfo{journal}{\bibinfo{title}{{Analytic Light Curves for
			Planetary Transit Searches}}}.
\newblock {\emph{\JournalTitle{Astrophysical Journal Letters}}}
\textbf{\bibinfo{volume}{580}}, \bibinfo{pages}{L171--L175}
(\bibinfo{year}{2002}).
\newblock \doiprefix 10.1086/345520.
\newblock \eprint{astro-ph/0210099}.

\bibitem{Schroter2011}
\bibinfo{author}{{Schr{\"o}ter}, S.} \emph{et~al.}
\newblock \bibinfo{journal}{\bibinfo{title}{{The corona and companion of
			CoRoT-2a. Insights from X-rays and optical spectroscopy}}}.
\newblock {\emph{\JournalTitle{Astronomy and Astrophysics}}}
\textbf{\bibinfo{volume}{532}}, \bibinfo{pages}{A3} (\bibinfo{year}{2011}).
\newblock \doiprefix 10.1051/0004-6361/201116961.
\newblock \eprint{1106.1522}.

\bibitem{2009A&A...493..193L}
\bibinfo{author}{{Lanza}, A.~F.} \emph{et~al.}
\newblock \bibinfo{journal}{\bibinfo{title}{{Magnetic activity in the
			photosphere of CoRoT-Exo-2a. Active longitudes and short-term spot cycle in a
			young Sun-like star}}}.
\newblock {\emph{\JournalTitle{Astronomy and Astrophysics}}}
\textbf{\bibinfo{volume}{493}}, \bibinfo{pages}{193--200}
(\bibinfo{year}{2009}).
\newblock \doiprefix 10.1051/0004-6361:200810591.
\newblock \eprint{0811.0461}.

\bibitem{2008ApJ...678L.129C}
\bibinfo{author}{{Cowan}, N.~B.} \& \bibinfo{author}{{Agol}, E.}
\newblock \bibinfo{journal}{\bibinfo{title}{{Inverting Phase Functions to Map
			Exoplanets}}}.
\newblock {\emph{\JournalTitle{Astrophysical Journal Letters}}}
\textbf{\bibinfo{volume}{678}}, \bibinfo{pages}{L129} (\bibinfo{year}{2008}).
\newblock \doiprefix 10.1086/588553.
\newblock \eprint{0803.3622}.

\bibitem{2005Charbonneau}
\bibinfo{author}{{Charbonneau}, D.} \emph{et~al.}
\newblock \bibinfo{journal}{\bibinfo{title}{{Detection of Thermal Emission from
			an Extrasolar Planet}}}.
\newblock {\emph{\JournalTitle{Astrophysical Journal}}}
\textbf{\bibinfo{volume}{626}}, \bibinfo{pages}{523--529}
(\bibinfo{year}{2005}).
\newblock \doiprefix 10.1086/429991.
\newblock \eprint{astro-ph/0503457}.

\bibitem{2012ApJ...754..136S}
\bibinfo{author}{{Stevenson}, K.~B.} \emph{et~al.}
\newblock \bibinfo{journal}{\bibinfo{title}{{Transit and Eclipse Analyses of
			the Exoplanet HD 149026b Using BLISS Mapping}}}.
\newblock {\emph{\JournalTitle{Astrophysical Journal}}}
\textbf{\bibinfo{volume}{754}}, \bibinfo{pages}{136} (\bibinfo{year}{2012}).
\newblock \doiprefix 10.1088/0004-637X/754/2/136.
\newblock \eprint{1108.2057}.

\bibitem{2016AJ....152...44I}
\bibinfo{author}{{Ingalls}, J.~G.} \emph{et~al.}
\newblock \bibinfo{journal}{\bibinfo{title}{{Repeatability and Accuracy of
			Exoplanet Eclipse Depths Measured with Post-cryogenic Spitzer}}}.
\newblock {\emph{\JournalTitle{Astronomical Journal}}}
\textbf{\bibinfo{volume}{152}}, \bibinfo{pages}{44} (\bibinfo{year}{2016}).
\newblock \doiprefix 10.3847/0004-6256/152/2/44.
\newblock \eprint{1601.05101}.

\bibitem{2017PASP..129a4001S}
\bibinfo{author}{{Schwartz}, J.~C.} \& \bibinfo{author}{{Cowan}, N.~B.}
\newblock \bibinfo{journal}{\bibinfo{title}{{Knot a Bad Idea: Testing BLISS
			Mapping for Spitzer Space Telescope Photometry}}}.
\newblock {\emph{\JournalTitle{Publications of the Astronomical Society of the
			Pacific}}} \textbf{\bibinfo{volume}{129}}, \bibinfo{pages}{014001}
(\bibinfo{year}{2017}).
\newblock \doiprefix 10.1088/1538-3873/129/971/014001.
\newblock \eprint{1607.01013}.

\bibitem{2015ApJ...805..132D}
\bibinfo{author}{{Deming}, D.} \emph{et~al.}
\newblock \bibinfo{journal}{\bibinfo{title}{{Spitzer Secondary Eclipses of the
			Dense, Modestly-irradiated, Giant Exoplanet HAT-P-20b Using Pixel-level
			Decorrelation}}}.
\newblock {\emph{\JournalTitle{Astrophysical Journal}}}
\textbf{\bibinfo{volume}{805}}, \bibinfo{pages}{132} (\bibinfo{year}{2015}).
\newblock \doiprefix 10.1088/0004-637X/805/2/132.
\newblock \eprint{1411.7404}.

\bibitem{2017ApJ...834..187B}
\bibinfo{author}{{Benneke}, B.} \emph{et~al.}
\newblock \bibinfo{journal}{\bibinfo{title}{{Spitzer Observations Confirm and
			Rescue the Habitable-zone Super-Earth K2-18b for Future Characterization}}}.
\newblock {\emph{\JournalTitle{Astrophysical Journal}}}
\textbf{\bibinfo{volume}{834}}, \bibinfo{pages}{187} (\bibinfo{year}{2017}).
\newblock \doiprefix 10.3847/1538-4357/834/2/187.
\newblock \eprint{1610.07249}.

\bibitem{2013PASP..125..306F}
\bibinfo{author}{{Foreman-Mackey}, D.}, \bibinfo{author}{{Hogg}, D.~W.},
\bibinfo{author}{{Lang}, D.} \& \bibinfo{author}{{Goodman}, J.}
\newblock \bibinfo{journal}{\bibinfo{title}{{emcee: The MCMC Hammer}}}.
\newblock {\emph{\JournalTitle{Publications of the Astronomical Society of the
			Pacific}}} \textbf{\bibinfo{volume}{125}}, \bibinfo{pages}{306}
(\bibinfo{year}{2013}).
\newblock \doiprefix 10.1086/670067.
\newblock \eprint{1202.3665}.

\bibitem{2015MNRAS.450.1879E}
\bibinfo{author}{{Espinoza}, N.} \& \bibinfo{author}{{Jord{\'a}n}, A.}
\newblock \bibinfo{journal}{\bibinfo{title}{{Limb darkening and exoplanets:
			testing stellar model atmospheres and identifying biases in transit
			parameters}}}.
\newblock {\emph{\JournalTitle{Monthly Notices of the Royal Astronomical
			Society}}} \textbf{\bibinfo{volume}{450}}, \bibinfo{pages}{1879--1899}
(\bibinfo{year}{2015}).
\newblock \doiprefix 10.1093/mnras/stv744.
\newblock \eprint{1503.07020}.

\bibitem{2013MNRAS.435.2152K}
\bibinfo{author}{{Kipping}, D.~M.}
\newblock \bibinfo{journal}{\bibinfo{title}{{Efficient, uninformative sampling
			of limb darkening coefficients for two-parameter laws}}}.
\newblock {\emph{\JournalTitle{Monthly Notices of the Royal Astronomical
			Society}}} \textbf{\bibinfo{volume}{435}}, \bibinfo{pages}{2152--2160}
(\bibinfo{year}{2013}).
\newblock \doiprefix 10.1093/mnras/stt1435.
\newblock \eprint{1308.0009}.

\bibitem{2017arXiv170903502K}
\bibinfo{author}{{Keating}, D.} \& \bibinfo{author}{{Cowan}, N.~B.}
\newblock \bibinfo{journal}{\bibinfo{title}{{Revisiting the Energy Budget of
			WASP-43b: Enhanced day-night heat transport}}}.
\newblock {\emph{\JournalTitle{Astrophysical Journal Letters}}} \textbf{\bibinfo{volume}{849}}, \bibinfo{pages}{L5}
(\bibinfo{year}{2017}).
\newblock \doiprefix 10.3847/2041-8213/aa8b6b.
\newblock \eprint{1709.03502}.

\bibitem{schwarz1978estimating}
\bibinfo{author}{Schwarz, G.} \emph{et~al.}
\newblock \bibinfo{journal}{\bibinfo{title}{Estimating the dimension of a
		model}}.
\newblock {\emph{\JournalTitle{The annals of statistics}}}.
\textbf{\bibinfo{volume}{6}}, \bibinfo{pages}{461--464}. (\bibinfo{year}{1978}).

\bibitem{2012BIC}
\bibinfo{author}{{Wit}, E.}, \bibinfo{author}{{Heuvel}, E.} \& \bibinfo{author}{{Romeijn}, J.}
\newblock \bibinfo{journal}{\bibinfo{title}{{‘All models are wrong...’: an introduction to model uncertainty}}}.
\newblock {\emph{\JournalTitle{Statistica Neerlandica}}}. \textbf{\bibinfo{volume}{66} \bibinfo{Number}{3}}, \bibinfo{pages}{217--236} (\bibinfo{year}{2012}).

\bibitem{Kass1995}
\bibinfo{author}{{Kass}, R.~E}, \& \bibinfo{author}{{Raftery}, A.~E}.
\newblock \bibinfo{journal}{\bibinfo{title}{Journal of the American Statistical Association}}.
\newblock {\emph{\JournalTitle{Bayes Factors}}}.
\textbf{\bibinfo{volume}{90} \bibinfo{Number}{30}}, \bibinfo{pages}{773--795}
(\bibinfo{year}{1995}).
\newblock \doiprefix 10.2307/2291091.

\bibitem{2017arXiv171007642Z}
\bibinfo{author}{{Zhang}, M.}, \bibinfo{author}{{Knutson}, H. ~A.}, \bibinfo{author}{{Kataria}, T.}, \bibinfo{author}{{Schwartz}, J.~C.}, \bibinfo{author}{{Cowan}, N.~B.},
\bibinfo{author}{{Showman}, A.~P.}, \bibinfo{author}{{Burrows}, A.}, \bibinfo{author}{{Fortney}, J.~J.}, \bibinfo{author}{{Todorov}, K.}, \bibinfo{author}{{Desert}, J.-M.}, \bibinfo{author}{{Agol}, E.} \& \bibinfo{author}{{Deming}, D.}
\newblock \bibinfo{journal}{\bibinfo{title}{{Phase curves of WASP-33b and HD 149026b and a New Correlation Between Phase Curve Offset and Irradiation Temperature}}}.
\newblock {\emph{\JournalTitle{ArXiv e-prints}}}  (\bibinfo{year}{2017}).
\newblock \eprint{1710.07642}.

\bibitem{2012ApJ...757...80C}
\bibinfo{author}{{Cowan}, N.~B.}, \bibinfo{author}{{Voigt}, A.} \&
\bibinfo{author}{{Abbot}, D.~S.}
\newblock \bibinfo{journal}{\bibinfo{title}{{Thermal Phases of Earth-like Planets: Estimating Thermal Inertia from Eccentricity, Obliquity, and Diurnal Forcing}}}.
\newblock {\emph{\JournalTitle{Astrophysical Journal}}}
\textbf{\bibinfo{volume}{757}}, \bibinfo{pages}{80}
(\bibinfo{year}{2012}).
\newblock \doiprefix 10.1088/0004-637X/757/1/80.
\newblock \eprint{1205.5034}.

\bibitem{2011ApJ...729...54C}
\bibinfo{author}{{Cowan}, N.~B.} \& \bibinfo{author}{{Agol}, E.}
\newblock \bibinfo{journal}{\bibinfo{title}{{The Statistics of Albedo and Heat
			Recirculation on Hot Exoplanets}}}.
\newblock {\emph{\JournalTitle{Astrophysical Journal}}}
\textbf{\bibinfo{volume}{729}}, \bibinfo{pages}{54} (\bibinfo{year}{2011}).
\newblock \doiprefix 10.1088/0004-637X/729/1/54.
\newblock \eprint{1001.0012}.

\bibitem{2013ApJ...776..134P}
\bibinfo{author}{{Perez-Becker}, D.} \& \bibinfo{author}{{Showman}, A.~P.}
\newblock \bibinfo{journal}{\bibinfo{title}{{Atmospheric Heat Redistribution on
			Hot Jupiters}}}.
\newblock {\emph{\JournalTitle{Astrophysical Journal}}}
\textbf{\bibinfo{volume}{776}}, \bibinfo{pages}{134} (\bibinfo{year}{2013}).
\newblock \doiprefix 10.1088/0004-637X/776/2/134.
\newblock \eprint{1306.4673}.


\bibitem{2017arXiv170705790S}
\bibinfo{author}{{Schwartz}, J.~C.}, \bibinfo{author}{{Kashner}, Z.},
\bibinfo{author}{{Jovmir}, D.} \& \bibinfo{author}{{Cowan}, N.~B.}
\newblock \bibinfo{journal}{\bibinfo{title}{{Phase Offsets and the Energy
			Budgets of Hot Jupiters}}}.
\newblock {\emph{\JournalTitle{ArXiv e-prints}}}  (\bibinfo{year}{2017}).
\newblock \eprint{1707.05790}.

\bibitem{2013ascl.soft03023S}
\bibinfo{author}{{STScI Development Team}}
\newblock \bibinfo{journal}{\bibinfo{title}{{pysynphot: Synthetic photometry software package}}}.
\newblock (\bibinfo{year}{2013}).
\newblock \eprint{1303.023}.

\bibitem{2014Hansen}
\bibinfo{author}{{Hansen}, C.~J.}, \bibinfo{author}{{Schwartz}, J.~C.} \&
\bibinfo{author}{{Cowan}, N.~B.}
\newblock \bibinfo{journal}{\bibinfo{title}{{Features in the broad-band eclipse
			spectra of exoplanets: signal or noise?}}}
\newblock {\emph{\JournalTitle{Monthly Notices of the Royal Astronomical
			Society}}} \textbf{\bibinfo{volume}{444}}, \bibinfo{pages}{3632--3640}
(\bibinfo{year}{2014}).
\newblock \doiprefix 10.1093/mnras/stu1699.
\newblock \eprint{1402.6699}.

\bibitem{2010ApJ...719.1421P}
\bibinfo{author}{{Perna}, R.}, \bibinfo{author}{{Menou}, K.} \&
\bibinfo{author}{{Rauscher}, E.}
\newblock \bibinfo{journal}{\bibinfo{title}{{Magnetic Drag on Hot Jupiter
			Atmospheric Winds}}}.
\newblock {\emph{\JournalTitle{Astrophysical Journal}}}
\textbf{\bibinfo{volume}{719}}, \bibinfo{pages}{1421--1426}
(\bibinfo{year}{2010}).
\newblock \doiprefix 10.1088/0004-637X/719/2/1421.
\newblock \eprint{1003.3838}.

\bibitem{Numpy}
\bibinfo{author}{Van~der Walt, S.}, \bibinfo{author}{Colbert, C.~C.} \&
\bibinfo{author}{Varoquaux, G.}
\newblock \bibinfo{journal}{\bibinfo{title}{{The NumPy Array: A Structure for
			Efficient Numerical Computation}}}.
\newblock {\emph{\JournalTitle{Computing in Science \& Engineering}}}
\textbf{\bibinfo{volume}{13}}, \bibinfo{pages}{22--30}
(\bibinfo{year}{2011}).
\newblock \doiprefix 10.1109/MCSE.2011.37.

\bibitem{2013A&A...558A..33A}
\bibinfo{author}{{Astropy Collaboration}} \emph{et~al.}
\newblock \bibinfo{journal}{\bibinfo{title}{{Astropy: A community Python
			package for astronomy}}}.
\newblock {\emph{\JournalTitle{Astronomy \& Astrophysics}}}
\textbf{\bibinfo{volume}{558}}, \bibinfo{pages}{A33} (\bibinfo{year}{2013}).
\newblock \doiprefix 10.1051/0004-6361/201322068.
\newblock \eprint{1307.6212}.

\bibitem{Matplotlib}
\bibinfo{author}{Hunter, J.~D.}
\newblock \bibinfo{journal}{\bibinfo{title}{{Matplotlib: A 2D Graphics
			Environment}}}.
\newblock {\emph{\JournalTitle{Computing in Science \& Engineering}}}
\textbf{\bibinfo{volume}{9}}, \bibinfo{pages}{90--95} (\bibinfo{year}{2007}).
\newblock \doiprefix 10.1109/MCSE.2007.58.

\bibitem{corner}
\bibinfo{author}{Foreman-Mackey, D.}
\newblock \bibinfo{journal}{\bibinfo{title}{corner.py: Scatterplot matrices in
		python}}.
\newblock {\emph{\JournalTitle{The Journal of Open Source Software}}}
\textbf{\bibinfo{volume}{24}} (\bibinfo{year}{2016}).
\newblock \urlprefix\url{http://dx.doi.org/10.5281/zenodo.45906}.
\newblock \doiprefix 10.21105/joss.00024.

\bibitem{IPython}
\bibinfo{author}{P\'erez, F.} \&
\bibinfo{author}{Granger, B.~E.}
\newblock \bibinfo{journal}{\bibinfo{title}{{IPython: A System for Interactive
			Scientific Computing}}}.
\newblock {\emph{\JournalTitle{Computing in Science \& Engineering}}}
\textbf{\bibinfo{volume}{9}}, \bibinfo{pages}{21--29} (\bibinfo{year}{2007}).
\newblock \doiprefix 10.1109/MCSE.2007.53.

\setcounter{firstbib}{\value{enumiv}}
\end{thebibliography}

\begin{thebibliography}{10}
		\expandafter\ifx\csname url\endcsname\relax
		\def\url#1{\texttt{#1}}\fi
		\expandafter\ifx\csname urlprefix\endcsname\relax\def\urlprefix{URL }\fi
		\expandafter\ifx\csname doiprefix\endcsname\relax\def\doiprefix{DOI }\fi
		\providecommand{\bibinfo}[2]{#2}
		\providecommand{\eprint}[2][]{\url{#2}}
	\setcounter{enumiv}{\value{firstbib}}

\bibitem{2011MSAIS..16...80B}
\bibinfo{author}{{Borsa}, F.} \& \bibinfo{author}{{Poretti}, E.}
\newblock \bibinfo{journal}{\bibinfo{title}{{REM photometry of the exoplanetary
			system CoRoT-2b}}}.
\newblock {\emph{\JournalTitle{Memorie della Societa Astronomica Italiana
			Supplementi}}} \textbf{\bibinfo{volume}{16}}, \bibinfo{pages}{80}
(\bibinfo{year}{2011}).

\bibitem{2016A&A...595A..89B}
\bibinfo{author}{{Bruno}, G.} \emph{et~al.}
\newblock \bibinfo{journal}{\bibinfo{title}{{Disentangling planetary and
			stellar activity features in the CoRoT-2 light curve}}}.
\newblock {\emph{\JournalTitle{Astronomy and Astrophysics}}}
\textbf{\bibinfo{volume}{595}}, \bibinfo{pages}{A89} (\bibinfo{year}{2016}).
\newblock \doiprefix 10.1051/0004-6361/201527699.
\newblock \eprint{1608.01855}.

\bibitem{2005Mighell}
\bibinfo{author}{{Mighell}, K.~J.}
\newblock \bibinfo{journal}{\bibinfo{title}{{Stellar photometry and astrometry
			with discrete point spread functions}}}.
\newblock {\emph{\JournalTitle{Monthly Notices of the Royal Astronomical
			Society}}} \textbf{\bibinfo{volume}{361}}, \bibinfo{pages}{861--878}
(\bibinfo{year}{2005}).
\newblock \doiprefix 10.1111/j.1365-2966.2005.09208.x.
\newblock \eprint{astro-ph/0505455}.

\bibitem{2014A&A...572A..73L}
\bibinfo{author}{{Lanotte}, A.~A.} \emph{et~al.}
\newblock \bibinfo{journal}{\bibinfo{title}{{A global analysis of Spitzer and
			new HARPS data confirms the loneliness and metal-richness of GJ 436 b}}}.
\newblock {\emph{\JournalTitle{Astronomy and Astrophysics}}}
\textbf{\bibinfo{volume}{572}}, \bibinfo{pages}{A73} (\bibinfo{year}{2014}).
\newblock \doiprefix 10.1051/0004-6361/201424373.
\newblock \eprint{1409.4038}.

\bibitem{2010A&A...510A..25S}
\bibinfo{author}{{Silva-Valio}, A.}, \bibinfo{author}{{Lanza}, A.~F.},
\bibinfo{author}{{Alonso}, R.} \& \bibinfo{author}{{Barge}, P.}
\newblock \bibinfo{journal}{\bibinfo{title}{{Properties of starspots on
			CoRoT-2}}}.
\newblock {\emph{\JournalTitle{Astronomy and Astrophysics}}}
\textbf{\bibinfo{volume}{510}}, \bibinfo{pages}{A25} (\bibinfo{year}{2010}).
\newblock \doiprefix 10.1051/0004-6361/200911904.
\newblock \eprint{0909.4055}.

\bibitem{2013MNRAS.434.2465C}
\bibinfo{author}{{Cowan}, N.~B.}, \bibinfo{author}{{Fuentes}, P.~A.} \&
\bibinfo{author}{{Haggard}, H.~M.}
\newblock \bibinfo{journal}{\bibinfo{title}{{Light curves of stars and
			exoplanets: estimating inclination, obliquity and albedo}}}.
\newblock {\emph{\JournalTitle{Monthly Notices of the Royal Astronomical
			Society}}} \textbf{\bibinfo{volume}{434}}, \bibinfo{pages}{2465--2479}
(\bibinfo{year}{2013}).
\newblock \doiprefix 10.1093/mnras/stt1191.
\newblock \eprint{1304.6398}.

\end{thebibliography}
\end{document}